\documentclass[twocolumn,amsmath,aps,nofootinbib,superscriptaddress]{revtex4-1}
 
\usepackage{amssymb}
\usepackage{amsmath}
\usepackage{epsfig}
\usepackage{subfigure}
\usepackage{mathrsfs}
\usepackage{longtable}
\usepackage{color}
\usepackage[usenames,dvipsnames]{xcolor}
\usepackage{mathtools}
\usepackage{relsize}
\usepackage{array}

\usepackage{hyperref}

\newcommand{\be}{\begin{equation}}
\newcommand{\ee}{\end{equation}}
\newcommand{\bea}{\begin{eqnarray}}
\newcommand{\eea}{\end{eqnarray}}

\newcommand{\Mp}{M_{\rm p}}

\newcommand{\vp}{\varphi}
\newcommand{\vps}{\varphi_*}
\newcommand{\chis}{\chi_*}

\newcommand{\taunl}{ \tau_{\rm NL} }
\newcommand{\gnl}{ g_{\rm NL} }

\newcommand{\fnl}{f_{\rm NL}}

\newcommand{\nz}{n_\zeta}
\newcommand{\Nchi}{N_{,\chi}}
\newcommand{\Nvp}{N_{,\vp}}
\newcommand{\Nchichi}{N_{,\chi\chi}}
\newcommand{\Nvpvp}{N_{,\vp\vp}}
\newcommand{\Nchivp}{N_{,\vp\chi}}
\def\bkone{{\bf k_1}}
\def\bktwo{{\bf k_2}}

\def\picube{(2\pi)^3}
\newcommand{\sdelta}[1]{\!\delta^{\,3}(\mathbf{#1})}

\newcommand{\calP}{{\cal P}}

\newcommand{\g}{\gamma}
\newcommand{\G}{\Gamma}
\newcommand{\Gvp}{\G_\vp}
\newcommand{\Gchi}{\G_\chi}
\newcommand{\p}{\partial}

\newcommand{\Ovpdec}{\Omega_{\vp,{\rm dec}}}
\newcommand{\Ochidec}{\Omega_{\chi,{\rm dec}}}
\newcommand{\drhochidphi}{\frac{\partial\bar\rho_\chi}{\partial\vp_*}}
\newcommand{\drhophidphi}{\frac{\partial\bar\rho_\vp}{\partial\vp_*}}
\newcommand{\drhochidchi}{\frac{\partial\bar\rho_\chi}{\partial\chi_*}}
\newcommand{\drhophidchi}{\frac{\partial\bar\rho_\vp}{\partial\chi_*}}
\newcommand{\ddrhophiddchi}{\frac{\partial^2\bar\rho_\vp}{\partial\chi^2_*}}
\newcommand{\ddrhophiddphi}{\frac{\partial^2\bar\rho_\vp}{\partial\vp^2_*}}
\newcommand{\ddrhochiddchi}{\frac{\partial^2\bar\rho_\chi}{\partial\chi^2_*}}
\newcommand{\ddrhochiddphi}{\frac{\partial^2\bar\rho_\chi}{\partial\vp^2_*}}
\newcommand{\ddrhophidphidchi}{\frac{\partial^2\bar\rho_\vp}{\partial\vp_*\partial\chi_*}}
\newcommand{\ddrhochidphidchi}{\frac{\partial^2\bar\rho_\chi}{\partial\vp_*\partial\chi_*}}

\newcommand{\mcA}{{\mathcal{A}}}
\newcommand{\mcB}{{\mathcal{B}}}
\newcommand{\mctA}{\tilde{\mathcal{A}}}
\newcommand{\mctB}{\stackrel{\sim}{\smash{\mathcal{B}}\rule{0pt}{1.1ex}}}

\usepackage{colortbl}
\makeatletter
	
    \def\CT@@do@color{
      \global\let\CT@do@color\relax
            \@tempdima\wd\z@
            \advance\@tempdima\@tempdimb
            \advance\@tempdima\@tempdimc
    \advance\@tempdimb\tabcolsep
    \advance\@tempdimc\tabcolsep
    \advance\@tempdima2\tabcolsep
            \kern-\@tempdimb
            \leaders\vrule

                    \hskip\@tempdima\@plus  1.7fill
            \kern-\@tempdimc
            \hskip-\wd\z@ \@plus -1.7fill }
    \makeatother

\definecolor{lg}{gray}{0.85}
\begin{document}

\title{Perturbative Reheating After Multiple--Field Inflation: The Impact on Primordial Observables}

\author{Joel Meyers}
\email{jmeyers@cita.utoronto.ca}
\affiliation{Canadian Institute for Theoretical Astrophysics, University of Toronto, Toronto, Ontario M5S 3H8, Canada}

\author{Ewan R. M. Tarrant}
\email{e.tarrant@sussex.ac.uk}
\affiliation{School of Physics and Astronomy, University of Nottingham, University Park, Nottingham, NG7 2RD, UK} 
\affiliation{Astronomy Centre, University of Sussex, Falmer, Brighton BN1 9QH, UK}

\date{\today}

\begin{abstract}
We study the impact of perturbative reheating on primordial observables in models of multiple--field inflation.
By performing a sudden decay calculation, we derive analytic expressions for the local--type non--linearity parameter $\fnl^{\rm local}$, the scalar spectral index $\nz$, and the tensor--to--scalar ratio $r_T$ as functions of the decay rates of the inflationary fields. 
We compare our analytic results to a fully numerical classical field theory simulation, finding excellent agreement.
We find that the sensitivity of $\fnl$, $\nz$, and $r_T$ to the reheating phase depends heavily on the underlying inflationary model. We quantify this sensitivity, and discuss conditions that must be satisfied if observable predictions are to be insensitive to the dynamics of reheating.
We demonstrate that upon completion of reheating, all observable quantities take values within finite ranges, the limits of which are determined completely by the conditions during inflation. Furthermore, fluctuations in both fields play an important role in determining the full dependence of the observables on the dynamics of reheating. 
By applying our formalism to two concrete examples, we demonstrate that variations in $\fnl$, $\nz$, and $r_T$ caused by changes in reheating dynamics are well within the sensitivity of \emph{Planck}, and as such the impact of reheating must be accounted for when making predictions for models of multiple--field inflation. 
Our final expressions are very general, encompassing a wide range of two--field inflationary models, including the standard curvaton scenario. We show that the curvaton scenario is a limiting case of two--field inflation, and recover the standard curvaton results in the appropriate limit.
Our results allow a much more unified approach to studying two--field inflation including the effects of perturbative reheating.  As such, entire classes of models can be studied together, allowing a more systematic approach to gaining insight into the physics of the early universe through observation.
\end{abstract}

\maketitle

%------------------------------------------------------------------
\section{Introduction}

The paradigm of inflation~\cite{Starobinsky:1979ty,Starobinsky:1980te,Kazanas:1980tx,Sato:1980yn,Guth:1980zm} has been spectacularly successful.  It solves some classical problems of the hot Big Bang scenario while also providing a natural mechanism for generating primordial cosmological fluctuations~\cite{Mukhanov:1981xt,Hawking:1982cz,Guth:1982ec,Starobinsky:1982ee,Bardeen:1983qw} with the properties that we observe~\cite{Ade:2013uln}.  General agreement with the broad predictions of the inflationary paradigm has encouraged further investigation into the details of the mechanism.  It is unlikely that data will ever reveal exactly which model of inflation accurately describes our past, however, we can use observation to learn about the physical principles that governed the early universe.  In order to understand what observational data can teach us about the physics underlying inflation, we need to fully understand the detailed predictions of our models of the early universe.

The task of making detailed predictions is simplified in the case of single--field inflation by the fact that the curvature perturbation and its correlation functions are conserved outside the Hubble radius during and after single--field inflation, because it produces purely adiabatic fluctuations \cite{Weinberg:2008zzc,Weinberg:2003sw,Weinberg:2004kr,Weinberg:2008nf}.  We mean by `adiabatic fluctuations' those for which the perturbation to any four--scalar in the system is proportional to the rate of change of the scalar, with the same proportionality for all scalars.  This implies that in the presence of purely adiabatic fluctuations, hypersurfaces of constant individual fluid or field densities correspond to hypersurfaces of constant total energy density.  The situation is more complicated in inflationary models with multiple dynamical fields, because these models naturally produce non--adiabatic fluctuations, whose presence allows the curvature perturbation and its correlation functions to evolve outside the Hubble radius.  Therefore, in order to make predictions in multiple--field models, it is necessary to understand the evolution of the system until the fluctuations become adiabatic, or until they are observed.  The fluctuations produced from multiple--field inflation can become adiabatic if the universe passes through a phase of effectively single--field inflation \cite{Meyers:2010rg,Meyers:2011mm} or through a phase of local thermal and chemical equilibrium with no non--zero conserved quantum numbers \cite{Weinberg:2004kf,Weinberg:2008si,Meyers:2012ni}.  In either case, calculating the evolution of the curvature perturbation outside the Hubble radius is necessary for making firm predictions about primordial observables.

All observations are currently consistent with the simplest single--field, slow--roll models of inflation \cite{Linde:1981mu,Linde:1982zj,Linde:1982uu,Albrecht:1982wi}; however, more complicated scenarios are not ruled out~\cite{Ade:2013uln}.  Even though the observational data does not currently compel us to include multiple dynamical fields in our models of the early universe, there are many reasons for considering the predictions of multiple--field inflation.  

First, the value of observational data is not fully realized without theoretical predictions with which to compare.  To be specific, it is difficult to interpret observational constraints on the local form of the primordial bispectrum, parametrized by $\fnl^{\rm local}$, without a better understanding of the range of $\fnl^{\rm local}$ predicted in physically motivated models of the early universe.  It is known that all single--clock inflationary models predict negligible local non--Gaussianity, but the predictions of multiple--field inflation have not been thoroughly explored.  While there exist a handful of specific examples involving multiple fields that have been worked out in detail, there does not yet exist a coherent framework for understanding what can be learned from observational constraints on local non--Gaussianity.  One aim of this work is to achieve a better understanding of the general lessons that can be learned from constraints on $\fnl^{\rm local}$.

Next, there is theoretical motivation for considering models of inflation with multiple dynamical fields.  Models of the early universe motivated from high--energy theory quite often contain multiple scalar fields.  For example, string theory models typically contain many moduli fields and/or axion fields, several of which could be relevant during inflation.  Additionally, the Higgs field which we know to exist~\cite{Aad:2012tfa,Chatrchyan:2012ufa}, could conceivably be relevant for cosmology.\footnote{The details of how to properly treat the standard model Higgs field and its potential during inflation is outside the scope of this paper.  We raise the point here simply as motivation.}

Future observational data may also force us to consider multiple--field inflation.  There are a number of `consistency relations' that hold in all single--field models of inflation.  The local bispectrum is proportional to the deviation from scale--invariance in all models of `single--clock'\footnote{We distinguish `single--clock' from `single--field' here to allow for exceptions to the non--Gaussianity consistency relation for models with a modified initial state specified at a fixed time in the past~\cite{Agullo:2010ws,Ganc:2011dy,Ganc:2012ae} or non--attractor type behavior~\cite{Namjoo:2012aa,Martin:2012pe,Chen:2013aj,Chen:2013eea} which violate the consistency condition and only involve a single scalar field. In both cases there is an additional relevant time scale in the system.} inflation~\cite{Maldacena:2002vr,Creminelli:2004yq,Ganc:2010ff,RenauxPetel:2010ty}.  Also, the primordial tensor--to--scalar ratio $r_T$ is related to the tensor spectral tilt $n_T$ and the sound speed of scalar fluctuations during inflation $c_s$ in all models of single--field inflation~\cite{Gruzinov:2004jx}.  A future detection of $\fnl^{\rm local}>1$, a violation of the single--field tensor consistency relation\footnote{There is a slight modification of this relation for models of Galilean inflation \cite{Kobayashi:2010cm,Unnikrishnan:2013rka}.} $r_T=-8c_sn_T$, or a detection of non--adiabatic fluctuations would rule out single--field inflation.

Regardless of the inflationary model, the universe must eventually evolve to the hot radiation dominated era of the standard Big Bang. By the time inflation ends the universe is typically in a highly non--thermal state, and so a consistent theory of the early universe must also explain how the cosmos was reheated. The reheating process, which involves a transfer of energy from the inflating field(s) to the standard model particles, can be very complex and may proceed via a number of different mechanisms (see for example~\cite{Dolgov:1982th,Traschen:1990sw,Enqvist:2012vx,Braden:2010wd,Barnaby:2009wr}) depending on the inflationary theory.  In the presence of non--adiabatic fluctuations, the phase of reheating will in general result in evolution of the curvature perturbation on super Hubble scales, and this must be taken into account when considering the predictions of multiple--field inflation.

In Ref.~\cite{Leung:2012ve} it was established, using numerical methods, that the statistics of the primordial curvature perturbation are sensitive to the perturbative decay of the fields that were active during inflation. In particular, the local type non--linearity parameter $\fnl^{\rm local}$ depends on the decay history, and the variations of this parameter are within the sensitivity of cosmic microwave background (CMB) experiments such as \emph{Planck}. It is therefore of the utmost importance to fully understand how reheating affects the statistics of the curvature perturbation in models of multiple--field inflation, such that we can reliably interpret current and future constraints on the primordial observables. 

In this paper, we derive analytic expressions for the local non--linearity parameter $\fnl^{\rm local}$, the scalar spectral index $\nz$, and the tensor--to--scalar ratio $r_T$ as functions of the perturbative decay rates of the fields. In the cases we consider, the impact of reheating is well captured by appealing to the sudden decay approximation~\cite{Lyth:2001nq,Lyth:2005fi,Sasaki:2006kq}. The sudden decay approximation has been used frequently in the past to calculate the statistics of the primordial curvature perturbation for various models of inflation~\cite{Lyth:2005fi,Sasaki:2006kq,Kawasaki:2011pd,Kawasaki:2012gg,Fonseca:2012cj}, the most widely known example being the curvaton scenario~\cite{Linde:1996gt,Enqvist:2001zp,Lyth:2001nq,Lyth:2002my}. In~\cite{Assadullahi:2007uw}, the sudden decay approximation was used to consistently treat the evolution of the curvature perturbation in the presence of two curvaton decays. Numerical studies have shown that for the case of the curvaton, sudden decay reproduces the gradual decay result remarkably well~\cite{Sasaki:2006kq}. Our final expressions for $\fnl^{\rm local}$, $\nz$, and $r_T$ are very general, encompassing a wide range of two--field inflationary models, including the standard curvaton scenario. We show that the curvaton scenario is a limiting case of two--field inflation, and recover the standard curvaton results in the appropriate limit.

\renewcommand*\arraystretch{1.4}
\begin{center}
\begin{table*}[!t]
\hfill{}
\begin{tabular}{c|l|c}
\hline
\hline
Symbol			&	 Meaning   & Reference Eq. \\
\hline
$\Gvp$, $\Gchi$     			&      decay rate of $\vp$ and $\chi$ fields   &  (\ref{eq:fieldEoM})\\
\hline
$R$				& 	$\equiv\Gchi/\Gvp$ (ratio of field decay rates) & (\ref{eq:Rdefn}) \\
\hline
$t_*\,,\,t^{\vp,\chi}_{\rm osc}\,,\,t^{\vp,\chi}_{\rm dec}$		&  key transition times: Hubble exit, start of $\vp$ and $\chi$ oscillations, decay of $\vp$ and $\chi$  & (\ref{eq:numEfoldsDef}), Fig.~\ref{fig:decay_schematic} \\
\hline
$\zeta$			& total curvature perturbation at the end of reheating	&   (\ref{eq:deltaN}) \\
\hline
$\zeta_{\rm dec}$			& $\equiv\zeta(t^\vp_{\rm dec})$ (total curvature perturbation at time $t^\vp_{\rm dec}$)	&   (\ref{eq:zeta1}) \\
\hline
$\zeta_I$ and $\zeta^I_\gamma$			& curvature perturbations on surfaces of constant $\rho_I$ and $\rho^I_\gamma$ with $I=\vp,\chi$    &   (\ref{eq:zeta1}) \\
\hline
$\Omega_{\vp,{\rm osc}}$			& fraction of total energy density stored in $\vp$ field at $t_{\rm osc}$	&   (\ref{eq:Omega_osc}) \\
\hline
$\Omega_{\vp,{\rm dec}}$			& fraction of total energy density stored in $\vp$ field at $t^\vp_{\rm dec}$ in regime $R\le1$	&   (\ref{eq:SD:Ovpdec}) \\
\hline
$\tilde{\Omega}_{\chi,{\rm dec}}$		& fraction of total energy density stored in $\chi$ field at $t^\chi_{\rm dec}$ in regime $R\ge1$	&   (\ref{eq:SD:tildeOmegachi}) \\
\hline
$r$							& $\equiv \left.\frac{3\bar\rho_\chi}{3\bar\rho_\chi + 4\bar\rho_\gamma^\vp}\right|_{t_{\rm dec}^{\chi}}$ in regime $R\le1$. $r\to1$ as $R\to0$.	    &   (\ref{eq:SD:r}) \\
\hline
$\tilde{r}$						&   $\equiv \left.\frac{3\bar\rho_\vp}{3\bar\rho_\vp + 4\bar\rho_\gamma^\chi}\right|_{t_{\rm dec}^{\vp}}$ in regime $R\ge1$. $\tilde{r}\to1$ as $R\to\infty$.	&  (\ref{eq:SD:tilder}) \\
\hline
$\mcA$, $\mcB$				&   reheating functions in regime $R\le1$. $\mcA\to1$ and $\mcB\to0$ as $R\to0$.	&  (\ref{eq:SD:A}), (\ref{eq:SD:B}) \\
\hline
$\mctA$, $\mctB$				&   reheating functions in regime $R\ge1$. $\mctA\to1$ and $\mctB\to0$ as $R\to\infty$.	&  (\ref{eq:SD:mctA}), (\ref{eq:SD:mctB})  \\
\hline
$G,F,K,J$		&  functions which determine $\zeta_I$ from inflation	&  (\ref{eq:f_G}), (\ref{eq:f_F}), (\ref{eq:f_K}), (\ref{eq:f_J})\\
\hline
\end{tabular}
\hfill{}
\caption{Symbols used in this paper.}
\label{tab:symbols}
\end{table*}
\end{center}

Our most important findings are summarized as follows:
\begin{itemize}
\item At the end of reheating, all primordial observables take values within finite ranges, the limits of which are determined completely by the conditions during inflation. The presence of fluctuations in \emph{both} fields are of crucial importance in establishing these bounds. The particular values that the observables acquire within these limits are determined by the reheating parameters.
\item The sensitivity of $\fnl$, $\nz$, and $r_T$ to the reheating phase depends heavily on the underlying inflationary model. We quantify this sensitivity and discuss the conditions that must be satisfied by the inflationary model if its observable predictions are to be insensitive to the physics of reheating.
\item The conditions at Hubble exit play an important role in determining the sensitivity of observables to reheating.
\item Local non--Gaussianity is not in general damped toward small values by reheating, as is often (but not always~\cite{Kim:2010ud}) the case during multiple--field inflation if the adiabatic limit is reached before inflation ends~\cite{Meyers:2010rg,Meyers:2011mm,Watanabe:2011sm}.
\end{itemize}

This paper is organized as follows: In Section~\ref{sec:TheModel}, we briefly recap the $\delta N$ formalism and the elementary theory of perturbative reheating. We present our sudden decay calculation in Section~\ref{sec:SDCalc}, and derive expressions for the observables $\fnl$, $\nz$, and $r_T$ which capture the dynamics of the reheating phase. In Section~\ref{sec:numericalSims} we introduce our numerical simulations and discuss the modifications that need to be made to the sudden decay approximation in order to account for the non--instantaneous nature of the decay, and reconcile the exact numerical results. In Section~\ref{sec:examples} we provide examples of how to apply our formalism to models of multiple--field inflation, and we discuss and conclude in Section~\ref{sec:conclusions}. The expert reader familiar with the elementary theory of perturbative reheating and the $\delta N$ formalism may wish to omit Section~\ref{sec:TheModel}.

Our detailed analysis requires the use of many equations and parameter definitions. As a guide to the reader, in Table~\ref{tab:symbols} we summarize our key parameters and the equations where they are defined or first used.  We will work in units where $\hbar=c=1$.

%------------------------------------------------------------------
\section{The Model}\label{sec:TheModel}

Throughout this paper we focus on canonical two--field inflation models described by the action 
\begin{align}
S_{\rm inf}=&\int d^4 x \sqrt{-g}\left[\Mp^2\frac{R}{2}-\frac12g^{\mu\nu}\p_\mu\vp\p_\nu\vp\right. \nonumber \\
& \left.-\frac12g^{\mu\nu}\p_\mu\chi\p_\nu\chi -W(\vp,\chi)\right]\,.
\label{eq:action}
\end{align}
We demand that the potential be sum--separable, $W(\vp,\chi)=U(\vp)+V(\chi)$, and require that $U(\vp)$ and $V(\chi)$ have quadratic minima, but allow the shape of the potentials far from the minimum to take arbitrary form. The restriction of working with potentials that are sum--separable in the fields considerably simplifies our analysis, as will become clear shortly.
In standard cosmic time, the field equations are given by
\bea
\ddot\vp+(3H+\Gvp)\dot\vp+\p_\vp W &=& 0\,, \nonumber \\
\ddot\chi+(3H+\Gchi)\dot\chi+\p_\chi W &=& 0\,,
\label{eq:fieldEoM}
\eea
where we have included the additional decay terms $\Gvp\dot\vp$ and $\Gvp\dot\chi$, which \emph{parametrize} the perturbative decay of the fields following the inflationary phase. We model these decay products as perfect radiation fluids
\bea
\dot\rho^\vp_\g +4H\rho^\vp_\g &=& \Gvp\rho_\vp\,, \nonumber \\
\dot\rho^\chi_\g +4H\rho^\chi_\g &=& \Gchi\rho_\chi\,,
\label{eq:decay_prods}
\eea
which we assume to be produced in thermal equilibrium. Equations~(\ref{eq:fieldEoM}) and~(\ref{eq:decay_prods}) are subject to the Friedmann constraint
\be
3H^2 \Mp^2 = \rho_\vp + \rho_\chi + \rho^\vp_\g + \rho^\chi_\g \,,
\label{eq:Friedman}
\ee
where $\rho_\vp = \frac12\dot\vp^2 + U(\vp)$ and $\rho_\chi= \frac12\dot\chi^2 + V(\chi)$ are the energy densities of the fields. The decay rates $\Gvp$ and $\Gchi$ (which we assume to be constant throughout this paper) do not follow from the inflationary action, Eq.~(\ref{eq:action}). Instead, they follow from a standard quantum field theory calculation describing the perturbative decay of the fields, and are set to zero throughout the inflationary stage. They are only introduced to the field equations after inflation has ended, and only when the fields are coherently oscillating about their minima. Furthermore, the conditions on the mass of the field $m_\chi\gg{\rm min}\{H,\Gchi\}$ (and similarly for $\vp$) are satisfied as the fields begin their coherent oscillations for all cases that we consider. This parametrization of reheating is often referred to as the `elementary theory of reheating' which was developed in~\cite{Dolgov:1982th,Abbott:1982hn}. 

The perturbative decay of the oscillating fields relies on the assumption that the decay rates $\Gvp$ and $\Gchi$ can be calculated by standard perturbative methods in quantum field theory. However, if the amplitude of the field oscillations are sufficiently large then the perturbative approach fails, and reheating proceeds in a different way, through parametric resonance~\cite{Traschen:1990sw,Kofman:1994rk,Shtanov:1994ce}. The inflaton field typically begins with a stage of explosive production of particles at a stage of a broad parametric resonance. Later the resonance becomes narrow, and finally shuts off altogether. Interactions of particles produced during this stage, their decay into other particles and subsequent thermalization typically require much more time than the preheat stage, since these processes are suppressed by the small values of coupling constants. In many cases, these processes can be described by the elementary theory of reheating. Thus, the elementary theory of reheating can be useful even in the theories where reheating begins at the stage of parametric resonance, or when the amplitude of the field oscillations is small. We finally note that it has been debated whether there exists a regime where the elementary theory of reheating is applicable~\cite{Lawrie:2002wm} at all. Despite these limitations, the elementary theory is appealing due to its simplicity, and we consider it a useful parametrization which allows us to make analytic progress in determining the impact that reheating has on the primordial observables.

If the inflationary model has a known embedding in high energy physics (such as string theory or supergravity) the decay rates $\Gchi$ and $\Gvp$ could in principle be calculated, though in practice these quantities receive complicated corrections due to plasma effects which in general make them time dependent~\cite{Drewes:2013iaa}.  For the purposes of this paper we will treat them as constant free parameters, with the only restriction being that their values must be compatible with cosmological constraints.  One essential constraint on the decay rates comes from the fact that both fields must have decayed and their decay products must have thermalized before Big Bang Nucleosynthesis.  In most scenarios a much more stringent constraint comes from requiring that decay and thermalization must have completed before dark matter decoupling, which presumably happened at a much higher temperature determined by the details of the dark matter model.  This constraint is necessary to guarantee that there do not persist non--adiabatic fluctuations which would allow for further super Hubble evolution of the curvature perturbation, and which are currently not observed in data~\cite{Ade:2013uln}.  Upper limits on the decay rates are set by requiring that the fields do not begin decaying until after the end of inflation.  Observational constraints on the amplitude of primordial tensor fluctuations set a maximum value for the inflationary energy density, which then gives an upper bound on the decay rates of the fields.  As we will see below, observational predictions depend only on the dimensionless ratio of decay rates
\be
R\equiv\frac{\Gchi}{\Gvp}\,.
\label{eq:Rdefn}
\ee 
The most conservative constraints on $R$, without making additional assumptions about dark matter decoupling or the rate of thermalization are given by\footnote{This range is derived by taking $[10^{16}\,\mathrm{GeV}]^4$ as the maximum energy density during inflation which is set by the current observational constraint on primordial tensor modes, and $[1\,\mathrm{MeV}]^4$ as the minimum energy density at the completion of reheating which is set by the energy density at Big Bang Nucleosynthesis.} $10^{-19}\ll R\ll 10^{19}$.  Particular models with a lower scale of inflation, with dark matter decoupling taking place significantly before Big Bang Nucleosynthesis, or a long thermalization period~\cite{Mazumdar:2013gya} will only be viable in a narrower range of $R$.  

These constraints allow values of $R$ which span many orders of magnitude, and hence reheating may take many $e$--foldings to complete.  Such long reheating phases can affect the expansion history after inflation, and as such can affect the region of the inflationary potential to which observations are sensitive. 

%------------------------------------------------------------------
\subsection{$\delta N$ and Observables}\label{sec:deltaN}

The $\delta$N formalism~\cite{Sasaki:1995aw,Sasaki:1998ug,Lyth:2005fi} has been used extensively throughout the literature to compute the primordial curvature perturbation and its statistics. The formalism relates $\zeta$ to the number of $e$--foldings of expansion
\be 
\label{eq:numEfoldsDef}
N({t_*}\,, {t_c})=\int^{t_c}_{t_*} H(t){\rm d}t,
\ee
which is evaluated from an initial flat hypersurface to a final uniform density hypersurface. The perturbation in the number of $e$--foldings, $\delta N$, is the difference between the curvature perturbations on the initial and final hypersurfaces. We take the initial time, denoted by $t_*$, to be Hubble exit during inflation and the final time, denoted by $t_c$, to be a time deep in the radiation dominated era when reheating has completed. The curvature perturbation is then given by~\cite{Lyth:2005fi} (see~\cite{Saffin:2012et} for the covariant approach)
\be
\label{eq:deltaN} 
\zeta=\delta N= \sum_I
N_{,I}\delta\phi_{I*}+\frac12\sum_{IJ}N_{,IJ}\delta\phi_{I*}\delta\phi_{J*}+\cdots\,, 
\ee
where $N,_I=\partial N/(\partial \phi^I_*)$ and the index $I$ runs over $\vp$ and $\chi$.  In general, $N(t_*,t_c)$ depends on the fields, $\phi_I(t)$, and their first time derivatives, $\dot{\phi}_I(t)$, however, if the slow--roll conditions, $3H\dot{\phi}_I \simeq - W_{,I}$, are satisfied at Hubble exit, then $N$ depends only on $\phi_{I*}$.  In order to make use of this formalism, we need to work on super Hubble scales, where the evolution of the universe at each position (the local evolution) is given by the evolution of some unperturbed universe up to small corrections. This is the `separate universe' assumption~\cite{Wands:2000dp,Lyth:2004gb}. The statistics of $\zeta$ can then be evaluated once the evolution of a family of such universes is known.

The decay products $\rho^\vp_\g$ and $\rho^\chi_\g$ remain effectively unperturbed at Hubble exit (since they do not yet exist) and so they do not feature in the Taylor expansion Eq.~(\ref{eq:deltaN}).  Furthermore, since we assume that the decay products $\rho^\vp_\g$ and $\rho^\chi_\g$ are produced in thermal equilibrium, each separate super Hubble sized patch remains perfectly homogeneous. The decay products do however contribute to the expansion rate (and hence to $\zeta$) through Eq.~(\ref{eq:numEfoldsDef}) once the fields $\vp$ and $\chi$ begin to decay -- specifically when $\Gvp$ and $\Gchi$ are allowed to be non--zero in Eqs.~(\ref{eq:fieldEoM}) and~(\ref{eq:decay_prods}) during oscillation of the fields. 

The power spectrum and bispectrum (in Fourier space) are given by
\begin{eqnarray}
\label{eq:powerspectrumdefn} 
\langle\zeta_{\bkone}\zeta_{\bktwo}\rangle &\equiv&
\picube\,
\sdelta{\bkone+\bktwo}\frac{2\pi^2}{k_1^3}\calP_{\zeta}(k_1) \, , \\
\langle\zeta_{{\mathbf k_1}}\,\zeta_{{\mathbf k_2}}\,
\zeta_{{\mathbf k_3}}\rangle &\equiv& \picube\, \sdelta{{\mathbf
k_1}+{\mathbf k_2}+{\mathbf k_3}} B_\zeta( k_1,k_2,k_3) \,. 
\end{eqnarray}
From this we can define three quantities of key observational interest, respectively the scalar spectral index, the tensor--to--scalar ratio, and the non--linearity parameter
\bea 
\label{eq:fnldefn} 
\label{eq:nzeta}
\nz-1&\equiv& \frac{\partial \log\calP_{\zeta}}{\partial\log k}, \\
\label{eq:scaltensR}
r_T&=&\frac{\calP_T}{\calP_{\zeta}}=\frac{8\calP_*}{\Mp^2\calP_{\zeta}}, \\
\label{eq:fnl}
\fnl&=&\frac56\frac{k_1^3k_2^3k_3^3}{k_1^3+k_2^3+k_3^3}
\frac{B_{\zeta}(k_1,k_2,k_3)}{4\pi^4\calP_{\zeta}^2}. 
\eea
Here $\calP_*$ is the power spectrum of the scalar field fluctuations and $\calP_T=8\calP_*/\Mp^2=8H_*^2/(4\pi^2\Mp^2)$ is the power spectrum of the tensor fluctuations. As defined above, $\fnl$ is shape dependent, but it has been shown that the shape dependent part is much less than one~\cite{Seery:2005gb,Maldacena:2002vr} for local non--Gaussianity. In this paper we will primarily be interested in models which can generate an observably large non--Gaussianity, and so we calculate only the shape independent part of $\fnl$, which is local in real space~\cite{Vernizzi:2006ve}. This can, along with $\nz$ and $r_T$, be calculated by appealing to the $\delta N$ formalism,
\bea
{\cal P}_\zeta&=&\sum_I N_{,I}^2 {\cal P}_*,\label{spectrum} \\
n_{\zeta}-1&=& -2\epsilon_* + \frac{2}{H_*}\frac{\sum_{IJ}\dot{\vp}_{J_*} N_{,JI}N_{,I}}{\sum_K N_{,K}^2},\label{index} \\
\fnl&=&\frac56 \frac{\sum_{IJ}N_{,IJ}N_{,I}N_{,J}}{\left(\sum_I N_{,I}^2\right)^2}.
\label{eq:deltaNobservables}
\eea
The tightest constraints on these three observables currently come from the \emph{Planck} CMB experiment~\cite{Ade:2013zuv,Ade:2013ydc}
\bea 
\nz &=&0.9624\pm0.0075\,\,\,\,(95\%\,\,\rm{CL})\,,
 \\
r_T&<&0.12\,\,\,\,(95\%\,\,\rm{CL})\,, \\ 
\fnl^{\textrm{local}} &=& 2.7\pm5.8
\,\,\,\,(68\%\,\,\rm{CL})\,. 
\eea
Looking ahead, complementary probes (such as cluster number counts as seen by the Dark Energy Survey (DES)~\cite{Cunha:2010zz}, or 21 cm measurements~\cite{Lidz:2013tra}) may provide constraints on local--type non--Gaussianity that are competitive with \emph{Planck}.

%------------------------------------------------------------------
\section{Sudden Decay Calculation}\label{sec:SDCalc}

In this section we will present the calculation of the effects of reheating using the sudden decay approximation.  This approximation is motivated by the fact that we are primarily interested in how the energy density of the fields and their decay products change with time.  In the presence of non--adiabatic fluctuations, the curvature perturbation on super Hubble scales evolves whenever the total pressure of the universe is not a function of the total energy density~\cite{Wands:2000dp}.  If all components of the universe have the same equation of state, then the curvature perturbation does not evolve outside the Hubble radius, even though non--adiabatic fluctuations may persist.  To a very a good approximation, this is the case for a period after inflation while both fields are coherently oscillating and before either field has decayed appreciably into radiation.  If one field decays into radiation before the other, then the pressure of the universe is not a function of its total energy density, and the curvature perturbation will evolve outside the Hubble radius during this period.  After the remaining field decays into radiation, all of the components of the universe redshift in the same way, and the curvature perturbation will no longer evolve on super Hubble scales.  If we make the further assumption that the decay products of both fields come to local thermal and chemical equilibrium with no non--zero quantum numbers, then the fluctuations become adiabatic and the curvature perturbation and its correlation functions will be conserved thereafter for all modes which are outside the Hubble radius~\cite{Weinberg:2008si,Meyers:2012ni}. 
This assumption is necessary to ensure that the curvature perturbation does not undergo further evolution during subsequent phases of the universe.  For example, if local thermal equilibrium were not achieved, and non--adiabatic perturbations persisted the curvature perturbation could evolve during and after the phase of dark matter decoupling.  In Ref.~\cite{Huston:2013kgl} the possible survival of a non--adiabatic pressure perturbation during reheating was considered. Even allowing for possibility that the inflationary fields decay into both radiation and matter, the isocurvature mode quickly became negligible in all models that the authors studied; however, this need not always be the case.

While the sudden decay approximation does not properly treat the gradual transition from an oscillating scalar field to a radiation bath, the behavior far from this transition period and thus the resulting ratios of energy densities are well captured by the approximation, and as we will see below, this is enough to quite accurately reproduce the effects of reheating on primordial observables.

\begin{figure}[!t]
\includegraphics[width=8.6cm]{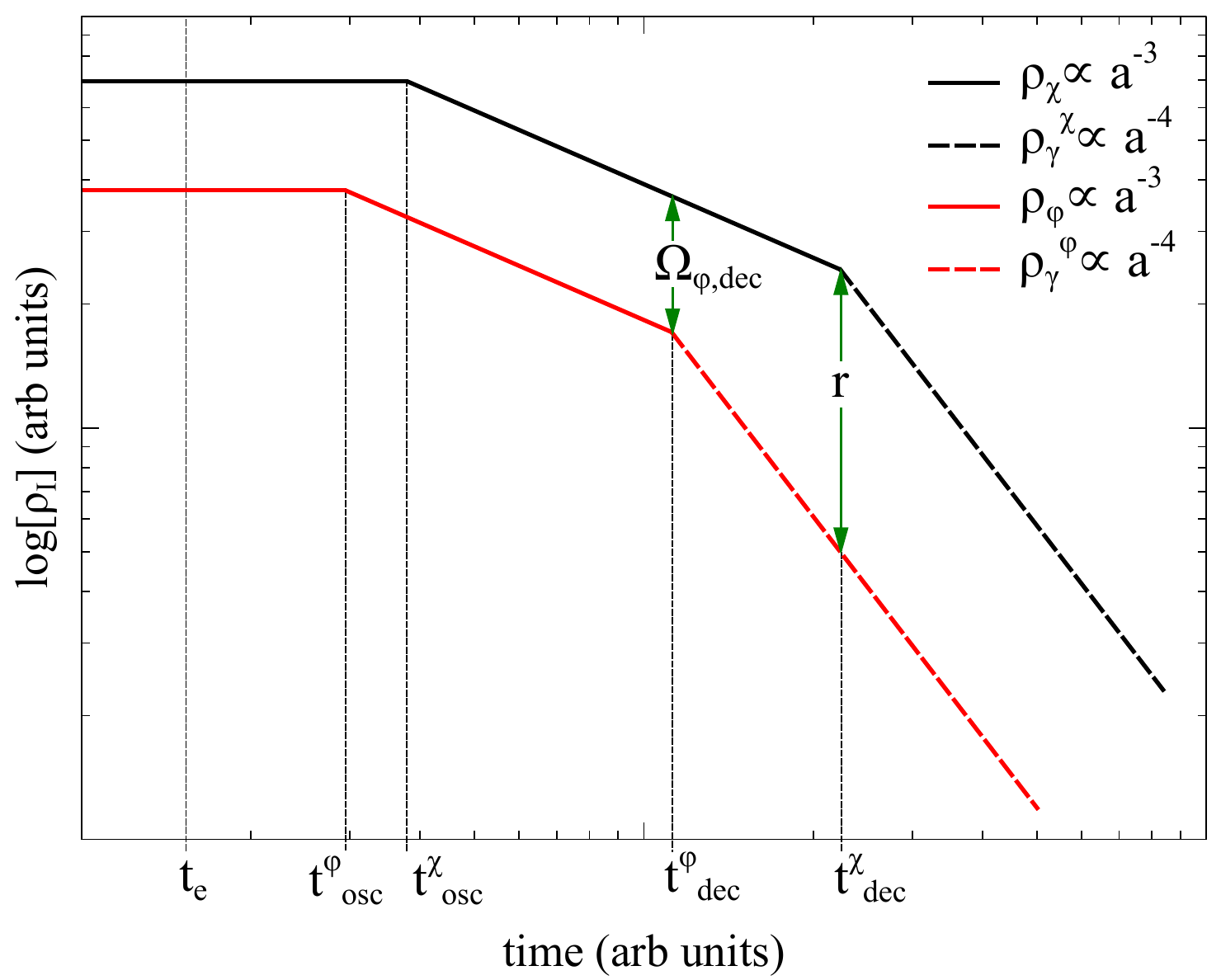}
\caption{A typical sudden decay energy diagram illustrating the dynamics of the reheating stage in the regime $R<1$. We label the end of inflation, $t_{\rm e}$, the start of $\vp$ and $\chi$ oscillations, $t_{\rm osc}^\vp$ and $t_{\rm osc}^\chi$, and the times at which these fields instantly decay, $t_{\rm dec}^\vp$ and $t_{\rm dec}^\chi$. Reheating completes immediately after $t_{\rm dec}^\chi$, when the universe is totally radiation dominated. We also label two important density ratios $\Omega_{\vp,{\rm dec}}\equiv 
\left.\frac{\bar\rho_\vp }{\bar\rho_\vp + \bar\rho_\chi} \right|_{ t_{\rm dec}^{\vp} }$ and $r\equiv \left.\frac{3\bar\rho_\chi}{3\bar\rho_\chi 
+ 4\bar\rho_\gamma^\vp}\right|_{t_{\rm dec}^{\chi}}$, which are first defined in Eqs.~(\ref{eq:SD:Ovpdec}) and~(\ref{eq:SD:r}) respectively.}
\label{fig:decay_schematic}
\end{figure}

Some time after the end of inflation, the fields $\vp$ and $\chi$ oscillate coherently about quadratic minima such that the energy density in each field redshifts like that of pressureless dust when averaged over several oscillations. We label the times at which $\vp$ and $\chi$ begin their oscillations as $t_{\rm osc}^{\vp}$ and $t_{\rm osc}^{\chi}$ respectively. The dynamics of each field are then described by a barotropic fluid with vanishing pressure. We assume that these matter fluids, $\rho_\vp$ and $\rho_\chi$, scale as $a^{-3}$ and do not interact with their decay products until they instantly decay at $t_{\rm dec}^{\vp}$ and $t_{\rm dec}^{\chi}$. These dynamics are illustrated in Fig.~\ref{fig:decay_schematic}.

In the absence of interactions, each fluid with barotropic equation of state, such as radiation or matter, has an individually conserved curvature perturbation~\cite{Lyth:2004gb,Sasaki:2006kq}:
\be
\zeta_I = \delta N +\frac13\int^{\rho_I(t,{\bf x})}_{\bar\rho_I(t)} 
\frac{{\rm d}\tilde\rho_I}{\tilde\rho_I + P_I(\tilde\rho_I)}\,.
\label{eq:SD:zeta_I}
\ee
Here, $\delta N$ is the perturbed amount of expansion, as defined in Eq.~(\ref{eq:deltaN}), and $\rho(t,{\bf x})$ and $P$ denote the local energy density and pressure respectively. In this notation, adiabatic fluctuations are those for which $\zeta_I=\zeta$ for all constituents of the universe.
Fluctuations in the energy density of each fluid (generated by the initial quantum fluctuations of the fields at Hubble exit) lead to a local energy density which is related to the background (unperturbed) density $\bar\rho(t)$ as follows:
\bea
\rho_\vp(t,{\bf x})&=&\bar\rho_\vp(t)+\delta\rho_\vp(t,{\bf x})\,, \nonumber \\
\rho_\chi(t,{\bf x})&=&\bar\rho_\chi(t)+\delta\rho_\chi(t,{\bf x})\,.
\label{eq:SD:fluctuations}
\eea
We will always use a bar to denote homogeneous, unperturbed quantities. 

The underlying assumption of the sudden decay approximation is that these fluids decay \emph{instantly} into radiation when the Hubble rate $H$ becomes equal to the decay rate of each field\footnote{Our results are not affected if instead we assume that each field decays when $H=c\Gamma$ for some constant $c$, since observables depend only on the ratio of decay rates.}:
\be
H(t_{\rm dec}^{\vp}) = \Gvp\,, \qquad  H(t_{\rm dec}^{\chi}) = \Gchi\,.
\label{eq:SD:decaytimesSD}
\ee
The approximation Eq.~(\ref{eq:SD:decaytimesSD}) is most reliable when the decay rates are weak (that is when $\Gvp\ll m_{\vp}$ and $\Gchi\ll m_{\chi}$ where $m_\vp$ and $m_\chi$ are the masses of the fields), such that each field has undergone several coherent oscillations before decaying. As we shall see shortly, it is only the \emph{ratio} of decay rates $R$ (as defined in Eq.~(\ref{eq:Rdefn})) that is important, and so we always choose to work in the weak coupling regime. In the derivation of this section, we restrict ourselves to the regime $t_{\rm dec}^{\chi}>t_{\rm dec}^{\vp}$, that is, the regime where $\vp$ decays before $\chi$, as illustrated in Fig.~\ref{fig:decay_schematic}, which is equivalent to $R<1$. The other regime where $t_{\rm dec}^{\chi}<t_{\rm dec}^{\vp}$ ($R>1$) follows straightforwardly, by exchanging the labels $\vp$ and $\chi$. For our sudden decay calculation to be valid, the first field to decay must do so when the other field is oscillating.  One could modify the formalism slightly to allow for the first field to decay before the onset of oscillations in the other field, but we will not treat that case explicitly here.

A key point in calculating observables in the sudden decay approximation is that the decay hypersurfaces are surfaces of \emph{uniform energy density}. Let us first consider the  $\vp$ decay hypersurface, on which
\be
\rho_\vp(t_{\rm dec}^{\vp},{\bf x}) 
+ \rho_\chi(t_{\rm dec}^{\vp},{\bf x}) = \bar\rho_{\rm tot}(t_{\rm dec}^{\vp})\,.
\label{eq:SD:uniformphi}
\ee
From Eq.~(\ref{eq:SD:decaytimesSD}), we see that $\bar\rho_{\rm tot}(t_{\rm dec}^{\vp})$ must be defined by the condition
\be
3\Mp^2\Gvp^2=\bar\rho_{\rm tot}(t_{\rm dec}^{\vp}) 
= \rho_\vp(t_{\rm dec}^{\vp}) + \rho_\chi(t_{\rm dec}^{\vp})\,.
\label{eq:SD:uniformphi2}
\ee
While the total energy density on the decay surface must be uniform, the individual $\vp$ and $\chi$ densities on the decay surface may be inhomogeneous. From Eq.~(\ref{eq:SD:zeta_I}) we can express the curvature perturbation on uniform $\rho_\vp$ and uniform $\rho_\chi$ hypersurfaces at $t_{\rm dec}^{\vp}$ as 
\bea
\zeta_\vp &=& \zeta_{\rm dec} + \frac13{\rm ln}\left( \frac{\rho_\vp}{\bar\rho_\vp} \right)_{t_{\rm dec}^{\vp}}\,, \nonumber \\
\zeta_\chi &=& \zeta_{\rm dec} + \frac13{\rm ln}\left( \frac{\rho_\chi}{\bar\rho_\chi} \right)_{t_{\rm dec}^{\vp}}\,,
\label{eq:zeta1}
\eea
where $\zeta_{\rm dec}$ is the total curvature perturbation on the decay surface, $\zeta_{\rm dec}\equiv\zeta(t_{\rm dec}^{\vp})=\delta N$. From this point on, all unbarred quantities will have an implicit dependence on position, while barred quantities have no spatial dependence. This allows us to express the energy densities of the fields on the $\vp$ decay hypersurface as
\be
\rho_\vp = \bar\rho_\vp e^{3(\zeta_\vp-\zeta_{\rm dec})}\,, \qquad
\rho_\chi = \bar\rho_\chi e^{3(\zeta_\chi-\zeta_{\rm dec})}\,.
\label{eq:SD:nonlinenergy}
\ee
From the requirement that the total density is uniform on the decay surface, i.e., Eq.~(\ref{eq:SD:uniformphi}), we have
\be
\Omega_{\vp,{\rm dec}}e^{3(\zeta_\vp-\zeta_{\rm dec})} + (1-\Omega_{\vp,{\rm dec}})e^{3(\zeta_\chi-\zeta_{\rm dec})}=1\,,
\label{eq:SD:omega-nonlin}
\ee
where we have defined the quantity 
\be
\Omega_{\vp,{\rm dec}}\equiv 
\left.\frac{\bar\rho_\vp }{\bar\rho_\vp + \bar\rho_\chi} \right|_{ t_{\rm dec}^{\vp} }\,.
\label{eq:SD:Ovpdec}
\ee
The dimensionless density parameter $\Omega_{\vp,{\rm dec}}$ is an important quantity in determining the dynamics of $\zeta$ during reheating. It represents a ratio of energy densities at the time of decay of the $\vp$ field, which is illustrated in the schematic Fig.~\ref{fig:decay_schematic}. The non--linear expression Eq.~(\ref{eq:SD:omega-nonlin}) relates the curvature perturbation on a constant energy hypersurface at the time of decay of $\vp$ to the curvature perturbations on surfaces of constant $\rho_\vp$ and $\rho_\chi$. 

In the sudden decay approximation, the energy density in the field $\vp$ immediately before $t_{\rm dec}^{\vp}$ is converted into radiation immediately after $t_{\rm dec}^{\vp}$, and so we have
\be
\rho_\gamma^{\vp}(t_{\rm dec}^{\vp},{\bf x})=\rho_\vp(t_{\rm dec}^{\vp},{\bf x})\,.
\ee
From Eq.~(\ref{eq:SD:zeta_I}) we can express the curvature perturbation on a surface of constant $\rho_\gamma^\vp$ just after $t_{\rm dec}^{\vp}$ as
\be
\zeta_\gamma^\vp = \zeta_{\rm dec}+ \frac14{\rm ln}\left( \frac{\rho_\gamma^\vp}{\bar\rho_\gamma^\vp} \right)_{t_{\rm dec}^{\vp}}\,,
\ee
which gives
\be
\bar\rho_\gamma^\vp (t_{\rm dec}^{\vp}) e^{4(\zeta_\gamma^\vp-\zeta_{\rm dec})}=
\bar\rho_\vp (t_{\rm dec}^{\vp})e^{3(\zeta_\vp-\zeta_{\rm dec})}\,.
\ee
This relation must be true even in the absence of fluctuations, and so $\bar\rho_\gamma^\vp(t_{\rm dec}^{\vp})=\bar\rho_\vp(t_{\rm dec}^{\vp})$, and we find
\be
\zeta_\gamma^\vp =\frac34\zeta_\vp + \frac14\zeta_{\rm dec}\,.
\label{eq:SD:matching_cond}
\ee
This expression provides the matching condition for the curvature perturbation on surfaces of uniform $\rho_\vp$ and uniform $\rho_\vp^\gamma$ on either side of the decay time $t_{\rm dec}^{\vp}$. From this time until the decay of the $\chi$ field at $t_{\rm dec}^{\chi}$, the universe is filled with two fluids whose energy densities scale differently: $\rho_\gamma^\vp$ which is scaling as radiation, and $\rho_\chi$, which is still scaling like pressureless dust. Hence during this time, $\zeta$ and its statistics will evolve in the presence of non--adiabatic fluctuations.  Remember however that between  $t_{\rm dec}^{\vp}$ and $t_{\rm dec}^{\chi}$, the \emph{individual} curvature perturbations $\zeta_\chi$ and $\zeta_\gamma^\vp$ describing each fluid are conserved. On the $\chi$ decay hypersurface, we have
\be
\rho_\chi(t_{\rm dec}^{\chi},{\bf x})+\rho_\gamma^\vp(t_{\rm dec}^{\chi},{\bf x})=\bar\rho_{\rm tot}(t_{\rm dec}^{\chi})\,,
\label{eq:SD:uniformchi1}
\ee
where $\bar\rho_{\rm tot}(t_{\rm dec}^{\chi})$ is defined by the condition
\be
3\Mp^2\Gchi^2=\bar\rho_{\rm tot}(t_{\rm dec}^{\chi}) 
= \rho_\chi(t_{\rm dec}^{\chi}) + \rho_\gamma^\vp(t_{\rm dec}^{\chi})\,.
\label{eq:SD:uniformchi2}
\ee
The curvature perturbation on uniform $\rho_\chi$ and uniform $\rho_\gamma^\vp$ energy density hypersurfaces read
\bea
\zeta_\chi &=& \zeta + \frac13{\rm ln}\left( \frac{\rho_\chi}{\bar\rho_\chi} \right)_{t_{\rm dec}^{\chi}}\,, \nonumber \\
\zeta_\gamma^\vp &=& \zeta + \frac14{\rm ln}\left( \frac{\rho_\gamma^\vp}{\bar\rho_\gamma^\vp} \right)_{t_{\rm dec}^{\chi}}\,,
\eea
where $\zeta$ is the curvature perturbation on the uniform energy density hypersurface at time $t_{\rm dec}^{\chi}$. It immediately follows that
\be
\rho_\chi = \bar\rho_\chi e^{3(\zeta_\chi-\zeta)}\,, \qquad
\rho_\gamma^\vp = \bar\rho_\gamma^\vp e^{4(\zeta_\gamma^\vp-\zeta)}\,,
\ee
which allows us to express Eq.~(\ref{eq:SD:uniformchi1}) as
\be
(1-\Omega_{\chi,{\rm dec}})e^{4(\zeta_\gamma^\vp-\zeta)} + \Omega_{\chi,{\rm dec}}e^{3(\zeta_\chi-\zeta)}=1\,,
\label{eq:SD:omega-nonlin2}
\ee
where we have defined the quantity 
\be
\Omega_{\chi,{\rm dec}}\equiv 
\left.\frac{\bar\rho_\chi }{\bar\rho_\chi + \bar\rho_\gamma^\vp} \right|_{ t_{\rm dec}^{\chi} }\,.
\label{eq:SD:omegachi2}
\ee
We have, by means of Eq.~(\ref{eq:SD:omega-nonlin2}), a non--linear relation that specifies $\zeta$ in terms of $\zeta_\chi$ and $\zeta_\gamma^\vp$. Since we assume that the decay products $\rho_\gamma^\vp$ and $\rho_\gamma^\chi$ are produced in local thermal and chemical equilibrium with no non--zero conserved quantum numbers, the fluctuations quickly become adiabatic, so $\zeta$ and its correlation functions will subsequently be conserved for all modes outside the Hubble radius \cite{Weinberg:2008si,Meyers:2012ni}. Therefore, the statistics of $\zeta$ appearing in Eq.~(\ref{eq:SD:omega-nonlin2}) are those which are relevant for primordial observables.  Now we can solve Eq.~(\ref{eq:SD:omega-nonlin2}) for $\zeta$, which we do order by order, i.e.,
\be
\zeta = \zeta^{(1)}+\sum_{n=2}^\infty \frac{1}{n!}\zeta^{(n)}\,.
\label{eq:SD:expand}
\ee
Since we are interested in calculating $\fnl$, we will work to second order in $\zeta$. This requires an expression for $\zeta_\gamma^\vp$ at second order, which in turn requires an expression for $\zeta_{\rm dec}$ at second order. 

By expanding Eq.~(\ref{eq:SD:omega-nonlin}) to first and second order we obtain
\be
\zeta_{\rm dec}^{(1)}=\Ovpdec\zeta_\vp^{(1)} + \left(1-\Ovpdec\right)\zeta_\chi^{(1)}\,,
\label{eq:SD:zeta1_1st}
\ee
and
\bea
\zeta_{\rm dec}^{(2)} &=& \Ovpdec\zeta_\vp^{(2)} + \left(1-\Ovpdec\right)\zeta_\chi^{(2)} \nonumber \\
	&+&3\Ovpdec \left( \zeta_\vp^{(1)} - \zeta_{\rm dec}^{(1)} \right)^2 \nonumber \\
	&+&3\left(1-\Ovpdec\right)\left( \zeta_\chi^{(1)} - \zeta_{\rm dec}^{(1)}  \right)^2
\label{eq:SD:zeta1_2nd}
\eea
respectively. These two equations enable us to express $\zeta_\gamma^\vp$ (Eq.~(\ref{eq:SD:matching_cond})) at first and second order as
\be
\zeta_\gamma^{\vp\,{(1)}} = \frac14\left(3+\Ovpdec\right)\zeta_\vp^{(1)}
+\frac14\left(1-\Ovpdec\right)\zeta_\chi^{(1)}\,,
\label{eq:SD:zetar_1st}
\ee
and
\bea
\zeta_\gamma^{\vp\,{(2)}} &=& \frac14\left(3+\Ovpdec\right)\zeta_\vp^{(2)}
+\frac14\left(1-\Ovpdec\right)\zeta_\chi^{(1)} \nonumber \\
  &+&\frac34(\Ovpdec - \Ovpdec^2)\left( \zeta_\vp^{(1)} - \zeta_\chi^{(1)}\right)^2\,,
\label{eq:SD:zetar_2nd}
\eea
respectively. We now expand Eq.~(\ref{eq:SD:omega-nonlin2}), to first order and substitute $\zeta_\gamma^{\vp\,{(1)}}$ from Eq.~(\ref{eq:SD:zetar_1st}) to find
\be
\zeta^{(1)}=(1-\mcA)\zeta_\vp^{(1)}+\mcA\zeta_\chi^{(1)}\,,
\label{eq:SD:zeta_1st}
\ee
where we have defined
\be
\mcA \equiv \frac14(1+3r-\Ovpdec+r\Ovpdec)\,,
\label{eq:SD:A}
\ee
and
\be
r\equiv \left.\frac{3\bar\rho_\chi}{3\bar\rho_\chi 
+ 4\bar\rho_\gamma^\vp}\right|_{t_{\rm dec}^{\chi}}\,.
\label{eq:SD:r}
\ee
The dimensionless density parameter $r$ represents a ratio of energy densities at the time of decay of the second field, which is illustrated in the schematic Fig.~\ref{fig:decay_schematic}.

At second order, after substituting for $\zeta_\gamma^{\vp\,{(1)}}$, $\zeta_\gamma^{\vp\,{(2)}}$ and  $\zeta^{(1)}$, (Eqs.~(\ref{eq:SD:zetar_1st}), ~(\ref{eq:SD:zetar_2nd}),  and~(\ref{eq:SD:zeta_1st}) respectively) we obtain
\be
\zeta^{(2)} = (1-\mcA)\zeta_\vp^{(2)}+\mcA\zeta_\chi^{(2)} 
+ \mcB\left(  \zeta_\vp^{(1)} -  \zeta_\chi^{(1)} \right)^2\,,
\label{eq:SD:zeta_2nd}
\ee
where
\bea
\mcB &\equiv& \frac{1}{16}(1-r)\left[9r(3+r)+\Ovpdec^2(r^2+3r-12) \right. \nonumber \\
&+&\left.6\Ovpdec(r^2+3r+2)\right]\,.
\label{eq:SD:B}
\eea

Eqs.~(\ref{eq:SD:zeta_1st}) and~(\ref{eq:SD:zeta_2nd}) are the expressions for the final conserved curvature perturbation at first and second order at the end of reheating. 
Since $\Ovpdec$ and $r$ are bounded between $0$ and $1$, the functions $\mcA$ and $\mcB$ are bounded between $0$ and $1$ also. 

Our use of the sudden decay approximation to follow the evolution of $\zeta$ is very similar to that found in Ref.~\cite{Assadullahi:2007uw}, where the authors considered the decay of two curvaton fields in a radiation bath. Indeed, if in Ref.~\cite{Assadullahi:2007uw}, one consistently takes the limit where the density of the radiation bath and its perturbation vanishes, Eqs.~(64) and~(96) of~\cite{Assadullahi:2007uw} are equivalent to Eqs.~(\ref{eq:SD:zeta_1st}) and~(\ref{eq:SD:zeta_2nd}) above. 
While our use of the sudden decay approximation is similar to Ref.~\cite{Assadullahi:2007uw}, our subsequent application of the approximation is slightly different. In particular, we place no constraint on the form of the potentials $U(\vp)$ and $V(\chi)$ away from the minimum, and we allow for the onset of oscillations to be influenced by the dynamics of both fields. Furthermore, we consistently account for fluctuations in the dominant field, which is necessary for establishing the maximum and minimum attainable values of primordial observables given a particular inflationary model.
%
%Specifically, in the notation of~\cite{Assadullahi:2007uw}, the coefficients of the individual curvature perturbations associated with the decaying fields are related to our functions $\mcA$ and $\mcB$ in the following way: $r_a=1-\mcA$, $r_b=\mcA$,  $F=1-\mcA$, $G=\mcA$, $\tilde{C}=\tilde{D}=\mcB$, and $E=-2\mcB$.
%
%Our expressions are more complicated than the equivalent result in the standard curvaton scenario, where there is only one decay parameter at the time of curvaton decay, $r_{\rm curv}$, which determines $\zeta$~\cite{Linde:1996gt,Enqvist:2001zp,Lyth:2001nq}.  

Notice that we recover the standard curvaton results in the limit $\Ovpdec\rightarrow 1$ and $\zeta_{\vp}\rightarrow 0$.

The case of $t_{\rm dec}^{\chi}<t_{\rm dec}^{\vp}$ follows through exactly the same way with the roles of $\vp$ and $\chi$ reversed; in this case we find
\begin{align}
	\zeta^{(1)}&=(1-\mctA)\zeta_\chi^{(1)}+\mctA\zeta_\vp^{(1)}\,, \nonumber \\
	\zeta^{(2)}&= (1-\mctA)\zeta_\chi^{(2)}+\mctA\zeta_\vp^{(2)} 
+ \mctB\left(  \zeta_\vp^{(1)} -  \zeta_\chi^{(1)} \right)^2\,.
\label{eq:SD:zeta_tilde}
\end{align}
Here, $\mctA$ and $\mctB$ are defined as
\be
\mctA \equiv \frac14(1+3\tilde r-\tilde\Omega_{\chi,{\rm dec}}+\tilde r\tilde\Omega_{\chi,{\rm dec}})\,,
\label{eq:SD:mctA}
\ee
and
\bea
\mctB &\equiv& \frac{1}{16}(1-\tilde r)\left[9\tilde r(3+\tilde r)
+\tilde\Omega_{\chi,{\rm dec}}^2(\tilde r^2+3\tilde r-12)\right. \nonumber \\
&+&\left.6\tilde\Omega_{\chi,{\rm dec}}(\tilde r^2+3\tilde r+2)\right]\,,
\label{eq:SD:mctB}
\eea
where
\be
\tilde r\equiv \left.\frac{3\bar\rho_\vp}{3\bar\rho_\vp 
+ 4\bar\rho_\gamma^\chi}\right|_{t_{\rm dec}^{\vp}}\,,
\label{eq:SD:tilder}
\ee
and
\be
\tilde\Omega_{\chi,{\rm dec}}\equiv \left.\frac{\bar\rho_\chi}{\bar\rho_\chi + \bar\rho_\vp}\right|_{t_{\rm dec}^{\chi}}\,.
\label{eq:SD:tildeOmegachi}
\ee

Notice that when $R=1$, such that $t^{\vp}_{\rm dec}=t^{\chi}_{\rm dec}$, we have $\tilde\Omega_{\chi,{\rm dec}}=1-\Ovpdec$, and so we find $\mctA=\Ovpdec=(1-\mcA)$ and $\mctB=3\Ovpdec(1-\Ovpdec)=\mcB$.  As a result Eqs.~(\ref{eq:SD:zeta_1st}) and~(\ref{eq:SD:zeta_2nd}) and Eq.~(\ref{eq:SD:zeta_tilde}) are equally valid at $R=1$.

As shown above, the energy ratios $r$ and $\tilde{r}$ play an important role in determining the final value of $\zeta$ after reheating.  On the other hand, these quantities are not input parameters of the theory, and instead are determined by the inflationary potential and the decay rates of the fields.  Within the confines of the sudden decay approximation, $r$ and $\tilde{r}$ can be directly related to the ratio of decay rates $R$, which we will now demonstrate. We find however that this is one area where the sudden decay approximation falls short. We provide more accurate expressions which relate these energy ratios to $R$ in Section~\ref{sec:numericalSims}.

We will focus on the case where $\vp$ decays first, and simply state the result for the opposite case.  We will make use of the parameter $\Ochidec$ which was defined in Eq.~(\ref{eq:SD:omegachi2}) as
\begin{equation}
	\Ochidec\equiv\frac{\bar\rho_\chi }{\bar\rho_\chi + \bar\rho_\gamma^\vp} \Bigg|_{ t_{\rm dec}^{\chi} }\,.
\end{equation}
Using the fact that in the sudden decay approximation, between the time $t_{\rm dec}^{\vp}$ and $t_{\rm dec}^{\chi}$, the energy density of the field $\chi$ redshifts as pressureless dust while the decay products of the field $\vp$ redshift as radiation, this expression can be rewritten as
\begin{equation}
	\Ochidec=\frac{\bar\rho_\chi(t_{\rm dec}^{\vp})\left(\frac{a(t_{\rm dec}^{\vp})}{a(t_{\rm dec}^{\chi})}\right)^3}{\bar\rho_\chi(t_{\rm dec}^{\vp})\left(\frac{a(t_{\rm dec}^{\vp})}{a(t_{\rm dec}^{\chi})}\right)^3+\bar\rho_\vp(t_{\rm dec}^{\vp})\left(\frac{a(t_{\rm dec}^{\vp})}{a(t_{\rm dec}^{\chi})}\right)^4} \, .
\end{equation}
We solve this expression for the ratio of the scale factors at times $t_{\rm dec}^{\vp}$ and $t_{\rm dec}^{\chi}$ which can be rewritten in terms of $r$ as
\begin{equation}\label{eq:ScaleFactorRatio}
	\frac{a(t_{\rm dec}^{\vp})}{a(t_{\rm dec}^{\chi})}=\frac{3(1-r)(1-\Ovpdec)}{4r\Ovpdec} \, .
\end{equation}

Now, since the Hubble rate is equal to $\Gvp$ at time $t_{\rm dec}^{\vp}$ and equal to $\Gchi$ at $t_{\rm dec}^{\chi}$, we can write
\begin{align}
	R^2=\frac{\bar\rho_\chi(t_{\rm dec}^{\vp})\left(\frac{a(t_{\rm dec}^{\vp})}{a(t_{\rm dec}^{\chi})}\right)^3+\bar\rho_\vp(t_{\rm dec}^{\vp})\left(\frac{a(t_{\rm dec}^{\vp})}{a(t_{\rm dec}^{\chi})}\right)^4}{\bar\rho_\chi(t_{\rm dec}^{\vp})+\bar\rho_\vp(t_{\rm dec}^{\vp})} \, ,
\end{align}
and using Eq.~(\ref{eq:ScaleFactorRatio}) gives
\begin{equation}
	R^2=\frac{27(1-\Ovpdec)^4(1-r)^3(3+r)}{256r^4\Ovpdec^3} \quad \rm{for} \quad R\le1 \, .
\label{eq:largeR_SD}
\end{equation}
This equation gives the relation between $R$ and $r$ for the sudden decay approximation in the regime $R\le1$.  The case of $R\ge1$ follows straightforwardly, and gives
\begin{equation}
	R^2=\left[\frac{27(1-\tilde\Omega_{\chi,{\rm dec}})^4(1-\tilde{r})^3(3+\tilde{r})}{256\tilde{r}^4\tilde\Omega_{\chi,{\rm dec}}^3}\right]^{-1} \quad \rm{for} \quad R\ge1 \, .
\label{eq:smallR_SD}
\end{equation}

These expressions are not exact when the fields decay gradually into radiation, however, they show that it is only the ratio of decay rates, and not the individual decay rates, which is important for calculating primordial observables.  A more accurate picture which captures the effect of the gradual decay of the fields is discussed in Section~\ref{sec:numericalSims}, and in that case also, only the ratio of decay rates plays a role for calculating observables.

Eqs.~(\ref{eq:SD:zeta_1st}) and~(\ref{eq:SD:zeta_2nd}) also depend upon the curvature perturbations $\zeta_\vp$ and $\zeta_\chi$ at first and second order, which are seeded by quantum fluctuations of the fields $\vp$ and $\chi$ at Hubble exit.  These quantities generically evolve during multiple--field inflation. For example, if inflation takes place close to a ridge or a valley in the potential $W(\vp,\chi)$, the classical background trajectory may undergo a `turn' in field space, which will cause $\zeta_\vp$ and $\zeta_\chi$ (and hence the total $\zeta$) to evolve outside of the Hubble radius. With appropriate levels of fine--tuning, this mechanism has been shown to generate a large $\fnl^{\rm local}$ during multiple--field inflation~\cite{Byrnes:2008wi}. The curvature perturbations $\zeta_\vp$ and $\zeta_\chi$ will generally continue to evolve until the adiabatic limit is reached, at which point they become equal and conserved~\cite{Meyers:2010rg,Meyers:2011mm,Elliston:2011dr}. Whether conservation is achieved before the end of inflation depends upon the specifics of the inflationary model. However, regardless of these specifics, it is guaranteed that $\zeta_\vp$ and $\zeta_\chi$ will (to a very good approximation) be conserved quantities during the period where both fields are coherently oscillating, and before either field has decayed appreciably into radiation. Within the framework of sudden decay, $\zeta_\vp$ and $\zeta_\chi$ are \emph{individually and exactly conserved} between $t_{\rm osc}^{\vp}$ and $t_{\rm dec}^{\vp}$, and $t_{\rm osc}^{\chi}$ and $t_{\rm dec}^{\chi}$ respectively, and hence it is sufficient to compute them at $t_{\rm osc}^{\vp}$ and $t_{\rm osc}^{\chi}$.

Unfortunately, it is not yet known how to analytically compute $\zeta_\vp$ and $\zeta_\chi$ for arbitrary potentials.  Some progress is possible for models with a separable potential during slow--roll; however, there are significant challenges with non--separable potentials and with moving beyond slow roll within the formalisms that currently exist, such as the $\delta N$ formalism~\cite{Sasaki:1995aw,Sasaki:1998ug,Lyth:2005fi}, moment transport methods~\cite{Mulryne:2009kh}, gradient expansion techniques~\cite{Rigopoulos:2004gr,Rigopoulos:2005xx}, etc.

Despite such technical limitations, we show in the following section that there exists a universal form for $\zeta_\vp$ and $\zeta_\chi$ under rather mild assumptions. Furthermore, the structure of our final expressions for  $\zeta_\vp$ and $\zeta_\chi$ allows us to make some very general statements about the impact of reheating on the primordial observables, without having to calculate their numerical values in all cases.

%------------------------------------------------------------------
\subsection{Analytic Form of $\zeta_\vp^{(1,2)}$ and $\zeta_\chi^{(1,2)}$}\label{sec:zeta12form}

In this section we will show that the forms of $\zeta_\vp^{(1,2)}$ and $\zeta_\chi^{(1,2)}$ are quite universal, even if we do not yet have a method to analytically calculate their numerical values in all cases.  We will assume only that the slow--roll approximation provides a valid description of the evolution within a few $e$--foldings of Hubble exit for the relevant modes.  For the sake of definiteness, let us define a time $t_a$ which is well after Hubble exit for all relevant modes, but early enough such that the slow--roll approximation is still valid for both fields.  We will not need to specify precisely the condition which determines $t_a$, but one could define it as the time when one of the slow--roll parameters becomes larger than some fixed value, for example.

For a potential of the form $W(\vp,\chi)=U(\vp)+V(\chi)$, the number of $e$--foldings from some initially flat hypersurface at time $t_*$ up to a hypersurface defined by $t_a$ is given in the slow--roll approximation by~\cite{Vernizzi:2006ve}
\begin{equation}\label{NaSlowRoll}
	N_a=-\frac{1}{\Mp^2}\int_{\vp_*}^{\vp_a}\frac{U(\vp)}{U'(\vp)}\, d\vp-\frac{1}{\Mp^2}\int_{\chi_*}^{\chi_a}\frac{V(\chi)}{V'(\chi)}\, d\chi \, ,
\end{equation}
and as long as the slow--roll approximation is valid, there exists a conserved quantity
\begin{equation} \label{Cdef}
	C\equiv -\Mp^2\int_{\vp_0}^{\vp}\frac{1}{U'(\vp')}\, d\vp'+\Mp^2\int_{\chi_0}^{\chi}\frac{1}{V'(\chi)} \, d\chi' \, ,
\end{equation}
where $(\vp_0,\chi_0)$ is an arbitrarily chosen point in field space.  By using the slow--roll equations of motion, one can see explicitly that $\dot{C}=0$.  Each trajectory in field space is then labeled by a value of $C$ which can be used to relate initial field values to those at a later time.\footnote{For a fuller discussion of how this conserved quantity is constructed in more general scenarios, see Ref.~\cite{Meyers:2010rg}.}  To do this, we note that we can write
\begin{align}
	d\vp_a&=\frac{d\vp_a}{dC}\left(\frac{\partial C}{\partial \vp_*}d\vp_*+\frac{\partial C}{\partial \chi_*}d\chi_*\right) \, , \nonumber \\
	d\chi_a&=\frac{d\chi_a}{dC}\left(\frac{\partial C}{\partial \vp_*}d\vp_*+\frac{\partial C}{\partial \chi_*}d\chi_*\right) \, .
\end{align}
By differentiating Eq. (\ref{Cdef}) we find
\begin{equation}
	\frac{\partial C}{\partial \vp_*}=-\frac{\Mp^2}{U_*'} \, , \qquad	\frac{\partial C}{\partial \chi_*}=+\frac{\Mp^2}{V_*'} \, .
\end{equation}
Now, let us write the condition which defines the time $t_a$ as
\begin{equation}
	f(\vp_a,\chi_a)=0 \, ,
\end{equation}
where $f(\vp,\chi)$ is some function which is local in the field values.\footnote{For example, we could define $f(\vp,\chi)\equiv z\Mp^2U''-(U+V)$ which would then fix $t_a$ as the time labeling the hypersurface on which $\eta^{\vp}=z$ for some constant $z$.}  Differentiating this condition with respect to $C$, we find
\begin{equation}\label{dphidC}
	\left.\frac{\partial f}{\partial \vp}\right|_a\frac{d \vp_a}{dC}=-\left.\frac{\partial f}{\partial \chi}\right|_a\frac{d \chi_a}{dC} \, .
\end{equation}
Differentiating (\ref{Cdef}) evaluated at $t_a$ with respect to $C$ then gives
\begin{equation}
	1=-\frac{\Mp^2}{U'_a}\frac{d \vp_a}{dC}+\frac{\Mp^2}{V'_a}\frac{d\chi_a}{dC} \, ,
\end{equation}
and using (\ref{dphidC}) allows us to write
\begin{align}
	\frac{d \vp_a}{dC}&=-\Mp^{-2}\left.\frac{\partial f}{\partial \chi}\right|_a\left(\frac{U'_aV'_a}{U'_a\left.\frac{\partial f}{\partial \vp}\right|_a+V'_a\left.\frac{\partial f}{\partial \chi}\right|_a}\right) \, , \nonumber \\
	\frac{d \chi_a}{dC}&=\Mp^{-2}\left.\frac{\partial f}{\partial \vp}\right|_a\left(\frac{U'_aV'_a}{U'_a\left.\frac{\partial f}{\partial \vp}\right|_a+V'_a\left.\frac{\partial f}{\partial \chi}\right|_a}\right) \, .
\end{align}
Putting this together, we find
\begin{align}\label{HCDerivatives}
	\frac{\partial \vp_a}{\partial \vp_*}&=\frac{1}{U_*'}\left.\frac{\partial f}{\partial \chi}\right|_a\left(\frac{U'_aV'_a}{U'_a\left.\frac{\partial f}{\partial \vp}\right|_a+V'_a\left.\frac{\partial f}{\partial \chi}\right|_a}\right) \, , \nonumber \\
	\frac{\partial \vp_a}{\partial \chi_*}&=-\frac{1}{V_*'}\left.\frac{\partial f}{\partial \chi}\right|_a\left(\frac{U'_aV'_a}{U'_a\left.\frac{\partial f}{\partial \vp}\right|_a+V'_a\left.\frac{\partial f}{\partial \chi}\right|_a}\right) \, , \nonumber \\
	\frac{\partial \chi_a}{\partial \vp_*}&=-\frac{1}{U_*'}\left.\frac{\partial f}{\partial \vp}\right|_a\left(\frac{U'_aV'_a}{U'_a\left.\frac{\partial f}{\partial \vp}\right|_a+V'_a\left.\frac{\partial f}{\partial \chi}\right|_a}\right) \, , \nonumber \\
	\frac{\partial \chi_a}{\partial \chi_*}&=\frac{1}{V_*'}\left.\frac{\partial f}{\partial \vp}\right|_a\left(\frac{U'_aV'_a}{U'_a\left.\frac{\partial f}{\partial \vp}\right|_a+V'_a\left.\frac{\partial f}{\partial \chi}\right|_a}\right) \, .
\end{align}
These expressions can be used to calculate the perturbation to the number of $e$-foldings during the slow--roll phase through the following expression
\begin{align}\label{SRDeltaN}
	dN_a=\frac{1}{\Mp^2}&\left[\left(\frac{U}{U'}\right)_*-\frac{\partial \vp_a}{\partial \vp_*}\left(\frac{U}{U'}\right)_a-\frac{\partial \chi_a}{\partial \vp_*}\left(\frac{V}{V'}\right)_a\right]d\vp_* \nonumber \\
	+\frac{1}{\Mp^2}&\left[\left(\frac{V}{V'}\right)_*-\frac{\partial \vp_a}{\partial \chi_*}\left(\frac{U}{U'}\right)_a-\frac{\partial \chi_a}{\partial \chi_*}\left(\frac{V}{V'}\right)_a\right]d\chi_* \, .
\end{align}

In order to calculate observable quantities, we need to derive expressions for $\zeta_\vp$ and $\zeta_\chi$ after the onset of coherent oscillations of each field, which are the curvature perturbations on surfaces of constant $\rho_\vp$ and $\rho_\chi$, respectively.  That is, if $N^\vp$ and $N^\chi$ are the number of $e$--foldings from some initially flat hypersurface to surfaces of constant $\rho_\vp$ and $\rho_\chi$, respectively, then we have $\zeta_\vp=\delta N^\vp$ and $\zeta_\chi=\delta N^\chi$.  These quantities are conserved during the phase of coherent oscillations in the absence of interactions between the fluids up until the decay of each field.  Let us write these quantities as 
\begin{align}
	N^\vp &=N_a+N_B \, , \nonumber \\
	N^\chi &=N_a+N_C  \, ,
\end{align}
where $N_B$ and $N_C$ are defined by 
\begin{align}
	N_B&=\int_{t_a}^{t_{\rm osc}^\vp}H \, dt \, , \nonumber \\
	N_C&=\int_{t_a}^{t_{\rm osc}^\chi}H \, dt \, .
\end{align}
While it is assumed that the slow--roll approximation is valid from Hubble exit to the time $t_a$ such that the quantity $N_a$ is given by Eq. (\ref{NaSlowRoll}), no such assumption is made for the period after the time $t_a$.  On the other hand, regardless of the specific form of $N_B$ and $N_C$, these quantities are determined entirely by the field values at the time $t_a$ and the condition which defines coherent oscillations for each field.  Specifically, there is no explicit dependence on the field values at Hubble exit in $N_B$ and $N_C$, and changes in $\vps$ and $\chis$ affect these quantities only implicitly through the dependence on $\vp_a$ and $\chi_a$.  In other words, we may write
\begin{align}
	\frac{\partial N_B}{\partial \vp_*}&=\frac{\partial N_B}{\partial \vp_a}\frac{\partial \vp_a}{\partial \vp_*}+\frac{\partial N_B}{\partial \chi_a}\frac{\partial \chi_a}{\partial \vp_*} \, , \nonumber \\
	\frac{\partial N_B}{\partial \chi_*}&=\frac{\partial N_B}{\partial \vp_a}\frac{\partial \vp_a}{\partial \chi_*}+\frac{\partial N_B}{\partial \chi_a}\frac{\partial \chi_a}{\partial \chi_*} \, ,
\end{align}
and likewise for $N_C$.  Since $N_B$ and $N_C$ do not depend explicitly on $\vp_*$ and $\chi_*$, the dependence on the conditions at Hubble exit enter only through the derivatives given in Eq. (\ref{HCDerivatives}).

We are now in a position to work out the form of $\zeta_\vp$ and $\zeta_\chi$.  At first order, we have
\begin{align}\label{zetarelations}
	\zeta_\vp^{(1)}=&\delta N^{\vp(1)}=\left[\frac{\partial N_a}{\partial \vp_*}+\frac{\partial N_B}{\partial \vp_*}\right]\delta \vp_* + \left[\frac{\partial N_a}{\partial \chi_*}+\frac{\partial N_B}{\partial \chi_*}\right]\delta \chi_* \, , \nonumber \\
	\zeta_\chi^{(1)}=&\delta N^{\chi(1)}=\left[\frac{\partial N_a}{\partial \vp_*}+\frac{\partial N_C}{\partial \vp_*}\right]\delta \vp_* + \left[\frac{\partial N_a}{\partial \chi_*}+\frac{\partial N_C}{\partial \chi_*}\right]\delta \chi_* \, ,
\end{align}
and using Eqs. (\ref{SRDeltaN}) and (\ref{HCDerivatives}) this can be written as
\begin{align}\label{FirstOrderPerturbations}
	\zeta_\vp^{(1)}=&\Bigg[\frac{1}{\Mp^2}\left[\left(\frac{U}{U'}\right)_*-\frac{\partial \vp_a}{\partial \vp_*}\left(\frac{U}{U'}\right)_a-\frac{\partial \chi_a}{\partial \vp_*}\left(\frac{V}{V'}\right)_a\right] \nonumber \\
	&+\left[\frac{\partial N_B}{\partial \vp_a}\frac{\partial \vp_a}{\partial \vp_*}+\frac{\partial N_B}{\partial \chi_a}\frac{\partial \chi_a}{\partial \vp_*}\right]\Bigg]\delta\vp_* \nonumber \\
	&+\Bigg[\frac{1}{\Mp^2}\left[\left(\frac{V}{V'}\right)_*-\frac{\partial \vp_a}{\partial \chi_*}\left(\frac{U}{U'}\right)_a-\frac{\partial \chi_a}{\partial \chi_*}\left(\frac{V}{V'}\right)_a\right] \nonumber \\
	&+\left[\frac{\partial N_B}{\partial \vp_a}\frac{\partial \chi_a}{\partial \chi_*}+\frac{\partial N_B}{\partial \chi_a}\frac{\partial \chi_a}{\partial \chi_*}\right]\Bigg]\delta\chi_* \nonumber \\
	=&\frac{1}{\Mp^2}\left[\frac{U_*}{U_*'}+\frac{1}{U_*'}G\right]\delta\vp_*+\frac{1}{\Mp^2}\left[\frac{V_*}{V_*'}-\frac{1}{V_*'}G\right]\delta\chi_* \, , \nonumber \\
	\zeta_\chi^{(1)}=&\Bigg[\frac{1}{\Mp^2}\left[\left(\frac{U}{U'}\right)_*-\frac{\partial \vp_a}{\partial \vp_*}\left(\frac{U}{U'}\right)_a-\frac{\partial \chi_a}{\partial \vp_*}\left(\frac{V}{V'}\right)_a\right] \nonumber \\
	&+\left[\frac{\partial N_C}{\partial \vp_a}\frac{\partial \vp_a}{\partial \vp_*}+\frac{\partial N_C}{\partial \chi_a}\frac{\partial \chi_a}{\partial \vp_*}\right]\Bigg]\delta\vp_* \nonumber \\
	&+\Bigg[\frac{1}{\Mp^2}\left[\left(\frac{V}{V'}\right)_*-\frac{\partial \vp_a}{\partial \chi_*}\left(\frac{U}{U'}\right)_a-\frac{\partial \chi_a}{\partial \chi_*}\left(\frac{V}{V'}\right)_a\right] \nonumber \\
	&+\left[\frac{\partial N_C}{\partial \vp_a}\frac{\partial \chi_a}{\partial \chi_*}+\frac{\partial N_C}{\partial \chi_a}\frac{\partial \chi_a}{\partial \chi_*}\right]\Bigg]\delta\chi_* \nonumber \\
	=&\frac{1}{\Mp^2}\left[\frac{U_*}{U_*'}+\frac{1}{U_*'}F\right]\delta\vp_*+\frac{1}{\Mp^2}\left[\frac{V_*}{V_*'}-\frac{1}{V_*'}F\right]\delta\chi_* \, ,
\end{align}
where we have made the definitions
\begin{align}
	G&=U_*'\left[\frac{\partial \vp_a}{\partial \vp_*}\left[\Mp^2\frac{\partial N_B}{\partial \vp_a}-\left(\frac{U}{U'}\right)_a\right] \right. \nonumber \\
	&\left.+\frac{\partial \chi_a}{\partial \vp_*}\left[\Mp^2\frac{\partial N_B}{\partial \chi_a}-\left(\frac{V}{V'}\right)_a\right]\right] \nonumber \\
	&=-V_*'\left[\frac{\partial \vp_a}{\partial \chi_*}\left[\Mp^2\frac{\partial N_B}{\partial \vp_a}-\left(\frac{U}{U'}\right)_a\right] \right. \nonumber \\
	&\left.+\frac{\partial \chi_a}{\partial \chi_*}\left[\Mp^2\frac{\partial N_B}{\partial \chi_a}-\left(\frac{V}{V'}\right)_a\right]\right] \, , \label{eq:f_G}\\
	F&=U_*'\left[\frac{\partial \vp_a}{\partial \vp_*}\left[\Mp^2\frac{\partial N_C}{\partial \vp_a}-\left(\frac{U}{U'}\right)_a\right] \right. \nonumber \\
	&\left.+\frac{\partial \chi_a}{\partial \vp_*}\left[\Mp^2\frac{\partial N_C}{\partial \chi_a}-\left(\frac{V}{V'}\right)_a\right]\right] \nonumber \\
	&=-V_*'\left[\frac{\partial \vp_a}{\partial \chi_*}\left[\Mp^2\frac{\partial N_C}{\partial \vp_a}-\left(\frac{U}{U'}\right)_a\right] \right. \nonumber \\
	&\left.+\frac{\partial \chi_a}{\partial \chi_*}\left[\Mp^2\frac{\partial N_C}{\partial \chi_a}-\left(\frac{V}{V'}\right)_a\right]\right]  \label{eq:f_F}\, ,
\end{align}
and we have used the relations
\begin{equation}
	\frac{\partial \vp_a}{\partial \vp_*}=\frac{\partial \vp_a}{\partial \chi_*}\left(-\frac{V_*'}{U_*'}\right) \, , \qquad \frac{\partial \chi_a}{\partial \vp_*}=\frac{\partial \chi_a}{\partial \chi_*}\left(-\frac{V_*'}{U_*'}\right) \, , 
\end{equation}
which can be seen from Eq. (\ref{HCDerivatives}).  It is then straightforward to calculate these quantities at second order
\begin{align}\label{SecondOrderPerturbations}
	\zeta_\vp^{(2)}=&\frac{1}{\Mp^2}\left[1-\frac{U_*U_*''}{{U_*'}^2}-\frac{U_*''}{{U_*'}^2}G+\frac{1}{{U_*'}^2}K\right]\delta\vp_*^2 \nonumber \\
	&+\frac{1}{\Mp^2}\left[1-\frac{V_*V_*''}{{V_*'}^2}+\frac{V_*''}{{V_*'}^2}G+\frac{1}{{V_*'}^2}K\right]\delta\chi_*^2 \nonumber \\
	&+\frac{1}{\Mp^2}\left[-\frac{2}{U_*'V_*'}K\right]\delta\vp_*\delta\chi_* \, , \nonumber \\
	\zeta_\chi^{(2)}=&\frac{1}{\Mp^2}\left[1-\frac{U_*U_*''}{{U_*'}^2}-\frac{U_*''}{{U_*'}^2}F+\frac{1}{{U_*'}^2}J\right]\delta\vp_*^2 \nonumber \\
	&+\frac{1}{\Mp^2}\left[1-\frac{V_*V_*''}{{V_*'}^2}+\frac{V_*''}{{V_*'}^2}F+\frac{1}{{V_*'}^2}J\right]\delta\chi_*^2 \nonumber \\ &+\frac{1}{\Mp^2}\left[-\frac{2}{U_*'V_*'}J\right]\delta\vp_*\delta\chi_* \, ,
\end{align}
where the quantities $K$ and $J$ are defined by
\begin{align}
	K\equiv&U_*'\left(\frac{\partial G}{\partial \vp_a}\frac{\partial \vp_a}{\partial \vp_*}+\frac{\partial G}{\partial \chi_a}\frac{\partial \chi_a}{\partial \vp_*}\right) \nonumber \\
	=&-V_*'\left(\frac{\partial G}{\partial \vp_a}\frac{\partial \vp_a}{\partial \chi_*}+\frac{\partial G}{\partial \chi_a}\frac{\partial \chi_a}{\partial \chi_*}\right)\,, \label{eq:f_K} \\
	J\equiv&U_*'\left(\frac{\partial F}{\partial \vp_a}\frac{\partial \vp_a}{\partial \vp_*}+\frac{\partial F}{\partial \chi_a}\frac{\partial \chi_a}{\partial \vp_*}\right) \nonumber \\
	=&-V_*'\left(\frac{\partial F}{\partial \vp_a}\frac{\partial \vp_a}{\partial \chi_*}+\frac{\partial F}{\partial \chi_a}\frac{\partial \chi_a}{\partial \chi_*}\right)  \label{eq:f_J}\, .
\end{align}

It is not possible to calculate the values of $F$, $G$, $J$, and $K$ without specifying a particular model. However, the forms of $\zeta_\vp^{(1,2)}$ and $\zeta_\chi^{(1,2)}$ are given by Eqs. (\ref{FirstOrderPerturbations}) and (\ref{SecondOrderPerturbations}) for all models with a potential of the form $W(\vp,\chi)=U(\vp)+V(\chi)$ as long as the slow--roll approximation holds for at least a short period following Hubble exit.  We will be able to use this information in the following sections to learn some general lessons about the effects of reheating on primordial observables.

One case in which the functions $F$, $G$, $J$, and $K$ can be computed is when the adiabatic limit is achieved during slow--roll inflation~\cite{Meyers:2010rg}.  When this occurs, the time $t_a$ can be chosen to define a hypersurface of constant energy density which is also a hypersurface of constant $\vp$ and constant $\chi$.  In this case, $\zeta_\vp=\zeta_\chi$, and primordial observables will not be sensitive to the details of reheating.

%---------------------------------------------------------

\subsection{Energy Densities and Curvature Perturbations}\label{sec:EnergyDensities}

In this section we will show how the curvature perturbations $\zeta_\vp^{(1,2)}$ and $\zeta_\chi^{(1,2)}$ are related to perturbations in the energy density.  We will work out how the functions $F$, $G$, $J$, and $K$ defined above are related to derivatives of the energy densities of the oscillating fields, which will be useful for comparing analytic results with numerical simulations, and also as a method for calculating the curvature perturbations in the analytic example presented in Section~\ref{sec:analytic_example}.

Recall from Eq.~(\ref{eq:SD:zeta_I}) that on a flat hypersurface (where $\delta N=0$), while both fields are coherently oscillating, the energy densities in each of the fields are given by
\begin{equation}\label{eq:FlatHypersurface}
	\rho_\vp=\bar\rho_\vp e^{3\zeta_{\vp}}\, , \qquad 	\rho_\chi=\bar\rho_\chi e^{3\zeta_{\chi}} \, .
\end{equation}
Expanding these expressions in terms of $\delta\vp_*$ and $\delta\chi_*$ gives
\begin{align}\label{ZetaDensity1}
	\zeta_\vp^{(1)}&=\frac{1}{3\bar\rho_\vp}\left(\drhophidphi\delta\vp_*+\drhophidchi\delta\chi_*\right) \, , \nonumber \\
	\zeta_\chi^{(1)}&=\frac{1}{3\bar\rho_\chi}\left(\drhochidphi\delta\vp_*+\drhochidchi\delta\chi_*\right) \, ,
\end{align}
at first order, and
\begin{align}
	&\ddrhophiddphi\delta\vp_*^2+\ddrhophiddchi\delta\chi_*^2 +2\ddrhophidphidchi\delta\vp_*\delta\chi_* \nonumber \\
	&\qquad=9\bar\rho_\vp{\zeta_\vp^{(1)}}^2+3\bar\rho_\vp\zeta_\vp^{(2)}  \, , \nonumber \\
	&\ddrhochiddphi\delta\vp_*^2+\ddrhochiddchi\delta\chi_*^2 +2\ddrhochidphidchi\delta\vp_*\delta\chi_* \nonumber \\
	&\qquad=9\bar\rho_\chi{\zeta_\chi^{(1)}}^2+3\bar\rho_\chi\zeta_\chi^{(2)}  \, ,
\end{align}
at second order, which by using Eq.~(\ref{ZetaDensity1}) can be put into the form
\begin{align}
	\zeta_\vp^{(2)}	&=\frac{1}{3\bar\rho_\vp}\left[\left(\ddrhophiddphi-\frac{1}{\bar\rho_\vp}\left(\drhophidphi\right)^2\right)\delta\vp_*^2 \right. \nonumber \\
	& \qquad +\left(\ddrhophiddchi-\frac{1}{\bar\rho_\vp}\left(\drhophidchi\right)^2\right)\delta\chi_*^2 \nonumber \\
	& \qquad \left.+2\left(\ddrhophidphidchi-\frac{1}{\bar\rho_\vp}\drhophidphi\drhophidchi\right)\delta\vp_*\delta\chi_*\right] \, , \nonumber \\
	\zeta_\chi^{(2)}&=\frac{1}{3\bar\rho_\chi}\left[\left(\ddrhochiddphi-\frac{1}{\bar\rho_\chi}\left(\drhochidphi\right)^2\right)\delta\vp_*^2 \right. \nonumber \\
	& \qquad +\left(\ddrhochiddchi-\frac{1}{\bar\rho_\chi}\left(\drhochidchi\right)^2\right)\delta\chi_*^2 \nonumber \\
	& \qquad \left.+2\left(\ddrhochidphidchi-\frac{1}{\bar\rho_\chi}\drhochidphi\drhochidchi\right)\delta\vp_*\delta\chi_*\right] \, .
\end{align}
Comparing these expressions with Eqs. (\ref{FirstOrderPerturbations}) and (\ref{SecondOrderPerturbations}), we find
\begin{align}\label{densityderivatives1}
		\frac{1}{3\bar\rho_\vp}\drhophidphi&=\frac{1}{\Mp^2}\left[\frac{U_*}{U_*'}+\frac{1}{U_*'}G\right] \, , \nonumber \\
		\frac{1}{3\bar\rho_\vp}\drhophidchi&=\frac{1}{\Mp^2}\left[\frac{V_*}{V_*'}-\frac{1}{V_*'}G\right] \, , \nonumber \\
		\frac{1}{3\bar\rho_\chi}\drhochidphi&=\frac{1}{\Mp^2}\left[\frac{U_*}{U_*'}+\frac{1}{U_*'}F\right] \, , \nonumber \\
		\frac{1}{3\bar\rho_\chi}\drhochidchi&=\frac{1}{\Mp^2}\left[\frac{V_*}{V_*'}-\frac{1}{V_*'}F\right] \, ,
\end{align}
and
\begin{align}\label{densityderivatives2}
	&\frac{1}{3\bar\rho_\vp}\left(\ddrhophiddphi-\frac{1}{\bar\rho_\vp}\left(\drhophidphi\right)^2\right) \nonumber \\
	& \qquad=\frac{1}{\Mp^2}\left[1-\frac{U_*U_*''}{{U_*'}^2}-\frac{U_*''}{{U_*'}^2}G+\frac{1}{{U_*'}^2}K\right] \, , \nonumber \\
	&\frac{1}{3\bar\rho_\vp}\left(\ddrhophiddchi-\frac{1}{\bar\rho_\vp}\left(\drhophidchi\right)^2\right) \nonumber \\
	& \qquad=\frac{1}{\Mp^2}\left[1-\frac{V_*V_*''}{{V_*'}^2}+\frac{V_*''}{{V_*'}^2}G+\frac{1}{{V_*'}^2}K\right] \, , \nonumber \\
	&\frac{1}{3\bar\rho_\vp}\left(\ddrhophidphidchi-\frac{1}{\bar\rho_\vp}\drhophidphi\drhophidchi\right)=\frac{1}{\Mp^2}\left[-\frac{1}{U_*'V_*'}K\right] \, , \nonumber \\
	&\frac{1}{3\bar\rho_\chi}\left(\ddrhochiddphi-\frac{1}{\bar\rho_\chi}\left(\drhochidphi\right)^2\right) \nonumber \\
	& \qquad=\frac{1}{\Mp^2}\left[1-\frac{U_*U_*''}{{U_*'}^2}-\frac{U_*''}{{U_*'}^2}F+\frac{1}{{U_*'}^2}J\right] \, , \nonumber \\
	&\frac{1}{3\bar\rho_\chi}\left(\ddrhochiddchi-\frac{1}{\bar\rho_\chi}\left(\drhochidchi\right)^2\right) \nonumber \\
	& \qquad=\frac{1}{\Mp^2}\left[1-\frac{V_*V_*''}{{V_*'}^2}+\frac{V_*''}{{V_*'}^2}F+\frac{1}{{V_*'}^2}J\right] \, , \nonumber \\
	&\frac{1}{3\bar\rho_\chi}\left(\ddrhochidphidchi-\frac{1}{\bar\rho_\chi}\drhochidphi\drhochidchi\right)=\frac{1}{\Mp^2}\left[-\frac{1}{U_*'V_*'}J\right] \, .
\end{align}
Eqs.~(\ref{densityderivatives1}) and~(\ref{densityderivatives2}) will be used below in comparing with numerical simulations and also in an analytic example as a method for computing the functions $F$, $G$, $J$, and $K$.

%------------------------------------------------------------------
\subsection{The $\delta N$ Derivatives}\label{sec:NDerivs}

We are now at a stage where we can compute the derivatives of $N$ which can be assembled into the observable quantities $\fnl$, $\nz$, and $r_T$. Recall that the curvature perturbation $\zeta$ can be expressed as the perturbation in the number of $e$--foldings $N$ from an initially flat hypersurface at $t_*$ up to a hypersurface of constant energy density at $t_c$:
\bea
\delta N = \zeta &=& \Nvp\delta\vp_* + \Nchi\delta\chi_* \nonumber \\
&+& \frac12\Nvpvp\delta\vp_*^2 + \frac12\Nchichi\delta\chi_*^2 +
\Nchivp\delta\vp_*\delta\chi_* \nonumber \\
&+& \ldots\,.
\eea
Hence, using the notation of Eq.~(\ref{eq:SD:expand}), we can write
\bea
\zeta^{(1)} &=& \Nvp\delta\vp_* + \Nchi\delta\chi_*\,, \nonumber \\
\zeta^{(2)} &=& \Nvpvp\delta\vp_*^2 + \Nchichi\delta\chi_*^2 +
2 \Nchivp\delta\vp_*\delta\chi_* \,.
\label{eq:SD:deltaN_expand}
\eea
These expressions are equivalent to Eqs.~(\ref{eq:SD:zeta_1st}) and~(\ref{eq:SD:zeta_2nd}). Substituting the results for $\zeta_\vp^{(1,2)}$ and $\zeta_\chi^{(1,2)}$ (Eqs.~(\ref{FirstOrderPerturbations}) and~Eqs.~(\ref{SecondOrderPerturbations})) in Eqs.~(\ref{eq:SD:zeta_1st}) and~(\ref{eq:SD:zeta_2nd}), and equating the coefficients of $\delta\vp_*$ and $\delta\chi_*$ to those in Eq.~(\ref{eq:SD:deltaN_expand}) we obtain
\begin{align}
	N_{,\vp}&=\frac{1}{\Mp^2}\left[\frac{U_*+\alpha}{U_*'}\right] \, , \nonumber \\
	N_{,\chi}&=\frac{1}{\Mp^2}\left[\frac{V_*-\alpha}{V_*'}\right] \, , \nonumber \\
	N_{,\vp\vp}&=\frac{1}{\Mp^2}\left[1+\frac{-U_*''(U_*+\alpha)+\beta}{{U_*'}^2}\right] \, , \nonumber \\
	N_{,\chi\chi}&=\frac{1}{\Mp^2}\left[1+\frac{-V_*''(V_*-\alpha)+\beta}{{V_*'}^2}\right] \, , \nonumber \\
	N_{,\vp\chi}&=\frac{-\beta}{\Mp^2U_*'V_*'} \, ,
\label{eq:SD:deltaNderivs}
\end{align}
where we have made the definitions
\begin{align}\label{alphabeta}
	\alpha&\equiv
	\begin{cases}
		&\mcA F+(1-\mcA)G  \quad \rm{for} \quad R\le1\\
		&\mctA G+(1-\mctA)F \quad \rm{for} \quad R\ge1 
	\end{cases} \, , \nonumber \\
	\beta&\equiv
	\begin{cases}
		&\mcA J+(1-\mcA)K + \frac{\mcB}{\Mp^2}(F-G)^2 \quad \rm{for} \quad R\le1\\
		&\mctA K+(1-\mctA)J + \frac{\mctB}{\Mp^2}(F-G)^2 \quad \rm{for} \quad R\ge1
	\end{cases} \, .
\end{align}
Eqs.~(\ref{eq:SD:deltaNderivs}) are the final expressions for the $\delta N$ derivatives at the end of reheating.

The quantities $\alpha$ and $\beta$ are sensitive only to the \emph{ratio} of the decay rates, $R=\Gchi/\Gvp$, which explains the insensitivity to reheating that was observed in \cite{Leung:2012ve} when $\Gchi=\Gvp$. Furthermore, we can see that these functions are completely insensitive to reheating whenever $F=G$ and $J=K$.  This occurs when the adiabatic limit is reached during inflation~\cite{Meyers:2010rg}, since the condition $\zeta_\vp=\zeta_\chi$ is one way of defining the adiabatic limit.  It also occurs, for example, when the potential is completely symmetric, and both fields are given the same initial conditions.  The insensitivity to reheating for this case can be understood by noticing that when $\zeta_\vp=\zeta_\chi$, reheating has the effect of mixing two fluids with identical properties.  Alternatively, we expect that reheating will have the most dramatic effects on $\alpha$ and $\beta$ when $F$ is either much greater or much less than $G$ and likewise for $J$ and $K$.  This will occur when the two fields have very different histories leading up to the onset of coherent oscillations.  In general, the range of values spanned by $\alpha$ and $\beta$ is determined by the conditions during inflation through the functions $F$, $G$, $J$, and $K$, and reheating has the effect of picking out a value from this range.  Finally, to understand the impact of reheating on observables, we must compare the magnitude of $\alpha$ to $U_*$ and $V_*$ and also the magnitude of $\beta$ to $U_*''(U_*+\alpha)$ and $V_*''(V_*-\alpha)$.  It is reasonable to expect that reheating will play a particularly important role when $U_*$ is either much less or much greater than $V_*$, with $\alpha$ taking intermediate values, and similarly for $\beta$.  The examples we present below exhibit this hierarchy of scales.

%------------------------------------------------------------------
\subsection{Observables}\label{sec:Observables}
In this section, we will construct primordial observables which can be used to compare models of two--field inflation with observational data.  The $\delta N$ formalism relates the derivatives of $N$ calculated above to primordial observables as described in Section~\ref{sec:deltaN}.

Using the results of Section~\ref{sec:NDerivs}, the primordial observables after reheating are given by
\begin{equation}
	\mathcal{P}_{\zeta}=\frac{\mathcal{P}_*}{\Mp^4}\left[\left(\frac{U_*+\alpha}{U_*'}\right)^2+\left(\frac{V_*-\alpha}{V_*'}\right)^2\right] \, ,
\label{eq:SD_Pzeta}
\end{equation}
\begin{equation}
	r_T=8\Mp^2\left[\left(\frac{U_*+\alpha}{U_*'}\right)^2+\left(\frac{V_*-\alpha}{V_*'}\right)^2\right]^{-1} \, ,
\label{eq:SD_rT}
\end{equation}
\begin{align}
	n_{\zeta}-1=&-2\epsilon_*-\frac{2\Mp^2}{W_*}\left[\left(\frac{U_*+\alpha}{U_*'}\right)^2+\left(\frac{V_*-\alpha}{V_*'}\right)^2\right]^{-1} \nonumber \\
	&\times\left[\left[1-\frac{U_*''\left(U_*+\alpha\right)}{{U_*'}^2}\right]\left(U_*+\alpha\right) \right. \nonumber \\
	&\left.\qquad+\left[1-\frac{V_*''(V_*-\alpha)}{{V_*'}^2}\right]\left(V_*-\alpha\right)\right] \, ,
\label{eq:SD_nz}
\end{align}
\begin{align}
	\frac{6}{5}f_{\mathrm{NL}}=&\Mp^2\left[\left(\frac{U_*+\alpha}{U_*'}\right)^2+\left(\frac{V_*-\alpha}{V_*'}\right)^2\right]^{-2} \nonumber \\
	&\times\Bigg[\left[1+\frac{-U_*''(U_*+\alpha)+\beta}{{U_*'}^2}\right]\left[\frac{U_*+\alpha}{U_*'}\right]^2 \nonumber \\
	&\qquad+2\left[\frac{-\beta}{U_*'V_*'}\right]\left[\frac{U_*+\alpha}{U_*'}\right]\left[\frac{V_*-\alpha}{V_*'}\right] \nonumber \\
	&\qquad+\left[1+\frac{-V_*''(V_*-\alpha)+\beta}{{V_*'}^2}\right]\left[\frac{V_*-\alpha}{V_*'}\right]^2\Bigg] \, .
\label{eq:SD_fnl}
\end{align}

These formulas capture the observable impact of reheating following two--field inflation in just two functions, $\alpha$ and $\beta$.  Even without detailed knowledge of how to calculate $\alpha$ and $\beta$, there are some lessons that can be drawn from the fact that the observables take this form.  Once the conditions at Hubble exit are fixed, the entire range of possible predictions can be explored, simply by adjusting the quantities $\alpha$ and $\beta$\footnote{Arbitrarily chosen values of $\alpha$ and $\beta$ may not be attainable in realistic models, but this procedure may point the way forward for constructing models with specific combinations of observables.}.  For example, if a future observation fixes the value of $r_T$, then Eq.~(\ref{eq:SD_rT}) gives a relation between the conditions at Hubble exit and the quantity $\alpha$, which also plays a role in determining the other primordial observables.  One can also turn this around and ask what conditions must be imposed on $\alpha$, $\beta$, and the potential at Hubble exit to produce large values of $\fnl$.  It can be seen from Eqs.~(\ref{eq:SD_nz}) and~(\ref{eq:SD_fnl}), for example, that $\fnl$ can be made large while keeping $\nz-1$ small if
\begin{equation}
	\frac{\beta\Mp^2\left[\left(\frac{U_*+\alpha}{{U'_*}^2}\right)-\left(\frac{V_*-\alpha}{{V'_*}^2}\right)\right]^2}{\left[\left(\frac{U_*+\alpha}{U_*'}\right)^2+\left(\frac{V_*-\alpha}{V_*'}\right)^2\right]^2}\gg1 \, .
	\label{eq:Largefnl}
\end{equation}
Then, together with Eq.~(\ref{alphabeta}), these sorts of relations can be used to gain a better understanding of the general properties of models that produce large local non-Gaussianity.  We intend to explore this in more detail in future work.

%------------------------------------------------------------------

%------------------------------------------------------------------
\section{Numerical Simulations}\label{sec:numericalSims}

Up to this point, we have worked entirely within the confines of the sudden decay formalism, approximating the decay of the fields as instantaneous events which occur on hypersurfaces of constant energy density. Furthermore, we have treated the oscillating fields as perfect fluids by averaging over many oscillations. In this section we discuss how the observables $\fnl$, $\nz$, and $r_T$ can be calculated \emph{exactly}, by numerically evolving the field Eqs.~(\ref{eq:fieldEoM}) and~(\ref{eq:decay_prods}). By comparing our sudden decay calculation with numerical simulations, we describe the modifications that need to be made to the sudden decay approximation in order to account for the non--instantaneous nature of the decay, and reconcile the numerical results. \\

While still assuming slow--roll at Hubble exit, one can go beyond the slow--roll approximation (and the condition of separability of the potential) by computing the $\delta N$ derivatives of Eq.~(\ref{eq:deltaN}) numerically. We have written a finite--difference code in \texttt{fortran90} which is able to quickly and accurately compute two, three (and four) point statistics of $\zeta$ following a phase of perturbative reheating for any canonical two--field model.\footnote{An earlier version of the same code was used in Refs.~\cite{Leung:2012ve,Leung:2013rza}.} The numerical recipe is as follows.  First, the fiducial trajectory emanating from $\{\vps,\chis\}$ is constructed by solving the full second order field Eqs.~(\ref{eq:fieldEoM}) and~(\ref{eq:decay_prods}) subject to the Friedmann constraint Eq.~(\ref{eq:Friedman}). We assume that the slow--roll conditions hold at Hubble exit, which uniquely specify $\dot\vp_*$ and $\dot\chi_*$. A `bundle' containing forty--nine trajectories is then formed by evolving neighboring trajectories with slightly perturbed initial conditions, $\vps\rightarrow\vps+\delta\vps$ and $\chis\rightarrow\chis+\delta\chis$. Each trajectory in the bundle is brought to a common energy hypersurface at a time $t_c$ when the universe is completely radiation dominated. Partial derivatives of $N({t_*}\,, {t_c})$ with respect to the field values at Hubble exit are then taken using a seven--point `stencil' finite difference method~\cite{Abramowitz1965Handbook}. This provides a fast, efficient method for computing $\nz$, $r_T$, and $\fnl$ for an arbitrary two--field model, valid beyond slow--roll and through a phase of reheating. Numerical codes based on the moment transport equations have also been developed~\cite{Mulryne:2010rp}.

While conceptually simple, taking the $\delta N$ derivatives is technically a challenging computation to perform to a sufficient level of accuracy. This is particularly true for calculations at the end of reheating, which require integrating the field equations over many $e$--foldings and over many field oscillations. Cross derivatives such as $N_{,\vp\chi}$ require two finite--difference step--sizes, $h_\vp=\delta\vp_*$ and $h_\chi=\delta\chi_*$, which in general do not take the same numerical values. One must also ensure that the derivatives are insensitive to the values of $h_\vp$ and $h_\chi$, while at the same time minimizing truncation and round--off error. We find that, in general, a minimum of five points are required in the finite--difference stencils in order to keep numerical error to a minimum. We have verified our code against the only known exact solution beyond slow--roll, based on a sum separable ansatz for the Hubble parameter~\cite{Byrnes:2009qy}.

The reheating parameters $\Gvp$ and $\Gchi$ are set to zero during inflation. It is only when each individual trajectory in the bundle passes through the minimum of its potential $\{\chi_{\rm min},\vp_{\rm min}\}$ for the first time that $\Gvp$ and $\Gchi$ are introduced to the field equations, sourcing the radiation fluids $\rho_\gamma^\vp$, $\rho_\gamma^\chi$. In general, for any given trajectory, $\vp$ will not reach the minimum of its potential at the same time as $\chi$, and so $\Gvp$ and $\Gchi$ are `switched on' at different times along the same trajectory. Furthermore, for each field, the foliation of the entire bundle of trajectories as determined by each trajectory reaching $\chi_{\rm min}$ (and likewise $\vp_{\rm min}$) does not in general occur on a surface of constant energy as it does in the sudden decay formalism, but rather at a surface of constant $\chi_{\rm min}$ (and $\vp_{\rm min}$). This choice of reheating surface in our simulations is not unique, since the definition of the onset of \emph{coherent} oscillations is ambiguous. For example, we could have instead introduced the decay terms after, say, the third complete oscillation, instead of the first pass of the minimum. To an excellent approximation, we find that our results are insensitive to this choice, as should be expected from our sudden decay result, which only depends upon the ratio $R\equiv\Gchi/\Gvp$.

In our numerical simulations, the fields \emph{gradually} decay into radiation, and this renders the notion of the sudden decay `times' $t^{\chi}_{\rm dec}$ and $t^{\vp}_{\rm dec}$ ambiguous. We remove this ambiguity by defining $t^{\chi}_{\rm dec}$ and $t^{\vp}_{\rm dec}$ in our simulations to be the times when the energy density stored in the oscillating fields become equal to that of their decay products
\be
\rho_\vp(t^{\vp}_{\rm dec}) = \rho_\gamma^\vp(t^{\vp}_{\rm dec})\,, \qquad  \rho_\chi(t^{\chi}_{\rm dec}) = \rho_\gamma^\chi(t^{\chi}_{\rm dec})\,.
\ee
Taking the above definition for $t^{\chi}_{\rm dec}$ and $t^{\vp}_{\rm dec}$, we have confirmed that the underlying assumption of the sudden decay formalism, Eq.~(\ref{eq:SD:decaytimesSD}), (which states that $H(t_{\rm dec}^{\vp}) = \Gvp$ and $H(t_{\rm dec}^{\chi}) = \Gchi$) holds to a high level of accuracy. 

Recall that our sudden decay result depends upon two energy ratios: $r$ for $R\le1$ (Eq.~(\ref{eq:SD:r})) and $\tilde{r}$ for $R\ge1$ (Eq.~(\ref{eq:SD:tilder})). These energy ratios can be directly related to $R$ by appealing to the sudden decay approximation (see the end of Section~\ref{sec:SDCalc}). We find, however, that Eqs.~(\ref{eq:largeR_SD}) and~(\ref{eq:smallR_SD}) do not accurately reproduce the numerical solutions for $r(R)$ and $\tilde{r}(R)$, and they should not be used in the calculation of the observables. We include their derivation in Section~\ref{sec:SDCalc} only for completeness. The reason for their failure is that they do not account for the gradual decay of the fields.

\begin{figure*}[!t]
	\begin{tabular}{cc}
		\includegraphics[width=8.8cm]{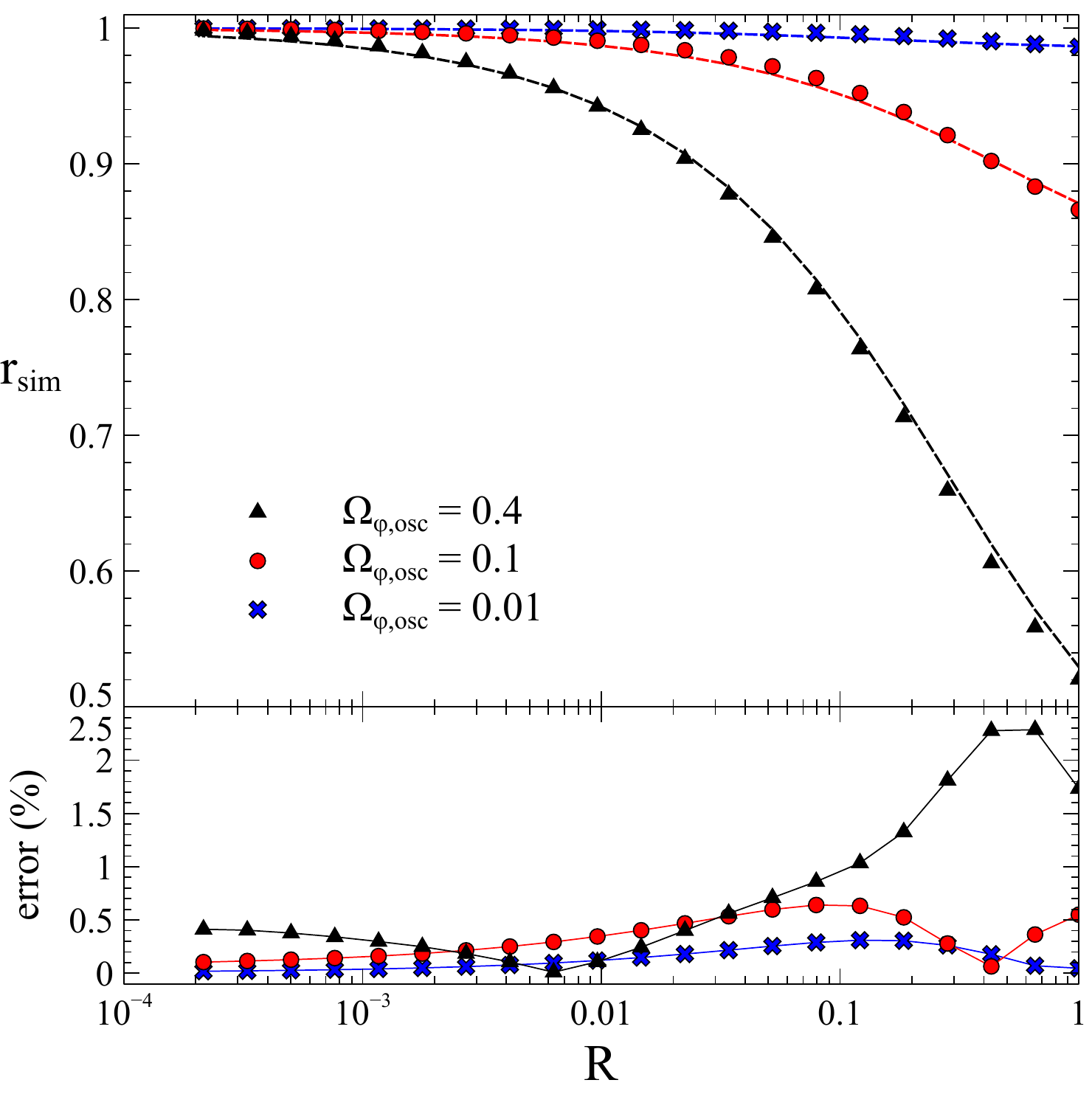} &
		\includegraphics[width=8.8cm]{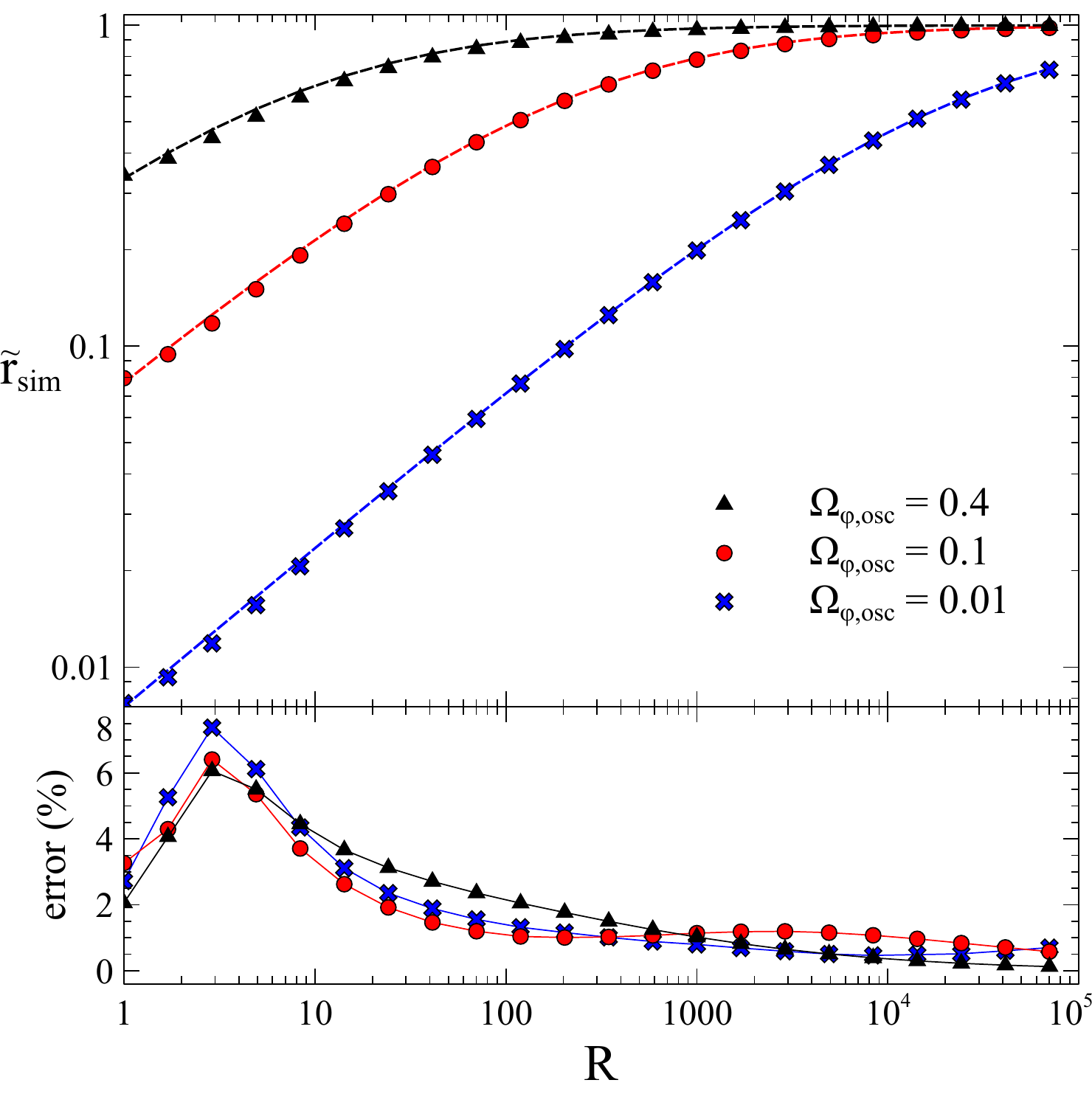} 
	\end{tabular}
	\caption{Comparison between the fitting functions for $r_{\rm sim}$ and $\tilde{r}_{\rm sim}$ (dashed lines) and the numerical values for $r_{\rm sim}$ and $\tilde{r}_{\rm sim}$ obtained from our field theory simulations (data points). The fitting function for $r_{\rm sim}$ is accurate to better than $3\%$ over the entire range of $R$ and $\left.\Omega_{\phi}\right|_{\rm osc}$ probed by our simulations, while the fitting function for $\tilde{r}_{\rm sim}$ is accurate to better than $8\%$.}
	\label{fig:fitting_functions}
\end{figure*}

It is at this point that we encounter the modifications that are required of the sudden decay formalism.  According to the sudden decay formalism,  $\chi$ does not interact with its decay products until $t^{\chi}_{\rm dec}$ where it then decays instantly into radiation (and similarly for $\vp$). Since in reality the fields have been decaying gradually since $t^{\vp,\chi}_{\rm osc}$, some fraction of radiation density will be present prior to $t^{\chi}_{\rm dec}$. With this in mind, we find that the sudden decay energy ratio $r$ should be modified to \emph{include the density of the decay products} in the following way:
\be
r\,\to\, r_{\rm sim}\equiv 
\left.\frac{3(\bar\rho_\chi+\bar\rho^\chi_\gamma)}{3(\bar\rho_\chi+\bar\rho^\chi_\gamma) + 4(\bar\rho_\vp + \bar\rho_\gamma^\vp)}\right|_{t^{\chi}_{\rm dec}}\,,
\label{eq:r_sim}
\ee
where we have used the subscript `sim' (for simulation) to make clear that this modification is required in order to reconcile the results of our numerical simulations, which account for the gradual decay of the fields. Similarly, the sudden decay energy ratio $\tilde r$ becomes
\be
\tilde r\,\to\, \tilde{r}_{\rm sim} \equiv 
\left.\frac{3(\bar\rho_\vp+\bar\rho_\gamma^\vp)}{3(\bar\rho_\vp+\bar\rho_\gamma^\vp) + 4(\bar\rho_\chi + \bar\rho_\gamma^\chi)}\right|_{t^{\vp}_{\rm dec}}\,.
\label{eq:tilder_sim}
\ee
It is clear that $r_{\rm sim}$ and $\tilde{r}_{\rm sim}$ cannot be related to $R$ through a sudden decay calculation. While this is true, we can construct fitting functions that relate $r_{\rm sim}$ and $\tilde{r}_{\rm sim}$ to $R$. We do this by writing the field equations for $\chi$ and $\vp$ as perfect fluids
\be
\dot\rho_\chi + (3H+\Gchi)\rho_\chi = 0\,, \qquad \dot\rho_\vp + (3H+\Gvp)\rho_\vp = 0\,, \nonumber
\ee
and solving them numerically along side Eqs.~(\ref{eq:decay_prods}) and~(\ref{eq:Friedman}) to determine $r_{\rm sim}$ and $\tilde{r}_{\rm sim}$. This fluid approximation is valid so long as both fields are coherently oscillating about quadratic minima. We make use of this fluid description (rather than evolving the field equations) as it enables us to construct the fitting functions with relative ease. In this fluid system, $r_{\rm sim}$ and $\tilde{r}_{\rm sim}$ depend only on two variables: the ratio of decay rates $R$, and the energy ratio
\be
\Omega_{\vp,{\rm osc}}\equiv\left.\frac{\bar\rho_\vp}{\bar\rho_\vp+\bar\rho_\chi}\right|_{\rm osc}\,,
\label{eq:Omega_osc}
\ee
where the subscript `osc' denotes the initial conditions on the fluids. The density of the decay products $\rho_\gamma^\vp$ and $\rho_\gamma^\chi$ are zero initially.  We find that $r_{\rm sim}$ is well fit as a function of $R$ and $\Omega_{\vp,{\rm osc}}$ (for $R\le1)$ by
\be
r_{\rm sim}(R,\Omega_{\vp,{\rm osc}}) = 1-\left[p+\frac{q}{R} \right]^{-v}\,,
\label{eq:r-fit}
\ee
where:
\bea
v &=& 0.60\,, \qquad q= \left.0.63\,\frac{{\rm ln}\,\Omega_{\vp}}{{\rm ln}\,(1-\Omega_{\vp})}\right|_{\rm osc} \nonumber \\
p &=& \left [\frac{4\Omega_{\vp}}{3+\Omega_{\vp}} \right]_{\rm osc}^{-1/v} - q\,.
\eea
For $R\ge1$, $\tilde{r}_{\rm sim}$ is well fit by
\be
\tilde{r}_{\rm sim}(R,\Omega_{\vp,{\rm osc}}) = 1-\left[1+\tilde{q}\sqrt{R} \right]^{-\tilde{v}}\,,
\label{eq:tilder-fit}
\ee
where:
\be
\tilde{v} = 1.666\,, \qquad \tilde{q} = \left [\frac{4(1-\Omega_{\vp})}{4-\Omega_{\vp}} \right]_{\rm osc}^{-1/\tilde{v}} - 1\,.
\ee
While these fitting functions have been obtained by comparing to a fluid description, we stress that this is the \emph{only} time in this paper that we use a fluid approximation in our numerical work. When numerically computing observables in the following sections, we evolve the field equations as given by Eq.~(\ref{eq:fieldEoM}). 

Importantly, Eqs.~(\ref{eq:r-fit}) and~(\ref{eq:tilder-fit}) reproduce the values of $r_{\rm sim}$ and $\tilde{r}_{\rm sim}$ obtained from our field theory simulations. This is illustrated in Fig.~\ref{fig:fitting_functions}, where the numerical data represents the values of $r_{\rm sim}$ and $\tilde{r}_{\rm sim}$ obtained from our field theory simulations for the potential $W(\vp,\chi)=\frac{1}{2}m^2\vp^2+\frac{1}{2}M^2\chi^2$. The dashed lines indicate the corresponding fit using Eqs.~(\ref{eq:r-fit}) and~(\ref{eq:tilder-fit}). As can be seen from the Figure, the fitting function for $r_{\rm sim}$ is accurate to better than $3\%$ over the entire range of $R$ and $\Omega_{\vp,{\rm osc}}$ probed by our simulations, while the fitting function for $\tilde{r}_{\rm sim}$ is accurate to better than $8\%$.  For the examples illustrated in Fig.~\ref{fig:fitting_functions}, the value of $\Omega_{\vp,{\rm osc}}$ was calculated numerically. In the following section we present an inflationary model where $\Omega_{\vp,{\rm osc}}$ may be computed analytically.

Similar modifications are required of the energy ratios $\Omega_{\vp,{\rm dec}}$ and $\tilde{\Omega}_{\chi,{\rm dec}}$ (Eqs.~(\ref{eq:SD:Ovpdec}) and~(\ref{eq:SD:tildeOmegachi}) respectively). For $\Omega_{\vp,{\rm dec}}$, we let
\bea
\Omega_{\vp,{\rm dec}} \,&\to&\, \Omega_{\vp,{\rm dec}}^{\,\rm sim}  \equiv
\left.\frac{(\bar\rho_\vp + \bar\rho^\vp_\gamma) }{ (\bar\rho_\vp + \bar\rho^\vp_\gamma) + (\bar\rho_\chi + \bar\rho^\chi_\gamma)} \right|_{ t_{\rm dec}^{\vp} } \nonumber \\
 &\approx& \left.\frac{2\bar\rho_\vp}{ 2\bar\rho_\vp + \bar\rho_\chi } \right|_{ t_{\rm dec}^{\vp} } \approx
\left.\frac{2\Omega_\vp}{1+\Omega_\vp}\right|_{\rm osc}\,.
\label{eq:Omvpdec_sim}
\eea
The first approximate equality is obtained by neglecting $\bar\rho^\chi_\gamma(t_{\rm dec}^{\vp})$ which is a good approximation in the limit of small $R$. The second approximate equality is obtained by assuming that $\bar\rho_\vp$ and $\bar\rho_\chi$ redshift at the same rate between $t_{\rm osc}$ and $t_{\rm dec}^{\vp}$. By similar reasoning, the energy ratio $\tilde{\Omega}_{\chi,{\rm dec}}$ becomes
\bea
\tilde{\Omega}_{\chi,{\rm dec}} \,&\to&\, \Omega_{\chi,{\rm dec}}^{\,\rm sim}  \equiv 
\left.\frac{(\bar\rho_\chi + \bar\rho^\chi_\gamma) }{ (\bar\rho_\chi + \bar\rho^\chi_\gamma) + (\bar\rho_\vp + \bar\rho^\vp_\gamma)} \right|_{ t_{\rm dec}^{\chi} } \nonumber \\
 &\approx& \left.\frac{2\bar\rho_\chi}{ 2\bar\rho_\chi + \bar\rho_\vp } \right|_{ t_{\rm dec}^{\chi} }\approx
\frac{2(1-\Omega_\vp)}{2-\Omega_\vp}\Bigg|_{\rm osc}\,.
\label{eq:Omchidec_sim}
\eea
As we shall see in the next section, these two approximations work very well when computing the primordial observables.\\

Before applying our formalism to specific examples, we take a moment to summarize our sudden decay calculation. Our final expressions for the primordial observables at the completion of reheating are given by Eqs.~(\ref{eq:SD_Pzeta}-\ref{eq:SD_fnl}). These expressions are valid for any inflationary potential that can be written in the form $W(\vp,\chi)=U(\vp)+V(\chi)$, where $U(\vp)$ and $V(\chi)$ have quadratic minima. Our analytic results will be most reliable in the regime where the first field to decay does so when the other field is oscillating. The details of the inflationary model and the impact of the reheating phase are captured in two functions, $\alpha$ and $\beta$ which are defined in Eq.~(\ref{alphabeta}). The details of the inflationary model are encoded in four functions $F$, $G$, $J$, and $K$, which appear in $\alpha$ and $\beta$ and do not depend in any way upon reheating. The effects of reheating are fully described by the reheating functions $\mcA$ and $\mcB$ for $R\le1$ (and $\mctA$ and $\mctB$ for $R\ge1$) which also appear in $\alpha$ and $\beta$. These reheating functions are only sensitive to the ratio $R\equiv\Gchi/\Gvp$ of decay rates, and are defined in Eqs.~(\ref{eq:SD:A}), (\ref{eq:SD:B}), (\ref{eq:SD:mctA}), and (\ref{eq:SD:mctB}). They take values between $0$ and $1$ depending upon the energy ratios $r$ and $\tilde{r}$. We have provided fitting functions, Eqs.~(\ref{eq:r-fit}) and~(\ref{eq:tilder-fit}), which relate these ratios to $R$. As $R\to0$, $r\to1$ and so $\mcA\to1$ and $\mcB\to0$. As $R\to\infty$, $\tilde{r}\to1$ and so $\mctA\to1$ and $\mctB\to0$.  These reheating functions also depend upon the energy ratios $\Omega_{\vp,{\rm dec}}$ and $\tilde{\Omega}_{\chi,{\rm dec}}$. We have shown in Eqs.~(\ref{eq:Omvpdec_sim}) and~(\ref{eq:Omchidec_sim}) that these energy ratios can be written in terms of a single parameter $\Omega_{\vp,{\rm osc}}$ that is defined in Eq.~(\ref{eq:Omega_osc}), and represents the partition of energy density between the fields at the start of their oscillatory phase. The complete list of quantities upon which our calculation depends is thus: the inflationary model functions $F$, $G$, $J$, and $K$ and the initial field configuration at Hubble exit, the partition of energy at the start of oscillations $\Omega_{\vp,{\rm osc}}$, and the ratio of decay rates $R$.

%------------------------------------------------------------------
\section{Examples}\label{sec:examples}
We now apply the formalism developed in Sections~\ref{sec:SDCalc} and~\ref{sec:numericalSims}, to two concrete examples. The first example we discuss contains terms that we have thus far been unable to evaluate analytically, and so we instead rely on numerical techniques. For the second example, we construct a fully analytic solution, though we stress that it should be possible to find other fully analytic examples. For both examples studied here, we compare the sudden decay result for $r_T$, $\nz$, and $\fnl$ (Eqs.~(\ref{eq:SD_rT}),~(\ref{eq:SD_nz}),~(\ref{eq:SD_fnl})), to the predictions obtained from the numerical simulation introduced in Section~\ref{sec:numericalSims}.

%------------------------------------------------------------------
\subsection{Example I: Axion N--flation}\label{sec:ex1}

In the scenario known as N--flation~\cite{Dimopoulos:2005ac}, many axion fields are used to cooperatively source inflation even if their potentials are individually too steep.\footnote{For earlier related work, see \cite{Liddle:1998jc,Copeland:1999cs}} The collective potential is comprised of a sum of $N_f$ uncoupled axions $\vp_i$:
\be
\label{eq:Nflation}
W(\vp)=\sum_{i=1}^{N_f}\Lambda_i^4\left[1 -{\rm cos}\left(\frac{2\pi}{f_i}\vp_i\right)\right]\,.
\ee
With only a single field present, this model is more commonly known as natural inflation~\cite{Adams:1992bn}. Each axion is fully described by its decay constant $f_i$ and its potential energy scale $\Lambda_i^4$. The standard arguments show that we should expect $f_i\sim10^{16}{\rm GeV}$. The mass of each field in vacuum satisfies $m_{\vp(i)}^2=4\pi^2\Lambda_i^4/f_i^2$. Due to the shift symmetry $\vp_i\rightarrow\vp_i+n f_i$, we can without loss of generality set the initial conditions $\vp_{*(i)}\in[0,\,f_i]$. We assume that the initial conditions are chosen so that only a single axion $\vp$ populates the hilltop region. The remaining $N_f-1$ axions, which begin away from the hilltop, contribute only to the expansion rate, and may be replaced by a single effective field $\chi$ with a quadratic potential. With $f_i=f$ for all axions, the effective two--field potential then reads:
\be
\label{eq:effectiveNflation}
W(\vp,\chi)=W_0\left[\frac{1}{2}m^2\chi^2 + \Lambda^4\left(1 -{\rm cos}\left(\frac{2\pi}{f}\vp\right)\right)\right]\,.
\ee
Replacing the collective potential with an effective two--field potential is well motivated, see for example~\cite{Battefeld:2008bu}, where it was shown that the energy density of the universe is dominated by fields with comparable masses even if one starts with thousands of fields. Reheating in models of N--flation has also been shown to proceed preferentially via perturbative decay channels as opposed to via parametric resonance and preheating~\cite{Battefeld:2008bu,Battefeld:2009xw}.

We now choose two sets of model parameters (Case A, and Case B), and examine the impact of reheating on $r_T$, $\nz$, and $\fnl$.

%------------------------------------------------------------------
\subsubsection{Case A}\label{sec:caseA}

In this first example, we fix the model parameters to be
\be 
\chis=15.0\,, \quad \vps=0.499\,, \quad m=f=1\,, \quad \Lambda^4=\frac{11}{20\pi^2}\,. \nonumber
\ee
With these values, $\vp$ is very light at Hubble exit and remains frozen throughout inflation, which is driven by the $\chi$ field. The fields reach their minima at roughly the same time, $t_{\rm osc}^\vp \approx t_{\rm osc}^\chi$. 

\begin{figure}[!t]
\centering
\includegraphics[width=8.5cm]{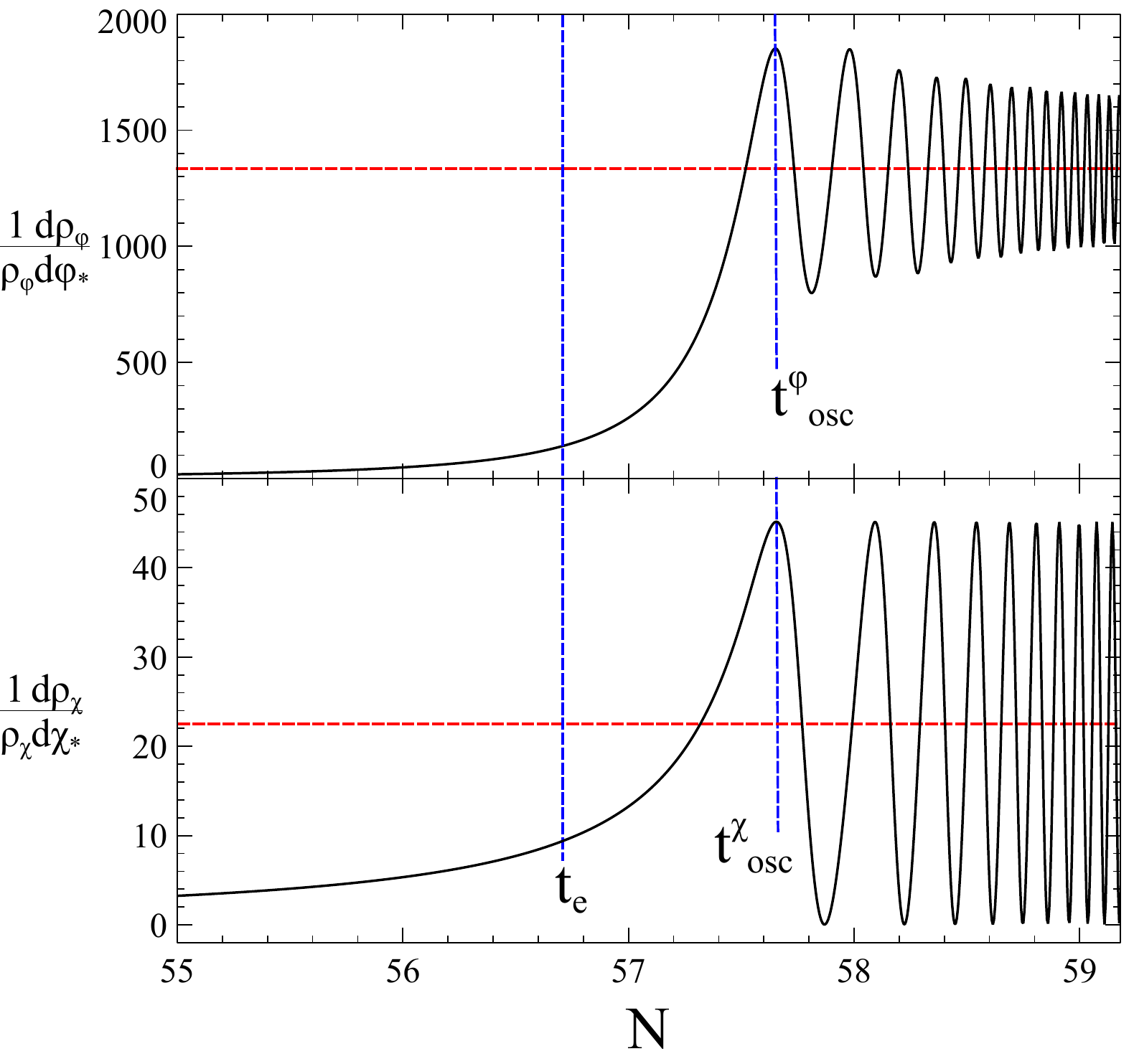}
\caption{Evolution of $\frac{1}{\bar\rho_\chi}\drhochidchi$ and $\frac{1}{\bar\rho_\vp}\drhophidphi$ during the very last stages of inflation, and into the period of field oscillations. These terms oscillate about a central, \emph{constant} value, which is indicated by the horizontal red line. This is the value that would be attained if the fields were modeled as pressureless dust.}
\label{fig:coeffs}
\end{figure}
\begin{figure*}[!t]
\centering
\includegraphics[width=16.0cm]{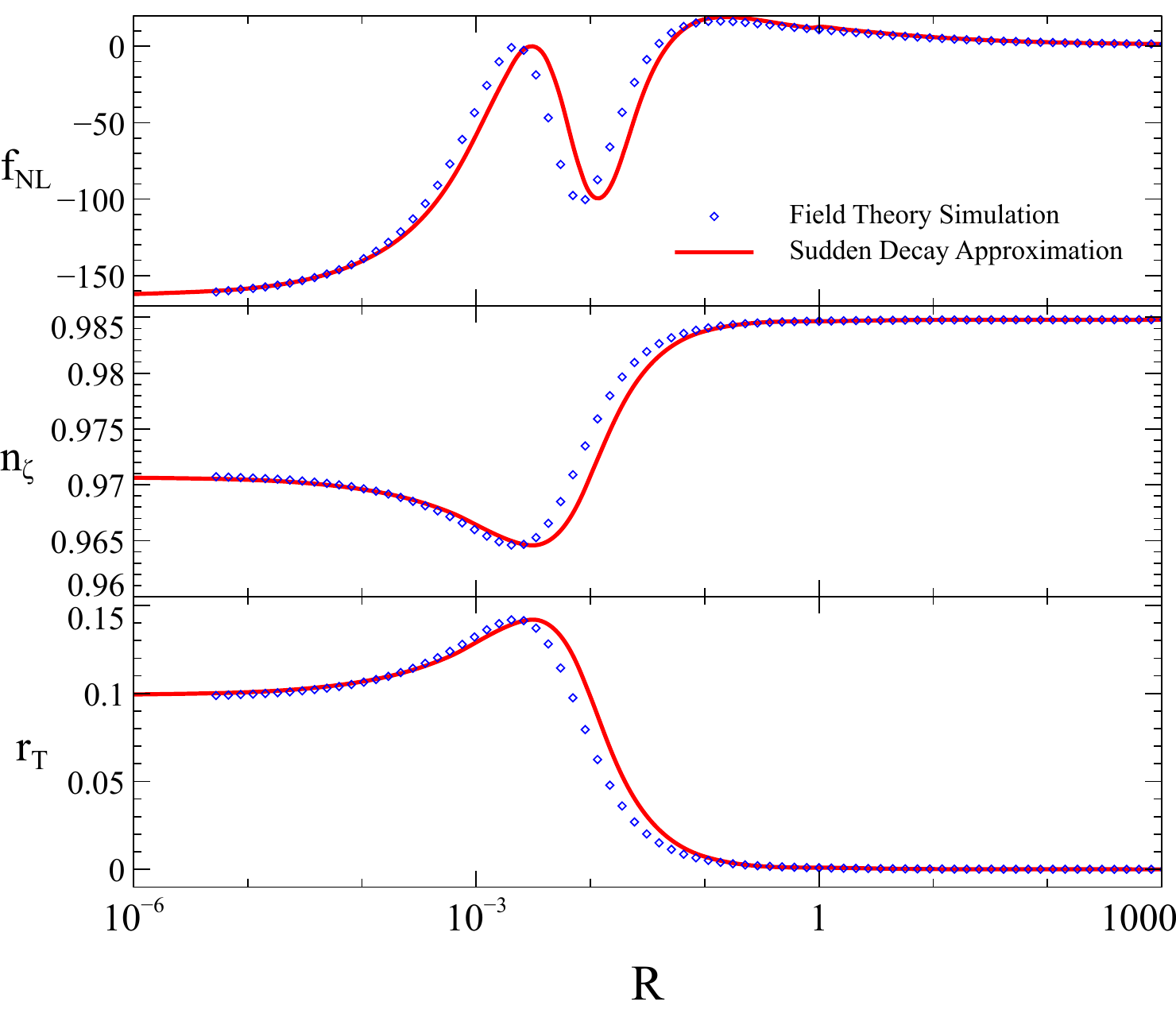}
\caption{The observables $\fnl$, $\nz$, and $r_T$ as a function of the ratio of decay rates $R$ for the model of Section~\ref{sec:caseA}. We compare the sudden decay approximation (red lines) with the corresponding result from our numerical simulations (blue diamonds).}
\label{fig:example1}
\end{figure*}

Our first task is to compute the functions $F$, $G$, $J$, and $K$, which appear in the expressions for the $\delta N$ derivatives (Eq.~(\ref{eq:SD:deltaNderivs})).  For the technical reasons discussed in Section~\ref{sec:zeta12form}, we do not yet have a method (for this particular potential) to calculate these quantities analytically, and so we rely on numerical techniques. We make use of Eqs.~(\ref{densityderivatives1}) and~(\ref{densityderivatives2}) to relate these functions to perturbations in the energy density, evaluating the derivatives which appear in these expressions with a seven--point finite difference method. To illustrate this calculation, we plot in Fig.~\ref{fig:coeffs} the evolution of the first order derivatives $(\p\bar\rho_\chi/\p\chi_*)/\bar\rho_\chi$ and $(\p\bar\rho_\vp/\p\vp_*)/\bar\rho_\vp$ during the very last stages of inflation, and into the period of oscillation. In the sudden decay approximation, the fields behave like pressureless dust during this oscillating phase, and do not interact with their decay products.  To respect this approximation in our numerical simulations, we set $\Gchi=\Gvp=0$ when calculating the $\p\bar\rho_{\phi_I}/\p\phi_{I_*}$ derivatives. As can be seen from Fig.~\ref{fig:coeffs}, $(\p\bar\rho_\chi/\p\chi_*)/\bar\rho_\chi$ and $(\p\bar\rho_\vp/\p\vp_*)/\bar\rho_\vp$ oscillate with (roughly) constant amplitude about a central, \emph{constant} value, which is indicated by the horizontal red line. This is the value that would be attained if the fields were modeled as pressureless dust. The red line represents this conserved value, which may be calculated by averaging these terms over several oscillations, which we do numerically.

We find that for this particular inflationary model, the second order derivative terms $\p^2\bar\rho_{\phi_I}/\p\phi^2_{I_*}$ oscillate with a very large amplitude about a central value. The amplitude of these oscillations are so large that the averaging procedure described above becomes unreliable, and so we rely on a slightly different approach. This involves running our field theory simulations (which calculates the $\delta N$ derivatives directly) in the regime of very large and very small $R$. Recall that in these limits, $\mcA\to1$ and $\mcB\to0$, and so from Eqs.~(\ref{alphabeta}) and~(\ref{eq:SD:deltaNderivs}) we see that $J$ and $K$ (and hence $\p^2\bar\rho_{\phi_I}/\p\phi^2_{I_*}$) can be expressed in terms of $\Nvpvp$ and $\Nchichi$.

The first order derivatives can be calculated using either of the above methods (averaging and small/large $R$ limits) and we have verified that they yield the same results.  Having evaluated all of the required derivative terms, we can evaluate the statistics of $\zeta$. \\

\begin{figure*}[!t]
\centering
\includegraphics[width=16.0cm]{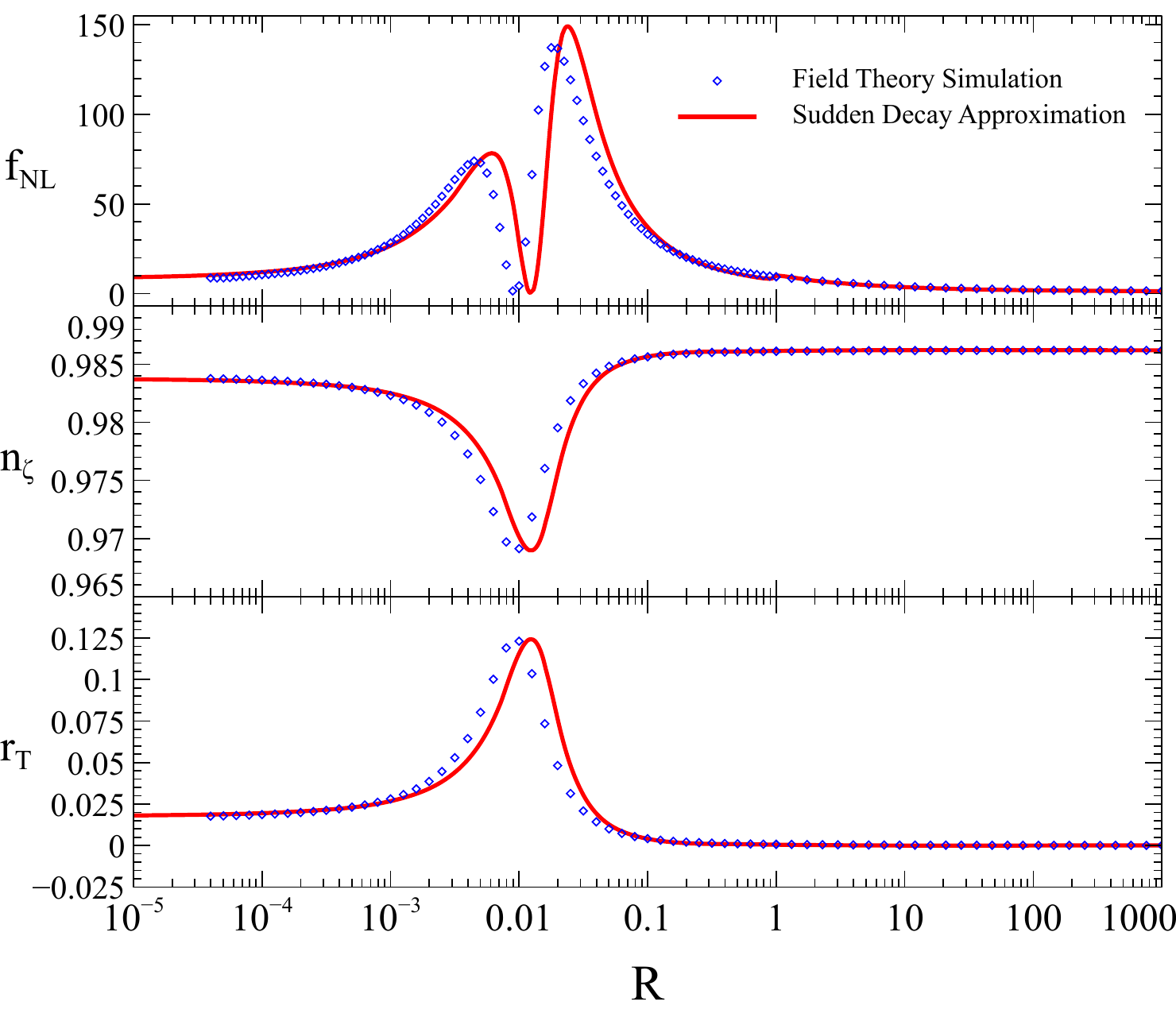}
\caption{The observables $\fnl$, $\nz$, and $r_T$ as a function of the ratio of decay rates $R$ for the model of Section~\ref{sec:caseB}. We compare the sudden decay approximation (red lines) with the corresponding result from our numerical simulations (blue diamonds).}
\label{fig:example2}
\end{figure*}

In Fig.~\ref{fig:example1} we compare $\fnl$, $\nz$, and $r_T$ calculated numerically against the sudden decay approximation. We use Eqs.~(\ref{eq:r-fit}) and~(\ref{eq:tilder-fit}) to compute $r_{\rm sim}$ and $\tilde{r}_{\rm sim}$, and we obtain $\Omega_{\vp,{\rm osc}}$ (which is used to compute $\Omega^{\,{\rm sim}}_{\vp,{\rm dec}}$ and $\tilde{\Omega}^{\,{\rm sim}}_{\chi,{\rm dec}}$) directly from our simulations. As is evident from Fig.~\ref{fig:example1}, our sudden decay calculation captures the dependence of $\fnl$, $\nz$, and $r_T$ on $R$ remarkably well.  We note that the behavior around $R=1$ is smooth, despite having used different fitting functions to compute observables on either side of this point. As anticipated, the observables asymptote to constant values in the small/large $R$ limits, while the behavior of $\fnl$ for intermediate values of $R$ is dramatic. In particular as one moves from large $R$ to small $R$, the sign of $\fnl$ becomes negative. This seemingly complicated dependence of $\fnl$ on $R$ may be explained by considering the dominant term in $\fnl$ (Eq.~(\ref{eq:deltaNobservables})): $\fnl\approx\Nvpvp\Nvp^2/(\Nvp^2+\Nchi^2)^2$.  Starting at $R=1$, as $R$ is deceased $\fnl$ reaches a maximum value, after which there comes a point where $\Nvpvp$ changes sign and hence so does $\fnl$. During this time, $\Nvp$ has been steadily decreasing with $R$, and there exists a value of $R$ for which $\Nvp=\Nchi$. This induces the minimum seen in $\fnl$. As $R$ becomes even smaller,  $\Nvp\to0$, and $\fnl$ decays towards zero. At $R\sim3\times10^{-3}$, $\Nvp$ passes through zero and becomes negative, which causes $|\fnl|$ to grow once more, finally asymptoting to a constant as $R\to0$. 

It is the value of the $\Nchi$ derivative in $\fnl$ which determines the approximate position of the minimum. This derivative is sourced by fluctuations of the inflaton field $\chi$, and it remains approximately constant at the Hubble exit value $\Nchi\approx8\Mp^{-1}$ across the whole range of $R$. This is because for this particular model, $\alpha\ll V_*$ (see Eq.~(\ref{eq:SD:deltaNderivs})).  Since in the small $R$ regime $\Nchi$ is larger than $\Nvp$, inflaton fluctuations cannot be neglected: including $\Nchi$ is vital if one would like to compute the maximum attainable value of $|\fnl|$. The presence of fluctuations in \emph{both} fields means that all observable quantities take values in a finite range whose limits are set entirely by the conditions during inflation.  Specifically, these conditions are the field configuration at Hubble exit and the functions $F$, $G$, $J$, and $K$.  The same conditions also determine whether $\fnl$ maintains the same sign over the entire range of $R$.

Similar reasoning may be used to describe the dependence of $\nz$ and $r_T$ on $R$.  We point out that we are not attempting to present a model which is compatible with observational constraints, but rather to illustrate the impact that reheating can have on the observables. As is clear from Fig.~\ref{fig:example1}, variations in $\fnl$, $\nz$, and $r_T$ for different values of $R$ are well within the sensitivity of current experiments such as \emph{Planck}.

%------------------------------------------------------------------
\subsubsection{Case B}\label{sec:caseB}

In this second example, the model parameters are
\be 
\chis=16.0\,, \quad \vps=0.499\,, \quad m=f=1\,, \quad \Lambda^4=0.04533\,.  \nonumber
\ee
We calculate the functions $F$, $G$, $J$, and $K$, and the value of $\Omega_{\vp,{\rm osc}}$ numerically using the same methods as above. The observables are plotted in Fig.~\ref{fig:example2}. For this example, $\Nvpvp>0$ for all $R$, and hence $\fnl$ remains positive. The first derivative $\Nvp$ however does change sign, and sends $\fnl$ momentarily toward zero. The first maximum in $\fnl$ corresponds approximately to $|\Nvp|=|\Nchi|$ when $\Nvp$ is positive, and a second maximum appears as $|\Nvp|=|\Nchi|$ when $\Nvp<0$. This behavior also generates the peaks that are observed in $\nz$ and $r_T$.

%------------------------------------------------------------------
\subsection{A Fully Analytic Example}\label{sec:analytic_example}

The effective axion model presented above contained terms that we have so far been unable to evaluate analytically, and instead we had to rely on numerical techniques. In this example, we show how to apply the formalism we developed above in a specific case which can be treated fully analytically.  This is not the only model where analytic progress can be made, though this particularly simple example will allow us to demonstrate our method and easily compare to previous studies.  Let us take the potential to be of the form
\begin{equation}
	W(\vp,\chi)=\frac{1}{2}m^2\vp^2+\frac{1}{2}M^2\chi^2 \, .
\end{equation}
We will study the case where the field $\chi$ dominates the energy density during inflation, while the field $\vp$ remains essentially frozen until $\chi$ begins to oscillate about the minimum of its potential.  These conditions require that $M\gg m$ and $\chi_*\gg \vp_*$.  In addition to demonstrating our formalism, this example is of particular interest, since it becomes remarkably similar to the standard curvaton scenario in a limiting case.

In this model, during the period following inflation, the dominant contribution to the energy density of the universe is a coherently oscillating scalar field, which on average mimics the behavior of pressureless dust.  In this phase, we have $\rho_\chi\gg\rho_\vp$, while $\rho_\chi\propto a^{-3}$, and so $a(t)\propto t^{2/3}$, which gives $H(t)=\frac{2}{3t}$.  We note that with this definition of time, we are implicitly working on surfaces of constant $\rho_\chi$, a fact that will become important when we calculate the fluctuations in these fields.  The equation of motion for $\vp$ during this period is then
\begin{equation}
	\ddot{\vp}+\frac{2}{t}\dot{\vp}+m^2\vp=0 \, .
\end{equation}
Restricting the solution to be real and constant in the limit $t\rightarrow0$, we find
\begin{equation}
	\vp(t)=\vp_*\frac{\sin(mt)}{mt} \, ,
\end{equation}
where $\vp_*$ is as usual the value of $\vp$ at Hubble exit.  One can then easily see that the energy density of the field $\vp$ is given by
\begin{equation}
	\rho_\vp=\frac{\vp_*^2}{2t^2}\left(1-\frac{2\cos(mt)\sin(mt)}{mt}+\frac{\sin^2(mt)}{m^2t^2}\right) \, .
\end{equation}
At late times when $mt\gg1$, we find
\begin{equation}\label{rhophisol}
	\rho_\vp=\frac{\vp_*^2}{2t^2} \, ,
\end{equation}
which is proportional to $a^{-3}$ as expected.

In order to compute observables, we need to know the ratio of energy densities while both fields are oscillating.  This ratio remains constant until one of the fields decays in the sudden decay approximation.  Using Eq.~(\ref{rhophisol}), we find
\begin{equation}
	\Omega_{\vp,{\rm osc}}=\frac{\rho_\vp}{3\Mp^2H^2}=\frac{3\vp_*^2}{8\Mp^2} \, .
	\label{eq:OmegaPhiEx}
\end{equation}

Next, we need to compute the perturbations to the fluid densities during the oscillating phase, $\zeta_\vp$ and $\zeta_\chi$.  Let us begin with the perturbations to $\rho_\chi$.  We are working in a regime where $\chi$ dominates the energy density during inflation, and $\vp$ remains essentially frozen until after inflation.  Under these assumptions, we can write the number of $e$--foldings during inflation, beginning from a flat hypersurface and ending on a surface of constant $\chi$, as
\begin{equation}
	N^\chi=\int_{\chi_*}^{\chi_e}\frac{H}{\dot{\chi}}\, d\chi=-\frac{1}{\Mp^2}\int_{\chi_*}^{\chi_e} \frac{U+V}{V'} \, d\chi \, ,
\end{equation}
where we have used the slow--roll equation of motion for $\chi$, which is $3H\dot{\chi}=-V'$.  Now, since $\vp$ is essentially frozen during inflation, $U$ is nearly constant over the range integration, and so we can replace $U$ with $U_*$.  This then gives
\begin{align}
	N^\chi&\approx -\frac{1}{\Mp^2}\int_{\chi_*}^{\chi_e}\frac{\frac{1}{2}m^2\vp_*^2+\frac{1}{2}M^2\chi^2}{M^2\chi}\, d\chi \nonumber \\
	&=-\frac{1}{\Mp^2}\left[\frac{m^2\vp_*^2}{2M^2}\ln \frac{\chi_e}{\chi_*}+\frac{1}{4}\left(\chi_e^2-\chi_*^2\right)\right] \, .
	\label{eq:NchiExample}
\end{align}
We will make the assumption that the perturbed number of $e$--foldings up to a surface of constant $\chi$ gives a good representation of the perturbed number of $e$--foldings up to a surface of constant $\rho_\chi$ during the phase of coherent oscillation.  In the absence of any contribution to the energy density from $\vp$, this approximation would be exact.  Since we are working in a regime where $\rho_\vp\ll\rho_\chi$ until well after inflation, this is quite a good approximation for our case.  We can then take a derivative of Eq.~(\ref{eq:NchiExample}) with respect to $\chi_*$ to find
\begin{equation}
	\frac{\partial N^\chi}{\partial \chi_*}=\frac{1}{\Mp^2}\left[\frac{m^2\vp_*^2}{2M^2\chi_*}+\frac{\chi_*}{2}\right] \, .
\end{equation}
Then, taking a derivative with respect to $\vp_*$ gives
\begin{equation}
	\frac{\partial^2 N^\chi}{\partial \phi_* \partial \chi_*}=\frac{1}{\Mp^2}\left[\frac{m^2\vp_*}{M^2\chi_*}\right] \, .
\end{equation}
Now, comparing this with Eqs.~(\ref{zetarelations}) and~(\ref{FirstOrderPerturbations}), we find
\begin{align}
	\frac{1}{\Mp^2}\left[\frac{m^2\vp_*^2}{2M^2\chi_*}+\frac{\chi_*}{2}\right]&=\frac{1}{\Mp^2}\left[\frac{\chi_*}{2}-\frac{F}{M^2\chi_*}\right] \, ,
\end{align}
which allows us to determine the quantity $F$
\begin{equation}
	F=-\frac{1}{2}m^2\vp_*^2 \, .
\end{equation}
Similarly, at second order we find
\begin{align}
	\frac{1}{\Mp^2}\left[\frac{m^2\vp_*}{M^2\chi_*}\right]&=\frac{1}{\Mp^2}\left[\frac{-J}{m^2\vp_* M^2 \chi_*}\right] \, ,
\end{align}
which gives 
\begin{equation}
	J=-m^4\vp_*^2
\end{equation}
for this model.\footnote{Our choice of which derivatives of $N^\chi$ to use in determining these quantities is motivated by maximizing the accuracy while minimizing additional assumptions.  Calculating $\frac{\partial N^\chi}{\partial \chi_*}$ rather than $\frac{\partial N^\chi}{\partial \vp_*}$ allows us to make progress without specifying how $\chi_e$ depends upon $\vp_*$.  Using the cross-derivative allows for the most direct determination of $J$, thus allowing us to avoid compounding any error we may have made in determining $F$, which would have been unavoidable had we compared to $\frac{\partial^2 N^\chi}{\partial \chi_*^2}$.  Furthermore, one can easily verify that the values of $F$ and $J$ which we determined here give $\drhochidphi=0$ and $\ddrhochiddphi=0$ using Eqs.~(\ref{densityderivatives1}) and~(\ref{densityderivatives2}) as one would expect from the fact that $\vp$ gives a negligible contribution to the energy density during inflation in the model under consideration.}

Next, we will calculate the perturbations to $\rho_\vp$.  For this task, we will make use of the results of Section~\ref{sec:EnergyDensities}.  However, in the case we are considering here, the value of $\rho_\vp$ given in Eq.~(\ref{rhophisol}) was computed on a surface of constant $\rho_\chi$ rather than on a flat hypersurface, as was assumed in Section~\ref{sec:EnergyDensities}.  This means that the factor of $\delta N$ appearing in Eq.~(\ref{eq:SD:zeta_I}) is given in this case by $\zeta_\chi$, and so Eq.~(\ref{eq:FlatHypersurface}) should be modified to read
\begin{equation}
	\rho_\vp=\bar\rho_\vp e^{3(\zeta_\vp-\zeta_\chi)}
\end{equation}
on this hypersurface.  However, using the values of $F$ and $J$ computed above, one finds that the part of $\zeta_\chi^{(1)}$ proportional to $\delta \vp_*$ and the part of $\zeta_\chi^{(2)}$ proportional to $\delta \vp_*^2$ both vanish.  Therefore, as long as we focus on derivatives of $\rho_\vp$ with respect to $\vp_*$, we can use the results of Section~\ref{sec:EnergyDensities} without modification to compute $G$ and $K$.  Given Eq.~(\ref{rhophisol}) it is straightforward to compute these derivatives with respect to $\vp_*$
\begin{equation}
	\frac{\partial  \rho_\vp}{\partial \vp_*}=\frac{\vp_*}{t^2} \, , \qquad \qquad \frac{\partial^2 \rho_\vp}{\partial \vp_*^2}=\frac{1}{t^2} \, .
\end{equation}
The relevant ratios for computing $\zeta_\vp$ are then
\begin{equation}
	\frac{1}{3\rho_\vp}\frac{\partial \rho_\vp}{\partial \vp_*}=\frac{2}{3\vp_*} \, , \qquad \qquad \frac{1}{3\rho_\vp}\frac{\partial^2 \rho_\vp}{\partial \vp_*^2}=\frac{2}{3\vp_*^2} \, .
\end{equation}
Recalling Eq. (\ref{densityderivatives1}), we find
\begin{align}
	\frac{2}{3\vp_*}&=\frac{1}{\Mp^2}\left[\frac{\vp_*}{2}+\frac{G}{m^2\vp_*}\right] \, ,
\end{align}
which allows us to determine the quantity $G$
\begin{equation}
	G=\frac{2}{3}m^2\Mp^2-\frac{1}{2}m^2\vp_*^2 \, .
\end{equation}
Using Eq. (\ref{densityderivatives2}), we find
\begin{equation}
	-\frac{2}{3\vp_*^2}=\frac{1}{\Mp^2}\left[1-\frac{1}{2}-\frac{G}{m^2\vp_*^2}+\frac{K}{m^4\vp_*^2}\right] \, ,
\end{equation}
which fixes the quantity $K$ to be
\begin{equation}
	K=-m^4\vp_*^2 \, .
\end{equation}

For comparison, we computed the functions $F$, $G$, $J$, and $K$ with our numerical field theory simulations using the methods (averaging and small/large $R$ limits) described in Section~\ref{sec:caseA}.  For $F$, $G$, and $J$, we found excellent agreement (within about 3\% for $\vp_*<0.3\Mp$ and even better for smaller $\vp_*$), but the function $K$ was smaller than our analytic prediction by about a factor of 0.6 for all values of $\vp_*$.  This disagreement of $K$ should not introduce large errors in computing observables since $K\ll Gm^2\vp_*^2$ for small $\vp_*$.

Now that we have computed the functions $F$, $G$, $J$, and $K$, we can begin to assemble the predictions for primordial observables.  Using Eq.~(\ref{alphabeta}) we find for this case
\begin{align}\label{eq:alphabetaexample}
	\alpha&\equiv
	\begin{cases}
		(1-\mcA)\left(\frac{2}{3}m^2\Mp^2\right)-\frac{1}{2}m^2\vp_*^2  \quad &\rm{for} \quad R\le1\\
		\mctA \left(\frac{2}{3}m^2\Mp^2\right)-\frac{1}{2}m^2\vp_*^2 \quad &\rm{for} \quad R\ge1 
	\end{cases} \, , \nonumber \\
	\beta&\equiv
	\begin{cases}
		-m^4\vp_*^2 + \mcB\left(\frac{4}{9}m^4\Mp^2\right) \quad &\rm{for} \quad R\le1\\
		-m^4\vp_*^2 + \mctB\left(\frac{4}{9}m^4\Mp^2\right) \quad &\rm{for} \quad R\ge1
	\end{cases} \, .
\end{align}
To achieve $\sim60$ $e$--foldings of inflation in this model, we require that $\chi_*\sim16 \Mp$, and so regardless of the values of $\mcA$, $\mcB$, $\mctA$, and $\mctB$, we always have $V_*\gg\alpha$ and $V_*''V_*\gg\beta$.  This means that up to small corrections, we have
\begin{align}\label{eq:ChiDerivs}
	N_{,\chi}&=\frac{\chi_*}{2\Mp^2} \, , \nonumber \\
	N_{,\chi\chi}&=\frac{1}{2\Mp^2} \, ,
\end{align}
regardless of the value of $R$.  The other derivatives of $N$ are affected by reheating, and are given by
\begin{align}
	N_{\vp}&=
	\begin{cases}
		(1-\mcA)\left(\frac{2}{3\vp_*}\right)  \quad &\rm{for} \quad R\le1\\
		\mctA \left(\frac{2}{3\vp_*}\right) \quad &\rm{for} \quad R\ge1 
	\end{cases} \, , \nonumber \\
	N_{,\vp\vp}&=
	\begin{cases}
		\left(\frac{2}{3}\mcB-(1-\mcA)\right)\left(\frac{2}{3\vp_*^2}\right) \quad &\rm{for} \quad R\le1\\
		\left(\frac{2}{3}\mctB-\mctA\right)\left(\frac{2}{3\vp_*^2}\right) \quad &\rm{for} \quad R\ge1
	\end{cases} \, , \nonumber \\
	N_{,\vp\chi}&=
	\begin{cases}
		\frac{m^2\vp_*}{\Mp^2M^2\chi_*}-\mcB\frac{4m^2}{9M^2\vp_*\chi_*} \quad &\rm{for} \quad R\le1\\
		\frac{m^2\vp_*}{\Mp^2M^2\chi_*}-\mctB\frac{4m^2}{9M^2\vp_*\chi_*} \quad &\rm{for} \quad R\ge1\\
	\end{cases} \, .
\end{align}

In the regime $R\le1$, we have $r\approx 1$ which means $\mcA\approx 1$ and $\mcB\approx 0$, which makes $N_{,\vp}\ll N_{,\chi}$ and $N_{,\vp\vp},N_{\vp,\chi}\ll N_{,\chi\chi}$, thus giving us essentially the same predictions as single--field inflation (up to small corrections which are at most of order $m^2/M^2$).  To be specific, the primordial observables in the range $R\le1$ are given by
\begin{equation}
	\mathcal{P}_{\zeta}=\frac{M^2\chi_*^4}{96\pi^2\Mp^6} \, ,
\label{eq:PzetaEx}
\end{equation}
\begin{equation}
	r_T=\frac{32\Mp^2}{\chi_*^2} \, ,
\label{eq:rTEx}
\end{equation}
\begin{equation}
	n_{\zeta}-1=-\frac{8\Mp^2}{\chi_*^2} \, ,
\label{eq:nzEx}
\end{equation}
\begin{equation}
	\frac{6}{5}f_{\mathrm{NL}}=\frac{2\Mp^2}{\chi_*^2} \, ,
\label{eq:fnlEx}
\end{equation}
just as expected for a single--field inflationary model with $V(\chi)=\frac{1}{2}M^2\chi^2$.  Note that in this regime, we should also include the shape dependent part of $\fnl$ since it is comparable in size to the shape independent part we calculated here.  In the local limit, the shape dependent part~\cite{Vernizzi:2006ve} adds an additional factor of $\frac{2\Mp^2}{\chi_*^2}$, which must be included to verify the single--field consistency relation.

Now let us turn to the regime $R>1$, where the effects of the field $\vp$ are not negligible.  In this regime,  the observables are given by
\begin{equation}
	\mathcal{P}_{\zeta}=\frac{M^2\chi_*^2}{24\pi^2\Mp^4}\left[\frac{\chi_*^2}{4\Mp^2}+\mctA^2\left(\frac{4\Mp^2}{9\vp_*^2}\right)\right] \, ,
\label{eq:PzetaEx2}
\end{equation}
\begin{equation}
	r_T=8\left[\frac{\chi_*^2}{4\Mp^2}+\mctA^2\left(\frac{4\Mp^2}{9\vp_*^2}\right)\right]^{-1} \, ,
\label{eq:rTEx2}
\end{equation}
\begin{align}
	n_{\zeta}-1=&-\frac{4\Mp^2}{\chi_*^2}-\frac{4\Mp^4}{M^2\chi_*^2}\left[\frac{\chi_*^2}{4\Mp^2}+\mctA^2\left(\frac{4\Mp^2}{9\vp_*^2}\right)\right]^{-1} \nonumber \\
	&\times\left[\left(\frac{M^2\chi_*^2}{4\Mp^4}\right)+\mctA\left(\frac{2m^2}{3\Mp^2}\right)-\mctA^2\left(\frac{4m^2}{9\vp_*^2}\right)\right] \nonumber \\
	\simeq&-\frac{4\Mp^2}{\chi_*^2}-\frac{1-\mctA^2\left(\frac{16m^2\Mp^4}{9M^2\chi_*^2\vp_*^2}\right)}{\frac{\chi_*^2}{4\Mp^2}+\mctA^2\left(\frac{4\Mp^2}{9\vp_*^2}\right)} \, ,
\label{eq:nzEx2}
\end{align}
\begin{align}
	\frac{6}{5}f_{\mathrm{NL}}=&\left[\frac{\chi_*^2}{4\Mp^2}+\mctA^2\left(\frac{4\Mp^2}{9\vp_*^2}\right)\right]^{-2} \nonumber \\
	&\times\Bigg[\frac{\chi_*^2}{8\Mp^2} +\mctA\frac{2m^2}{3M^2}-\mctA\mctB\frac{8m^2\Mp^2}{27M^2\vp_*^2} \nonumber \\
	&\qquad+\left(\frac{2}{3}\mctA^2\mctB-\mctA^3\right)\left(\frac{8\Mp^4}{27\vp_*^4}\right)\Bigg] \nonumber \\
	\simeq&\frac{\frac{\chi_*^2}{8\Mp^2}-\mctA\mctB\left(\frac{8m^2\Mp^2}{27M^2\vp_*^2}\right)+\left(\frac{2}{3}\mctA^2\mctB-\mctA^3\right)\left(\frac{8\Mp^4}{27\vp_*^4}\right)} {\left[\frac{\chi_*^2}{4\Mp^2}+\mctA^2\left(\frac{4\Mp^2}{9\vp_*^2}\right)\right]^2} \, .
\label{eq:fnlEx2}
\end{align}

\begin{figure*}[t]
\vspace{-20mm}
	\begin{tabular}{cc}
		\includegraphics[width=8.8cm]{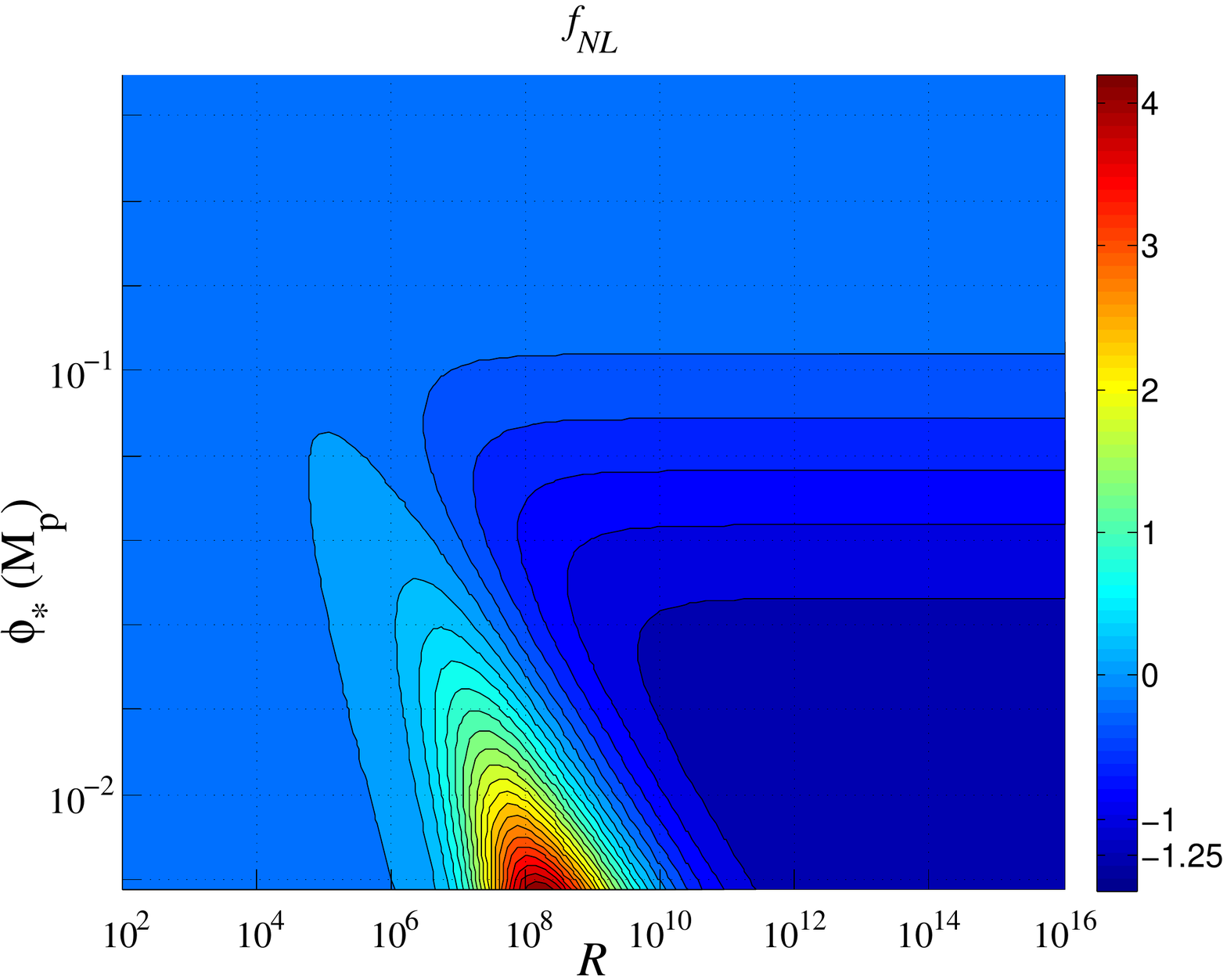} &
		\includegraphics[width=8.8cm]{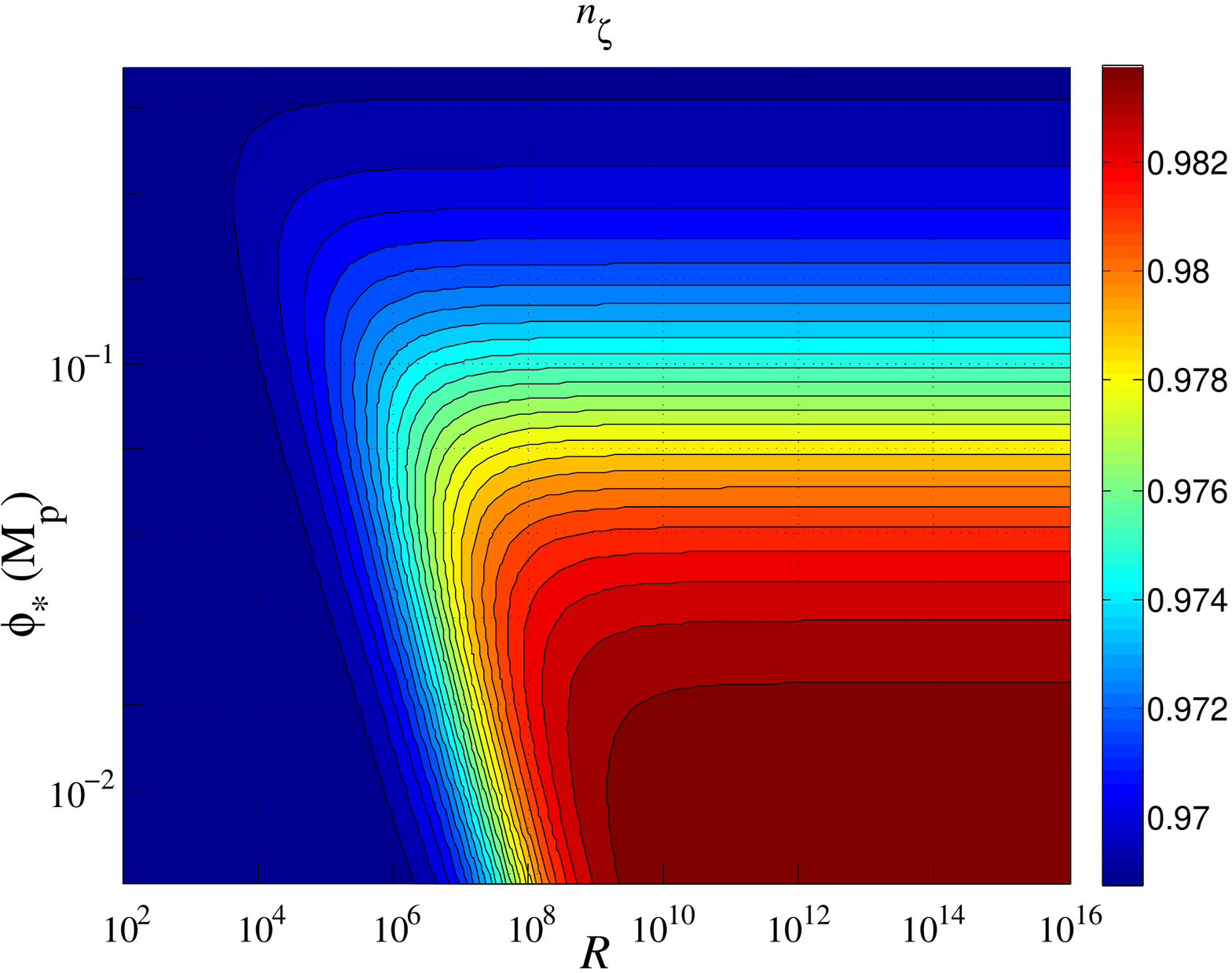} 
	\end{tabular}
\vspace{-20mm}
	\caption{Heatmaps of $\fnl$ (left panel) and $\nz$ (right panel) as a function of $R\equiv\Gchi/\Gvp$ and $\vp_*$ for the quadratic curvaton example of Section~\ref{sec:analytic_example}. As $R\to\infty$, we recover the standard curvaton limit $\fnl=-1.25$ for small $\vp_*$. These heatmaps were produced using the fully analytic solution presented in Section~\ref{sec:analytic_example}.}
	\label{fig:curvaton_heatmaps}
\end{figure*}

\begin{figure}[t]
\vspace{-20mm}
	\begin{tabular}{c}
		\includegraphics[width=8.8cm]{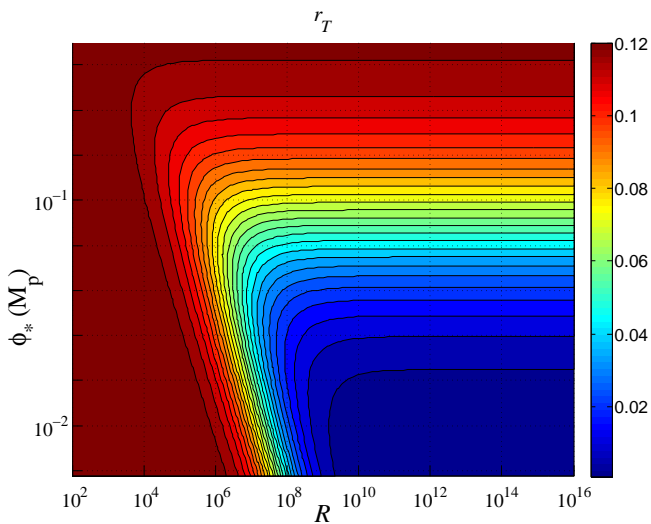}
	\end{tabular}
\vspace{-20mm}
	\caption{Heatmap of $r_T$ as a function of $R\equiv\Gchi/\Gvp$ and $\vp_*$ for the quadratic curvaton example of Section~\ref{sec:analytic_example}. This heatmap was produced using the fully analytic solution presented in Section~\ref{sec:analytic_example}.}
	\label{fig:curvaton_heatmaps2}
\end{figure}

There are several important things to notice here.  First, the effect of the field $\vp_*$ is to enhance $\mathcal P_\zeta$ relative to the single--field inflation case, thereby suppressing $r_T$, regardless of the value of $\mctA$.  This is to be expected, since the tensor power is fixed by the scale of inflation, while the scalar power receives independent contributions from each of the fields.  Next, the deviation of the spectral tilt from the single--field result is suppressed by $m^2/M^2$, and so for $m^2\ll M^2$, the spectral tilt is not particularly sensitive to reheating for this example.  However, the dependence of $\fnl$ on reheating is more complicated, and we see that $\fnl$ can deviate quite significantly from the single--field case.  In the regime $\vp_*\rightarrow 0$ and $\tilde r\rightarrow 1$ (such that $\mctA\rightarrow 1$ and $\mctB\rightarrow 0$), we find that $\fnl\rightarrow-5/4$, which is just what is predicted for a curvaton model where the curvaton comes to dominate the energy density of the universe before decaying~\cite{Lyth:2002my,Sasaki:2006kq}.  As we move from $\tilde r\simeq 1$ to the regime $\tilde r<1$, at first we reproduce the $\fnl\approx5/4r$ behavior of the curvaton scenario, but we soon enter a regime where the perturbations in the $\chi$ field cannot be neglected, and $\fnl$ begins to decrease toward the value it would have in single--field inflation.  This regime produces results similar to the predictions of the mixed inflaton--curvaton scenario \cite{Fonseca:2012cj}.  Had we dropped the contributions from the $\chi$ field entirely, we would have found that $\fnl$ increased without bound as we decreased $\tilde r$.  The fact that the contribution from $\chi$ limits the maximum value of $\fnl$ reinforces the point that the fluctuations of each field must in general be taken into account.

The full set of predictions of this model are shown in Figs.~\ref{fig:curvaton_heatmaps} and~\ref{fig:curvaton_heatmaps2} for a range of $\vp_*$ and $R$ with $\chi_*=16\Mp$ and $M^2/m^2=200$.  These plots are produced using Eq.~(\ref{eq:OmegaPhiEx}) to calculate $\tilde \Omega_{\chi,\mathrm{dec}}=1-\Omega_{\vp,\mathrm{osc}}$ and the fitting formula Eq.~(\ref{eq:tilder-fit}) for $\tilde{r}$. We can see from these plots that this model provides a smooth transition between the predictions of two--field inflation and the curvaton scenario.  Note that there is a small region of parameter space that predicts a detectable level of local non--Gaussianity.

Before ending this section, let us briefly comment on the physically viable range of $R$ for this example.  The amplitude of fluctuations is fixed by observation, and so we can use Eq.~(\ref{eq:PzetaEx2}) to determine the scale of inflation necessary to match data.  Specifically, with $V_{\mathrm{inf}}\simeq\frac{1}{2}M^2\chi^2$ we will fix $\chi_*=16 \Mp$ to ensure around 60 $e$--foldings of inflation, and rescale $M$ to match observations.  The condition that fixes $M$ is then~\cite{Ade:2013zuv}
\begin{equation}
	2.215\times10^{-9}=\frac{M^2\chi_*^2}{24\pi^2\Mp^4}\left[\frac{\chi_*^2}{4\Mp^2}+\mctA^2\left(\frac{4\Mp^2}{9\vp_*^2}\right)\right] \, .
\label{eq:PzetaObs}
\end{equation}
For $\mctA\ll 1$, this requires $V_{\mathrm{inf}}\approx[0.008\,\Mp]^4$ for $\chi_*=16\Mp$. In the regime $\mctA\approx1$ and $\vp_*\ll\Mp$, we have
\begin{equation}
 V_{\mathrm{inf}}\approx\left[0.03\,\Mp\left(\frac{\vp_*}{\Mp}\right)^{1/2}\right]^4 \, .
 \label{eq:VinfEx}
\end{equation}
Recalling that the maximum value of $R$ is set by requiring that reheating begins after inflation and completes before dark matter decoupling, we find
\begin{equation}
 R_{\rm max}\approx6\times10^{19}\left(\frac{\vp_*}{\Mp}\right)^{1/2}\left(\frac{1\,\rm{ MeV}}{k_{\rm B}T_{\rm DM}}\right) \, ,
 \label{eq:RmaxEx}
\end{equation}
where $T_{\rm DM}$ is the temperature of dark matter decoupling.  For small values of $\vp_*$ and large $T_{\rm DM}$, this constraint begins to become important for the range of $R$ that we plotted in Figs.~\ref{fig:curvaton_heatmaps} and~\ref{fig:curvaton_heatmaps2}.

%------------------------------------------------------------------
\section{Discussion and Conclusions}\label{sec:conclusions}

We have shown how perturbative reheating impacts primordial observables following two--field inflation.  Our main results are summarized in Eqs.~(\ref{eq:SD_Pzeta}--\ref{eq:SD_fnl}).  These expressions were derived analytically using the sudden decay approximation, and were shown to agree remarkably well with the results of a fully numerical classical field theory simulation.  Our results apply to any two--field model of inflation with a potential of the form $W(\vp,\chi)=U(\vp)+V(\chi)$ where each field has a quadratic minimum. Our analytic results will be most reliable in the regime where the first field to decay does so when the other field is oscillating. Since observable quantities depend only upon the \emph{ratio} of decay rates $R$, this is not a particularly restrictive condition. We have shown that this class of models includes the results of the standard curvaton scenario as a special case, but it also applies to a much broader set of inflationary models.

We have shown that primordial observables take values within finite ranges whose limits are set entirely by the conditions during inflation. The presence of fluctuations in \emph{both} fields are of crucial importance in determining these bounds. Since it is the details of the inflationary model alone which determine these ranges, it is possible to compute the maximum and minimum values of all primordial observables without specifying the details of the reheating phase. Through concrete examples, we have demonstrated that these ranges can lie well within the sensitivity of current experiments such as \emph{Planck}.  As was previously established in Ref.~\cite{Leung:2012ve}, the degree of sensitivity of primordial observables to the reheating phase depends heavily on the underlying inflationary model. In this work, we have quantified this sensitivity, and have discussed the conditions that must be satisfied by the inflationary model if its observable predictions are to be insensitive to the physics of reheating. The effects of reheating are important whenever $\zeta_\vp\neq\zeta_\chi$ at the end of inflation, as is generally the case when the adiabatic limit is not reached during inflation. If this is true of a particular two--field inflationary model then one \emph{must} account for the impact of reheating when computing observable quantities.

Furthermore, we have firmly established that local non--Gaussianity is not in general damped toward small values by reheating, as is often (but not always~\cite{Kim:2010ud}) the case during multiple--field inflation if the adiabatic limit is reached before the end of inflation~\cite{Meyers:2010rg,Meyers:2011mm,Watanabe:2011sm}.

Our results allow a much more unified approach to studying two--field inflation including the effects of perturbative reheating.  As such, entire classes of models can be studied together, allowing a much more systematic approach to gaining insight into the physics of the early universe through observation.

There are several ways in which this work can be extended.  We restricted ourselves to two fields in the analysis presented here, but it would be straightforward to extend the formalism to an arbitrary number of fields.  We could also carry out the calculations to higher order which would allow us to study higher order statistics of the curvature perturbation, such as $\taunl$ and $\gnl$. One could investigate consistency relations between observables such as the Suyama--Yamaguchi inequality~\cite{Suyama:2007bg}, which relates $\fnl$ in the squeezed limit to $\taunl$ in the collapsed limit: $\taunl\geq\left(\frac65\fnl\right)^2$.  Recent numerical work seems to suggest that the Suyama--Yamaguchi \emph{equality} cannot be strongly broken as a result of reheating~\cite{Leung:2013rza}, (i.e., $\taunl\gg\left(\frac65\fnl\right)^2$); however, the very small/large $R$ limits were not explored deeply. It was observed in Ref.~\cite{Fonseca:2012cj}, that an observation of scale--dependence of $\fnl$ may distinguish mixed inflaton--curvaton models from the pure curvaton limit for a quadratic curvaton potential. It would be very interesting to apply our results to $n_{\fnl}$ (as was recently studied numerically in~\cite{Leung:2013rza}) to generalize such claims.  It would also be interesting to study more general potentials.  Throughout this work, we assumed that the decay rates were constants, but it may be possible to modify our results in order to allow these decay rates to depend on the values of some scalar fields, which would allow us to study modulated reheating.  As briefly discussed at the end of Section \ref{sec:Observables}, we intend to carry out a more systematic study of the models which are capable of producing large local non--Gaussianity.  This will help us to understand exactly what observational constraints on $\fnl$ can teach us about models of the early universe.

\section*{Acknowledgements}

The authors would like to thank Christian Byrnes, Joseph Elliston, Takeshi Kobayashi, David J.E. Marsh, Donough Regan, David Seery, Navin Sivanandam, and David Wands for helpful conversations.  ERMT acknowledges support from the Leverhulme Trust, and from the University of Nottingham.

% ------------------------ APPENDIX ----------------------------------

% ---------------------- BIBLIOGRAPHY -----------------------------------------

\bibliographystyle{apsrev4-1}
\bibliography{bib_file}

%merlin.mbs apsrev4-1.bst 2010-07-25 4.21a (PWD, AO, DPC) hacked
%Control: key (0)
%Control: author (72) initials jnrlst
%Control: editor formatted (1) identically to author
%Control: production of article title (-1) disabled
%Control: page (0) single
%Control: year (1) truncated
%Control: production of eprint (0) enabled
\begin{thebibliography}{92}%
\makeatletter
\providecommand \@ifxundefined [1]{%
 \@ifx{#1\undefined}
}%
\providecommand \@ifnum [1]{%
 \ifnum #1\expandafter \@firstoftwo
 \else \expandafter \@secondoftwo
 \fi
}%
\providecommand \@ifx [1]{%
 \ifx #1\expandafter \@firstoftwo
 \else \expandafter \@secondoftwo
 \fi
}%
\providecommand \natexlab [1]{#1}%
\providecommand \enquote  [1]{``#1''}%
\providecommand \bibnamefont  [1]{#1}%
\providecommand \bibfnamefont [1]{#1}%
\providecommand \citenamefont [1]{#1}%
\providecommand \href@noop [0]{\@secondoftwo}%
\providecommand \href [0]{\begingroup \@sanitize@url \@href}%
\providecommand \@href[1]{\@@startlink{#1}\@@href}%
\providecommand \@@href[1]{\endgroup#1\@@endlink}%
\providecommand \@sanitize@url [0]{\catcode `\\12\catcode `\$12\catcode
  `\&12\catcode `\#12\catcode `\^12\catcode `\_12\catcode `\%12\relax}%
\providecommand \@@startlink[1]{}%
\providecommand \@@endlink[0]{}%
\providecommand \url  [0]{\begingroup\@sanitize@url \@url }%
\providecommand \@url [1]{\endgroup\@href {#1}{\urlprefix }}%
\providecommand \urlprefix  [0]{URL }%
\providecommand \Eprint [0]{\href }%
\providecommand \doibase [0]{http://dx.doi.org/}%
\providecommand \selectlanguage [0]{\@gobble}%
\providecommand \bibinfo  [0]{\@secondoftwo}%
\providecommand \bibfield  [0]{\@secondoftwo}%
\providecommand \translation [1]{[#1]}%
\providecommand \BibitemOpen [0]{}%
\providecommand \bibitemStop [0]{}%
\providecommand \bibitemNoStop [0]{.\EOS\space}%
\providecommand \EOS [0]{\spacefactor3000\relax}%
\providecommand \BibitemShut  [1]{\csname bibitem#1\endcsname}%
\let\auto@bib@innerbib\@empty
%</preamble>
\bibitem [{\citenamefont {Starobinsky}(1979)}]{Starobinsky:1979ty}%
  \BibitemOpen
  \bibfield  {author} {\bibinfo {author} {\bibfnamefont {A.~A.}\ \bibnamefont
  {Starobinsky}},\ }\href@noop {} {\bibfield  {journal} {\bibinfo  {journal}
  {JETP Lett.}\ }\textbf {\bibinfo {volume} {30}},\ \bibinfo {pages} {682}
  (\bibinfo {year} {1979})}\BibitemShut {NoStop}%
%%CITATION = JTPLA,30,682;%%
\bibitem [{\citenamefont {Starobinsky}(1980)}]{Starobinsky:1980te}%
  \BibitemOpen
  \bibfield  {author} {\bibinfo {author} {\bibfnamefont {A.~A.}\ \bibnamefont
  {Starobinsky}},\ }\href {\doibase 10.1016/0370-2693(80)90670-X} {\bibfield
  {journal} {\bibinfo  {journal} {Phys.Lett.}\ }\textbf {\bibinfo {volume}
  {B91}},\ \bibinfo {pages} {99} (\bibinfo {year} {1980})}\BibitemShut
  {NoStop}%
%%CITATION = PHLTA,B91,99;%%
\bibitem [{\citenamefont {Kazanas}(1980)}]{Kazanas:1980tx}%
  \BibitemOpen
  \bibfield  {author} {\bibinfo {author} {\bibfnamefont {D.}~\bibnamefont
  {Kazanas}},\ }\href@noop {} {\bibfield  {journal} {\bibinfo  {journal}
  {Astrophys.J.}\ }\textbf {\bibinfo {volume} {241}},\ \bibinfo {pages} {L59}
  (\bibinfo {year} {1980})}\BibitemShut {NoStop}%
%%CITATION = ASJOA,241,L59;%%
\bibitem [{\citenamefont {Sato}(1981)}]{Sato:1980yn}%
  \BibitemOpen
  \bibfield  {author} {\bibinfo {author} {\bibfnamefont {K.}~\bibnamefont
  {Sato}},\ }\href@noop {} {\bibfield  {journal} {\bibinfo  {journal}
  {Mon.Not.Roy.Astron.Soc.}\ }\textbf {\bibinfo {volume} {195}},\ \bibinfo
  {pages} {467} (\bibinfo {year} {1981})}\BibitemShut {NoStop}%
%%CITATION = MNRAA,195,467;%%
\bibitem [{\citenamefont {Guth}(1981)}]{Guth:1980zm}%
  \BibitemOpen
  \bibfield  {author} {\bibinfo {author} {\bibfnamefont {A.~H.}\ \bibnamefont
  {Guth}},\ }\href {\doibase 10.1103/PhysRevD.23.347} {\bibfield  {journal}
  {\bibinfo  {journal} {Phys.Rev.}\ }\textbf {\bibinfo {volume} {D23}},\
  \bibinfo {pages} {347} (\bibinfo {year} {1981})}\BibitemShut {NoStop}%
%%CITATION = PHRVA,D23,347;%%
\bibitem [{\citenamefont {Mukhanov}\ and\ \citenamefont
  {Chibisov}(1981)}]{Mukhanov:1981xt}%
  \BibitemOpen
  \bibfield  {author} {\bibinfo {author} {\bibfnamefont {V.~F.}\ \bibnamefont
  {Mukhanov}}\ and\ \bibinfo {author} {\bibfnamefont {G.~V.}\ \bibnamefont
  {Chibisov}},\ }\href@noop {} {\bibfield  {journal} {\bibinfo  {journal} {JETP
  Lett.}\ }\textbf {\bibinfo {volume} {33}},\ \bibinfo {pages} {532} (\bibinfo
  {year} {1981})}\BibitemShut {NoStop}%
%%CITATION = JTPLA,33,532;%%
\bibitem [{\citenamefont {Hawking}(1982)}]{Hawking:1982cz}%
  \BibitemOpen
  \bibfield  {author} {\bibinfo {author} {\bibfnamefont {S.}~\bibnamefont
  {Hawking}},\ }\href {\doibase 10.1016/0370-2693(82)90373-2} {\bibfield
  {journal} {\bibinfo  {journal} {Phys.Lett.}\ }\textbf {\bibinfo {volume}
  {B115}},\ \bibinfo {pages} {295} (\bibinfo {year} {1982})}\BibitemShut
  {NoStop}%
%%CITATION = PHLTA,B115,295;%%
\bibitem [{\citenamefont {Guth}\ and\ \citenamefont {Pi}(1982)}]{Guth:1982ec}%
  \BibitemOpen
  \bibfield  {author} {\bibinfo {author} {\bibfnamefont {A.~H.}\ \bibnamefont
  {Guth}}\ and\ \bibinfo {author} {\bibfnamefont {S.}~\bibnamefont {Pi}},\
  }\href {\doibase 10.1103/PhysRevLett.49.1110} {\bibfield  {journal} {\bibinfo
   {journal} {Phys.Rev.Lett.}\ }\textbf {\bibinfo {volume} {49}},\ \bibinfo
  {pages} {1110} (\bibinfo {year} {1982})}\BibitemShut {NoStop}%
%%CITATION = PRLTA,49,1110;%%
\bibitem [{\citenamefont {Starobinsky}(1982)}]{Starobinsky:1982ee}%
  \BibitemOpen
  \bibfield  {author} {\bibinfo {author} {\bibfnamefont {A.~A.}\ \bibnamefont
  {Starobinsky}},\ }\href {\doibase 10.1016/0370-2693(82)90541-X} {\bibfield
  {journal} {\bibinfo  {journal} {Phys.Lett.}\ }\textbf {\bibinfo {volume}
  {B117}},\ \bibinfo {pages} {175} (\bibinfo {year} {1982})}\BibitemShut
  {NoStop}%
%%CITATION = PHLTA,B117,175;%%
\bibitem [{\citenamefont {Bardeen}\ \emph {et~al.}(1983)\citenamefont
  {Bardeen}, \citenamefont {Steinhardt},\ and\ \citenamefont
  {Turner}}]{Bardeen:1983qw}%
  \BibitemOpen
  \bibfield  {author} {\bibinfo {author} {\bibfnamefont {J.~M.}\ \bibnamefont
  {Bardeen}}, \bibinfo {author} {\bibfnamefont {P.~J.}\ \bibnamefont
  {Steinhardt}}, \ and\ \bibinfo {author} {\bibfnamefont {M.~S.}\ \bibnamefont
  {Turner}},\ }\href {\doibase 10.1103/PhysRevD.28.679} {\bibfield  {journal}
  {\bibinfo  {journal} {Phys.Rev.}\ }\textbf {\bibinfo {volume} {D28}},\
  \bibinfo {pages} {679} (\bibinfo {year} {1983})}\BibitemShut {NoStop}%
%%CITATION = PHRVA,D28,679;%%
\bibitem [{\citenamefont {Ade}\ \emph {et~al.}(2013{\natexlab{a}})\citenamefont
  {Ade} \emph {et~al.}}]{Ade:2013uln}%
  \BibitemOpen
  \bibfield  {author} {\bibinfo {author} {\bibfnamefont {P.}~\bibnamefont
  {Ade}} \emph {et~al.} (\bibinfo {collaboration} {Planck Collaboration}),\
  }\href@noop {} {\  (\bibinfo {year} {2013}{\natexlab{a}})},\ \Eprint
  {http://arxiv.org/abs/1303.5082} {arXiv:1303.5082 [astro-ph.CO]} \BibitemShut
  {NoStop}%
%%CITATION = ARXIV:1303.5082;%%
\bibitem [{\citenamefont {Weinberg}(2008{\natexlab{a}})}]{Weinberg:2008zzc}%
  \BibitemOpen
  \bibfield  {author} {\bibinfo {author} {\bibfnamefont {S.}~\bibnamefont
  {Weinberg}},\ }\href@noop {} {\emph {\bibinfo {title} {{Cosmology}}}}\
  (\bibinfo  {publisher} {OUP Oxford},\ \bibinfo {year} {2008})\BibitemShut
  {NoStop}%
%%CITATION = ISBN-9780198526827 ETC.;%%
\bibitem [{\citenamefont {Weinberg}(2003)}]{Weinberg:2003sw}%
  \BibitemOpen
  \bibfield  {author} {\bibinfo {author} {\bibfnamefont {S.}~\bibnamefont
  {Weinberg}},\ }\href {\doibase 10.1103/PhysRevD.67.123504} {\bibfield
  {journal} {\bibinfo  {journal} {Phys.Rev.}\ }\textbf {\bibinfo {volume}
  {D67}},\ \bibinfo {pages} {123504} (\bibinfo {year} {2003})},\ \Eprint
  {http://arxiv.org/abs/astro-ph/0302326} {arXiv:astro-ph/0302326 [astro-ph]}
  \BibitemShut {NoStop}%
%%CITATION = ASTRO-PH/0302326;%%
\bibitem [{\citenamefont {Weinberg}(2004{\natexlab{a}})}]{Weinberg:2004kr}%
  \BibitemOpen
  \bibfield  {author} {\bibinfo {author} {\bibfnamefont {S.}~\bibnamefont
  {Weinberg}},\ }\href {\doibase 10.1103/PhysRevD.70.043541} {\bibfield
  {journal} {\bibinfo  {journal} {Phys.Rev.}\ }\textbf {\bibinfo {volume}
  {D70}},\ \bibinfo {pages} {043541} (\bibinfo {year} {2004}{\natexlab{a}})},\
  \Eprint {http://arxiv.org/abs/astro-ph/0401313} {arXiv:astro-ph/0401313
  [astro-ph]} \BibitemShut {NoStop}%
%%CITATION = ASTRO-PH/0401313;%%
\bibitem [{\citenamefont {Weinberg}(2008{\natexlab{b}})}]{Weinberg:2008nf}%
  \BibitemOpen
  \bibfield  {author} {\bibinfo {author} {\bibfnamefont {S.}~\bibnamefont
  {Weinberg}},\ }\href {\doibase 10.1103/PhysRevD.78.123521} {\bibfield
  {journal} {\bibinfo  {journal} {Phys.Rev.}\ }\textbf {\bibinfo {volume}
  {D78}},\ \bibinfo {pages} {123521} (\bibinfo {year} {2008}{\natexlab{b}})},\
  \Eprint {http://arxiv.org/abs/0808.2909} {arXiv:0808.2909 [hep-th]}
  \BibitemShut {NoStop}%
%%CITATION = ARXIV:0808.2909;%%
\bibitem [{\citenamefont {Meyers}\ and\ \citenamefont
  {Sivanandam}(2011{\natexlab{a}})}]{Meyers:2010rg}%
  \BibitemOpen
  \bibfield  {author} {\bibinfo {author} {\bibfnamefont {J.}~\bibnamefont
  {Meyers}}\ and\ \bibinfo {author} {\bibfnamefont {N.}~\bibnamefont
  {Sivanandam}},\ }\href {\doibase 10.1103/PhysRevD.83.103517} {\bibfield
  {journal} {\bibinfo  {journal} {Phys.Rev.}\ }\textbf {\bibinfo {volume}
  {D83}},\ \bibinfo {pages} {103517} (\bibinfo {year} {2011}{\natexlab{a}})},\
  \Eprint {http://arxiv.org/abs/1011.4934} {arXiv:1011.4934 [astro-ph.CO]}
  \BibitemShut {NoStop}%
%%CITATION = ARXIV:1011.4934;%%
\bibitem [{\citenamefont {Meyers}\ and\ \citenamefont
  {Sivanandam}(2011{\natexlab{b}})}]{Meyers:2011mm}%
  \BibitemOpen
  \bibfield  {author} {\bibinfo {author} {\bibfnamefont {J.}~\bibnamefont
  {Meyers}}\ and\ \bibinfo {author} {\bibfnamefont {N.}~\bibnamefont
  {Sivanandam}},\ }\href {\doibase 10.1103/PhysRevD.84.063522} {\bibfield
  {journal} {\bibinfo  {journal} {Phys.Rev.}\ }\textbf {\bibinfo {volume}
  {D84}},\ \bibinfo {pages} {063522} (\bibinfo {year} {2011}{\natexlab{b}})},\
  \Eprint {http://arxiv.org/abs/1104.5238} {arXiv:1104.5238 [astro-ph.CO]}
  \BibitemShut {NoStop}%
%%CITATION = ARXIV:1104.5238;%%
\bibitem [{\citenamefont {Weinberg}(2004{\natexlab{b}})}]{Weinberg:2004kf}%
  \BibitemOpen
  \bibfield  {author} {\bibinfo {author} {\bibfnamefont {S.}~\bibnamefont
  {Weinberg}},\ }\href {\doibase 10.1103/PhysRevD.70.083522} {\bibfield
  {journal} {\bibinfo  {journal} {Phys.Rev.}\ }\textbf {\bibinfo {volume}
  {D70}},\ \bibinfo {pages} {083522} (\bibinfo {year} {2004}{\natexlab{b}})},\
  \Eprint {http://arxiv.org/abs/astro-ph/0405397} {arXiv:astro-ph/0405397
  [astro-ph]} \BibitemShut {NoStop}%
%%CITATION = ASTRO-PH/0405397;%%
\bibitem [{\citenamefont {Weinberg}(2009)}]{Weinberg:2008si}%
  \BibitemOpen
  \bibfield  {author} {\bibinfo {author} {\bibfnamefont {S.}~\bibnamefont
  {Weinberg}},\ }\href {\doibase 10.1103/PhysRevD.79.043504} {\bibfield
  {journal} {\bibinfo  {journal} {Phys.Rev.}\ }\textbf {\bibinfo {volume}
  {D79}},\ \bibinfo {pages} {043504} (\bibinfo {year} {2009})},\ \Eprint
  {http://arxiv.org/abs/0810.2831} {arXiv:0810.2831 [hep-ph]} \BibitemShut
  {NoStop}%
%%CITATION = ARXIV:0810.2831;%%
\bibitem [{\citenamefont {Meyers}(2012)}]{Meyers:2012ni}%
  \BibitemOpen
  \bibfield  {author} {\bibinfo {author} {\bibfnamefont {J.}~\bibnamefont
  {Meyers}},\ }\href@noop {} {\  (\bibinfo {year} {2012})},\ \Eprint
  {http://arxiv.org/abs/1212.4438} {arXiv:1212.4438 [astro-ph.CO]} \BibitemShut
  {NoStop}%
%%CITATION = ARXIV:1212.4438;%%
\bibitem [{\citenamefont {Linde}(1982{\natexlab{a}})}]{Linde:1981mu}%
  \BibitemOpen
  \bibfield  {author} {\bibinfo {author} {\bibfnamefont {A.~D.}\ \bibnamefont
  {Linde}},\ }\href {\doibase 10.1016/0370-2693(82)91219-9} {\bibfield
  {journal} {\bibinfo  {journal} {Phys.Lett.}\ }\textbf {\bibinfo {volume}
  {B108}},\ \bibinfo {pages} {389} (\bibinfo {year}
  {1982}{\natexlab{a}})}\BibitemShut {NoStop}%
%%CITATION = PHLTA,B108,389;%%
\bibitem [{\citenamefont {Linde}(1982{\natexlab{b}})}]{Linde:1982zj}%
  \BibitemOpen
  \bibfield  {author} {\bibinfo {author} {\bibfnamefont {A.~D.}\ \bibnamefont
  {Linde}},\ }\href {\doibase 10.1016/0370-2693(82)90086-7} {\bibfield
  {journal} {\bibinfo  {journal} {Phys.Lett.}\ }\textbf {\bibinfo {volume}
  {B114}},\ \bibinfo {pages} {431} (\bibinfo {year}
  {1982}{\natexlab{b}})}\BibitemShut {NoStop}%
%%CITATION = PHLTA,B114,431;%%
\bibitem [{\citenamefont {Linde}(1982{\natexlab{c}})}]{Linde:1982uu}%
  \BibitemOpen
  \bibfield  {author} {\bibinfo {author} {\bibfnamefont {A.~D.}\ \bibnamefont
  {Linde}},\ }\href {\doibase 10.1016/0370-2693(82)90293-3} {\bibfield
  {journal} {\bibinfo  {journal} {Phys.Lett.}\ }\textbf {\bibinfo {volume}
  {B116}},\ \bibinfo {pages} {335} (\bibinfo {year}
  {1982}{\natexlab{c}})}\BibitemShut {NoStop}%
%%CITATION = PHLTA,B116,335;%%
\bibitem [{\citenamefont {Albrecht}\ and\ \citenamefont
  {Steinhardt}(1982)}]{Albrecht:1982wi}%
  \BibitemOpen
  \bibfield  {author} {\bibinfo {author} {\bibfnamefont {A.}~\bibnamefont
  {Albrecht}}\ and\ \bibinfo {author} {\bibfnamefont {P.~J.}\ \bibnamefont
  {Steinhardt}},\ }\href {\doibase 10.1103/PhysRevLett.48.1220} {\bibfield
  {journal} {\bibinfo  {journal} {Phys.Rev.Lett.}\ }\textbf {\bibinfo {volume}
  {48}},\ \bibinfo {pages} {1220} (\bibinfo {year} {1982})}\BibitemShut
  {NoStop}%
%%CITATION = PRLTA,48,1220;%%
\bibitem [{\citenamefont {Aad}\ \emph {et~al.}(2012)\citenamefont {Aad} \emph
  {et~al.}}]{Aad:2012tfa}%
  \BibitemOpen
  \bibfield  {author} {\bibinfo {author} {\bibfnamefont {G.}~\bibnamefont
  {Aad}} \emph {et~al.} (\bibinfo {collaboration} {ATLAS Collaboration}),\
  }\href {\doibase 10.1016/j.physletb.2012.08.020} {\bibfield  {journal}
  {\bibinfo  {journal} {Phys.Lett.}\ }\textbf {\bibinfo {volume} {B716}},\
  \bibinfo {pages} {1} (\bibinfo {year} {2012})},\ \Eprint
  {http://arxiv.org/abs/1207.7214} {arXiv:1207.7214 [hep-ex]} \BibitemShut
  {NoStop}%
%%CITATION = ARXIV:1207.7214;%%
\bibitem [{\citenamefont {Chatrchyan}\ \emph {et~al.}(2012)\citenamefont
  {Chatrchyan} \emph {et~al.}}]{Chatrchyan:2012ufa}%
  \BibitemOpen
  \bibfield  {author} {\bibinfo {author} {\bibfnamefont {S.}~\bibnamefont
  {Chatrchyan}} \emph {et~al.} (\bibinfo {collaboration} {CMS Collaboration}),\
  }\href {\doibase 10.1016/j.physletb.2012.08.021} {\bibfield  {journal}
  {\bibinfo  {journal} {Phys.Lett.}\ }\textbf {\bibinfo {volume} {B716}},\
  \bibinfo {pages} {30} (\bibinfo {year} {2012})},\ \Eprint
  {http://arxiv.org/abs/1207.7235} {arXiv:1207.7235 [hep-ex]} \BibitemShut
  {NoStop}%
%%CITATION = ARXIV:1207.7235;%%
\bibitem [{\citenamefont {Agullo}\ and\ \citenamefont
  {Parker}(2011)}]{Agullo:2010ws}%
  \BibitemOpen
  \bibfield  {author} {\bibinfo {author} {\bibfnamefont {I.}~\bibnamefont
  {Agullo}}\ and\ \bibinfo {author} {\bibfnamefont {L.}~\bibnamefont
  {Parker}},\ }\href {\doibase 10.1103/PhysRevD.83.063526} {\bibfield
  {journal} {\bibinfo  {journal} {Phys.Rev.}\ }\textbf {\bibinfo {volume}
  {D83}},\ \bibinfo {pages} {063526} (\bibinfo {year} {2011})},\ \Eprint
  {http://arxiv.org/abs/1010.5766} {arXiv:1010.5766 [astro-ph.CO]} \BibitemShut
  {NoStop}%
%%CITATION = ARXIV:1010.5766;%%
\bibitem [{\citenamefont {Ganc}(2011)}]{Ganc:2011dy}%
  \BibitemOpen
  \bibfield  {author} {\bibinfo {author} {\bibfnamefont {J.}~\bibnamefont
  {Ganc}},\ }\href {\doibase 10.1103/PhysRevD.84.063514} {\bibfield  {journal}
  {\bibinfo  {journal} {Phys.Rev.}\ }\textbf {\bibinfo {volume} {D84}},\
  \bibinfo {pages} {063514} (\bibinfo {year} {2011})},\ \Eprint
  {http://arxiv.org/abs/1104.0244} {arXiv:1104.0244 [astro-ph.CO]} \BibitemShut
  {NoStop}%
%%CITATION = ARXIV:1104.0244;%%
\bibitem [{\citenamefont {Ganc}\ and\ \citenamefont
  {Komatsu}(2012)}]{Ganc:2012ae}%
  \BibitemOpen
  \bibfield  {author} {\bibinfo {author} {\bibfnamefont {J.}~\bibnamefont
  {Ganc}}\ and\ \bibinfo {author} {\bibfnamefont {E.}~\bibnamefont {Komatsu}},\
  }\href {\doibase 10.1103/PhysRevD.86.023518} {\bibfield  {journal} {\bibinfo
  {journal} {Phys.Rev.}\ }\textbf {\bibinfo {volume} {D86}},\ \bibinfo {pages}
  {023518} (\bibinfo {year} {2012})},\ \Eprint {http://arxiv.org/abs/1204.4241}
  {arXiv:1204.4241 [astro-ph.CO]} \BibitemShut {NoStop}%
%%CITATION = ARXIV:1204.4241;%%
\bibitem [{\citenamefont {Namjoo}\ \emph {et~al.}(2013)\citenamefont {Namjoo},
  \citenamefont {Firouzjahi},\ and\ \citenamefont {Sasaki}}]{Namjoo:2012aa}%
  \BibitemOpen
  \bibfield  {author} {\bibinfo {author} {\bibfnamefont {M.~H.}\ \bibnamefont
  {Namjoo}}, \bibinfo {author} {\bibfnamefont {H.}~\bibnamefont {Firouzjahi}},
  \ and\ \bibinfo {author} {\bibfnamefont {M.}~\bibnamefont {Sasaki}},\ }\href
  {\doibase 10.1209/0295-5075/101/39001} {\bibfield  {journal} {\bibinfo
  {journal} {Europhys.Lett.}\ }\textbf {\bibinfo {volume} {101}},\ \bibinfo
  {pages} {39001} (\bibinfo {year} {2013})},\ \Eprint
  {http://arxiv.org/abs/1210.3692} {arXiv:1210.3692 [astro-ph.CO]} \BibitemShut
  {NoStop}%
%%CITATION = ARXIV:1210.3692;%%
\bibitem [{\citenamefont {Martin}\ \emph {et~al.}(2013)\citenamefont {Martin},
  \citenamefont {Motohashi},\ and\ \citenamefont {Suyama}}]{Martin:2012pe}%
  \BibitemOpen
  \bibfield  {author} {\bibinfo {author} {\bibfnamefont {J.}~\bibnamefont
  {Martin}}, \bibinfo {author} {\bibfnamefont {H.}~\bibnamefont {Motohashi}}, \
  and\ \bibinfo {author} {\bibfnamefont {T.}~\bibnamefont {Suyama}},\ }\href
  {\doibase 10.1103/PhysRevD.87.023514} {\bibfield  {journal} {\bibinfo
  {journal} {Phys.Rev.}\ }\textbf {\bibinfo {volume} {D87}},\ \bibinfo {pages}
  {023514} (\bibinfo {year} {2013})},\ \Eprint {http://arxiv.org/abs/1211.0083}
  {arXiv:1211.0083 [astro-ph.CO]} \BibitemShut {NoStop}%
%%CITATION = ARXIV:1211.0083;%%
\bibitem [{\citenamefont {Chen}\ \emph
  {et~al.}(2013{\natexlab{a}})\citenamefont {Chen}, \citenamefont {Firouzjahi},
  \citenamefont {Namjoo},\ and\ \citenamefont {Sasaki}}]{Chen:2013aj}%
  \BibitemOpen
  \bibfield  {author} {\bibinfo {author} {\bibfnamefont {X.}~\bibnamefont
  {Chen}}, \bibinfo {author} {\bibfnamefont {H.}~\bibnamefont {Firouzjahi}},
  \bibinfo {author} {\bibfnamefont {M.~H.}\ \bibnamefont {Namjoo}}, \ and\
  \bibinfo {author} {\bibfnamefont {M.}~\bibnamefont {Sasaki}},\ }\href
  {\doibase 10.1209/0295-5075/102/59001} {\bibfield  {journal} {\bibinfo
  {journal} {Europhys.Lett.}\ }\textbf {\bibinfo {volume} {102}},\ \bibinfo
  {pages} {59001} (\bibinfo {year} {2013}{\natexlab{a}})},\ \Eprint
  {http://arxiv.org/abs/1301.5699} {arXiv:1301.5699 [hep-th]} \BibitemShut
  {NoStop}%
%%CITATION = ARXIV:1301.5699;%%
\bibitem [{\citenamefont {Chen}\ \emph
  {et~al.}(2013{\natexlab{b}})\citenamefont {Chen}, \citenamefont {Firouzjahi},
  \citenamefont {Komatsu}, \citenamefont {Namjoo},\ and\ \citenamefont
  {Sasaki}}]{Chen:2013eea}%
  \BibitemOpen
  \bibfield  {author} {\bibinfo {author} {\bibfnamefont {X.}~\bibnamefont
  {Chen}}, \bibinfo {author} {\bibfnamefont {H.}~\bibnamefont {Firouzjahi}},
  \bibinfo {author} {\bibfnamefont {E.}~\bibnamefont {Komatsu}}, \bibinfo
  {author} {\bibfnamefont {M.~H.}\ \bibnamefont {Namjoo}}, \ and\ \bibinfo
  {author} {\bibfnamefont {M.}~\bibnamefont {Sasaki}},\ }\href@noop {} {\
  (\bibinfo {year} {2013}{\natexlab{b}})},\ \Eprint
  {http://arxiv.org/abs/1308.5341} {arXiv:1308.5341 [astro-ph.CO]} \BibitemShut
  {NoStop}%
%%CITATION = ARXIV:1308.5341;%%
\bibitem [{\citenamefont {Maldacena}(2003)}]{Maldacena:2002vr}%
  \BibitemOpen
  \bibfield  {author} {\bibinfo {author} {\bibfnamefont {J.~M.}\ \bibnamefont
  {Maldacena}},\ }\href@noop {} {\bibfield  {journal} {\bibinfo  {journal}
  {JHEP}\ }\textbf {\bibinfo {volume} {0305}},\ \bibinfo {pages} {013}
  (\bibinfo {year} {2003})},\ \Eprint {http://arxiv.org/abs/astro-ph/0210603}
  {arXiv:astro-ph/0210603 [astro-ph]} \BibitemShut {NoStop}%
%%CITATION = ASTRO-PH/0210603;%%
\bibitem [{\citenamefont {Creminelli}\ and\ \citenamefont
  {Zaldarriaga}(2004)}]{Creminelli:2004yq}%
  \BibitemOpen
  \bibfield  {author} {\bibinfo {author} {\bibfnamefont {P.}~\bibnamefont
  {Creminelli}}\ and\ \bibinfo {author} {\bibfnamefont {M.}~\bibnamefont
  {Zaldarriaga}},\ }\href {\doibase 10.1088/1475-7516/2004/10/006} {\bibfield
  {journal} {\bibinfo  {journal} {JCAP}\ }\textbf {\bibinfo {volume} {0410}},\
  \bibinfo {pages} {006} (\bibinfo {year} {2004})},\ \Eprint
  {http://arxiv.org/abs/astro-ph/0407059} {arXiv:astro-ph/0407059 [astro-ph]}
  \BibitemShut {NoStop}%
%%CITATION = ASTRO-PH/0407059;%%
\bibitem [{\citenamefont {Ganc}\ and\ \citenamefont
  {Komatsu}(2010)}]{Ganc:2010ff}%
  \BibitemOpen
  \bibfield  {author} {\bibinfo {author} {\bibfnamefont {J.}~\bibnamefont
  {Ganc}}\ and\ \bibinfo {author} {\bibfnamefont {E.}~\bibnamefont {Komatsu}},\
  }\href {\doibase 10.1088/1475-7516/2010/12/009} {\bibfield  {journal}
  {\bibinfo  {journal} {JCAP}\ }\textbf {\bibinfo {volume} {1012}},\ \bibinfo
  {pages} {009} (\bibinfo {year} {2010})},\ \Eprint
  {http://arxiv.org/abs/1006.5457} {arXiv:1006.5457 [astro-ph.CO]} \BibitemShut
  {NoStop}%
%%CITATION = ARXIV:1006.5457;%%
\bibitem [{\citenamefont {Renaux-Petel}(2010)}]{RenauxPetel:2010ty}%
  \BibitemOpen
  \bibfield  {author} {\bibinfo {author} {\bibfnamefont {S.}~\bibnamefont
  {Renaux-Petel}},\ }\href {\doibase 10.1088/1475-7516/2010/10/020} {\bibfield
  {journal} {\bibinfo  {journal} {JCAP}\ }\textbf {\bibinfo {volume} {1010}},\
  \bibinfo {pages} {020} (\bibinfo {year} {2010})},\ \Eprint
  {http://arxiv.org/abs/1008.0260} {arXiv:1008.0260 [astro-ph.CO]} \BibitemShut
  {NoStop}%
%%CITATION = ARXIV:1008.0260;%%
\bibitem [{\citenamefont {Gruzinov}(2005)}]{Gruzinov:2004jx}%
  \BibitemOpen
  \bibfield  {author} {\bibinfo {author} {\bibfnamefont {A.}~\bibnamefont
  {Gruzinov}},\ }\href {\doibase 10.1103/PhysRevD.71.027301} {\bibfield
  {journal} {\bibinfo  {journal} {Phys.Rev.}\ }\textbf {\bibinfo {volume}
  {D71}},\ \bibinfo {pages} {027301} (\bibinfo {year} {2005})},\ \Eprint
  {http://arxiv.org/abs/astro-ph/0406129} {arXiv:astro-ph/0406129 [astro-ph]}
  \BibitemShut {NoStop}%
%%CITATION = ASTRO-PH/0406129;%%
\bibitem [{\citenamefont {Kobayashi}\ \emph {et~al.}(2010)\citenamefont
  {Kobayashi}, \citenamefont {Yamaguchi},\ and\ \citenamefont
  {Yokoyama}}]{Kobayashi:2010cm}%
  \BibitemOpen
  \bibfield  {author} {\bibinfo {author} {\bibfnamefont {T.}~\bibnamefont
  {Kobayashi}}, \bibinfo {author} {\bibfnamefont {M.}~\bibnamefont
  {Yamaguchi}}, \ and\ \bibinfo {author} {\bibfnamefont {J.}~\bibnamefont
  {Yokoyama}},\ }\href {\doibase 10.1103/PhysRevLett.105.231302} {\bibfield
  {journal} {\bibinfo  {journal} {Phys.Rev.Lett.}\ }\textbf {\bibinfo {volume}
  {105}},\ \bibinfo {pages} {231302} (\bibinfo {year} {2010})},\ \Eprint
  {http://arxiv.org/abs/1008.0603} {arXiv:1008.0603 [hep-th]} \BibitemShut
  {NoStop}%
%%CITATION = ARXIV:1008.0603;%%
\bibitem [{\citenamefont {Unnikrishnan}\ and\ \citenamefont
  {Shankaranarayanan}(2013)}]{Unnikrishnan:2013rka}%
  \BibitemOpen
  \bibfield  {author} {\bibinfo {author} {\bibfnamefont {S.}~\bibnamefont
  {Unnikrishnan}}\ and\ \bibinfo {author} {\bibfnamefont {S.}~\bibnamefont
  {Shankaranarayanan}},\ }\href@noop {} {\  (\bibinfo {year} {2013})},\ \Eprint
  {http://arxiv.org/abs/1311.0177} {arXiv:1311.0177 [astro-ph.CO]} \BibitemShut
  {NoStop}%
%%CITATION = ARXIV:1311.0177;%%
\bibitem [{\citenamefont {Dolgov}\ and\ \citenamefont
  {Linde}(1982)}]{Dolgov:1982th}%
  \BibitemOpen
  \bibfield  {author} {\bibinfo {author} {\bibfnamefont {A.}~\bibnamefont
  {Dolgov}}\ and\ \bibinfo {author} {\bibfnamefont {A.~D.}\ \bibnamefont
  {Linde}},\ }\href {\doibase 10.1016/0370-2693(82)90292-1} {\bibfield
  {journal} {\bibinfo  {journal} {Phys.Lett.}\ }\textbf {\bibinfo {volume}
  {B116}},\ \bibinfo {pages} {329} (\bibinfo {year} {1982})}\BibitemShut
  {NoStop}%
%%CITATION = PHLTA,B116,329;%%
\bibitem [{\citenamefont {Traschen}\ and\ \citenamefont
  {Brandenberger}(1990)}]{Traschen:1990sw}%
  \BibitemOpen
  \bibfield  {author} {\bibinfo {author} {\bibfnamefont {J.~H.}\ \bibnamefont
  {Traschen}}\ and\ \bibinfo {author} {\bibfnamefont {R.~H.}\ \bibnamefont
  {Brandenberger}},\ }\href {\doibase 10.1103/PhysRevD.42.2491} {\bibfield
  {journal} {\bibinfo  {journal} {Phys.Rev.}\ }\textbf {\bibinfo {volume}
  {D42}},\ \bibinfo {pages} {2491} (\bibinfo {year} {1990})}\BibitemShut
  {NoStop}%
%%CITATION = PHRVA,D42,2491;%%
\bibitem [{\citenamefont {Enqvist}\ and\ \citenamefont
  {Rusak}(2013)}]{Enqvist:2012vx}%
  \BibitemOpen
  \bibfield  {author} {\bibinfo {author} {\bibfnamefont {K.}~\bibnamefont
  {Enqvist}}\ and\ \bibinfo {author} {\bibfnamefont {S.}~\bibnamefont
  {Rusak}},\ }\href {\doibase 10.1088/1475-7516/2013/03/017} {\bibfield
  {journal} {\bibinfo  {journal} {JCAP}\ }\textbf {\bibinfo {volume} {1303}},\
  \bibinfo {pages} {017} (\bibinfo {year} {2013})},\ \Eprint
  {http://arxiv.org/abs/1210.2192} {arXiv:1210.2192 [astro-ph.CO]} \BibitemShut
  {NoStop}%
%%CITATION = ARXIV:1210.2192;%%
\bibitem [{\citenamefont {Braden}\ \emph {et~al.}(2010)\citenamefont {Braden},
  \citenamefont {Kofman},\ and\ \citenamefont {Barnaby}}]{Braden:2010wd}%
  \BibitemOpen
  \bibfield  {author} {\bibinfo {author} {\bibfnamefont {J.}~\bibnamefont
  {Braden}}, \bibinfo {author} {\bibfnamefont {L.}~\bibnamefont {Kofman}}, \
  and\ \bibinfo {author} {\bibfnamefont {N.}~\bibnamefont {Barnaby}},\ }\href
  {\doibase 10.1088/1475-7516/2010/07/016} {\bibfield  {journal} {\bibinfo
  {journal} {JCAP}\ }\textbf {\bibinfo {volume} {1007}},\ \bibinfo {pages}
  {016} (\bibinfo {year} {2010})},\ \Eprint {http://arxiv.org/abs/1005.2196}
  {arXiv:1005.2196 [hep-th]} \BibitemShut {NoStop}%
%%CITATION = ARXIV:1005.2196;%%
\bibitem [{\citenamefont {Barnaby}\ \emph {et~al.}(2009)\citenamefont
  {Barnaby}, \citenamefont {Bond}, \citenamefont {Huang},\ and\ \citenamefont
  {Kofman}}]{Barnaby:2009wr}%
  \BibitemOpen
  \bibfield  {author} {\bibinfo {author} {\bibfnamefont {N.}~\bibnamefont
  {Barnaby}}, \bibinfo {author} {\bibfnamefont {J.~R.}\ \bibnamefont {Bond}},
  \bibinfo {author} {\bibfnamefont {Z.}~\bibnamefont {Huang}}, \ and\ \bibinfo
  {author} {\bibfnamefont {L.}~\bibnamefont {Kofman}},\ }\href {\doibase
  10.1088/1475-7516/2009/12/021} {\bibfield  {journal} {\bibinfo  {journal}
  {JCAP}\ }\textbf {\bibinfo {volume} {0912}},\ \bibinfo {pages} {021}
  (\bibinfo {year} {2009})},\ \Eprint {http://arxiv.org/abs/0909.0503}
  {arXiv:0909.0503 [hep-th]} \BibitemShut {NoStop}%
%%CITATION = ARXIV:0909.0503;%%
\bibitem [{\citenamefont {Leung}\ \emph {et~al.}(2012)\citenamefont {Leung},
  \citenamefont {Tarrant}, \citenamefont {Byrnes},\ and\ \citenamefont
  {Copeland}}]{Leung:2012ve}%
  \BibitemOpen
  \bibfield  {author} {\bibinfo {author} {\bibfnamefont {G.}~\bibnamefont
  {Leung}}, \bibinfo {author} {\bibfnamefont {E.~R.}\ \bibnamefont {Tarrant}},
  \bibinfo {author} {\bibfnamefont {C.~T.}\ \bibnamefont {Byrnes}}, \ and\
  \bibinfo {author} {\bibfnamefont {E.~J.}\ \bibnamefont {Copeland}},\ }\href
  {\doibase 10.1088/1475-7516/2012/09/008} {\bibfield  {journal} {\bibinfo
  {journal} {JCAP}\ }\textbf {\bibinfo {volume} {1209}},\ \bibinfo {pages}
  {008} (\bibinfo {year} {2012})},\ \Eprint {http://arxiv.org/abs/1206.5196}
  {arXiv:1206.5196 [astro-ph.CO]} \BibitemShut {NoStop}%
%%CITATION = ARXIV:1206.5196;%%
\bibitem [{\citenamefont {Lyth}\ and\ \citenamefont
  {Wands}(2002)}]{Lyth:2001nq}%
  \BibitemOpen
  \bibfield  {author} {\bibinfo {author} {\bibfnamefont {D.~H.}\ \bibnamefont
  {Lyth}}\ and\ \bibinfo {author} {\bibfnamefont {D.}~\bibnamefont {Wands}},\
  }\href {\doibase 10.1016/S0370-2693(01)01366-1} {\bibfield  {journal}
  {\bibinfo  {journal} {Phys.Lett.}\ }\textbf {\bibinfo {volume} {B524}},\
  \bibinfo {pages} {5} (\bibinfo {year} {2002})},\ \Eprint
  {http://arxiv.org/abs/hep-ph/0110002} {arXiv:hep-ph/0110002 [hep-ph]}
  \BibitemShut {NoStop}%
%%CITATION = HEP-PH/0110002;%%
\bibitem [{\citenamefont {Lyth}\ and\ \citenamefont
  {Rodriguez}(2005)}]{Lyth:2005fi}%
  \BibitemOpen
  \bibfield  {author} {\bibinfo {author} {\bibfnamefont {D.~H.}\ \bibnamefont
  {Lyth}}\ and\ \bibinfo {author} {\bibfnamefont {Y.}~\bibnamefont
  {Rodriguez}},\ }\href {\doibase 10.1103/PhysRevLett.95.121302} {\bibfield
  {journal} {\bibinfo  {journal} {Phys.Rev.Lett.}\ }\textbf {\bibinfo {volume}
  {95}},\ \bibinfo {pages} {121302} (\bibinfo {year} {2005})},\ \Eprint
  {http://arxiv.org/abs/astro-ph/0504045} {arXiv:astro-ph/0504045 [astro-ph]}
  \BibitemShut {NoStop}%
%%CITATION = ASTRO-PH/0504045;%%
\bibitem [{\citenamefont {Sasaki}\ \emph {et~al.}(2006)\citenamefont {Sasaki},
  \citenamefont {Valiviita},\ and\ \citenamefont {Wands}}]{Sasaki:2006kq}%
  \BibitemOpen
  \bibfield  {author} {\bibinfo {author} {\bibfnamefont {M.}~\bibnamefont
  {Sasaki}}, \bibinfo {author} {\bibfnamefont {J.}~\bibnamefont {Valiviita}}, \
  and\ \bibinfo {author} {\bibfnamefont {D.}~\bibnamefont {Wands}},\ }\href
  {\doibase 10.1103/PhysRevD.74.103003} {\bibfield  {journal} {\bibinfo
  {journal} {Phys.Rev.}\ }\textbf {\bibinfo {volume} {D74}},\ \bibinfo {pages}
  {103003} (\bibinfo {year} {2006})},\ \Eprint
  {http://arxiv.org/abs/astro-ph/0607627} {arXiv:astro-ph/0607627 [astro-ph]}
  \BibitemShut {NoStop}%
%%CITATION = ASTRO-PH/0607627;%%
\bibitem [{\citenamefont {Kawasaki}\ \emph {et~al.}(2011)\citenamefont
  {Kawasaki}, \citenamefont {Kobayashi},\ and\ \citenamefont
  {Takahashi}}]{Kawasaki:2011pd}%
  \BibitemOpen
  \bibfield  {author} {\bibinfo {author} {\bibfnamefont {M.}~\bibnamefont
  {Kawasaki}}, \bibinfo {author} {\bibfnamefont {T.}~\bibnamefont {Kobayashi}},
  \ and\ \bibinfo {author} {\bibfnamefont {F.}~\bibnamefont {Takahashi}},\
  }\href {\doibase 10.1103/PhysRevD.84.123506, 10.1103/PhysRevD.85.029905}
  {\bibfield  {journal} {\bibinfo  {journal} {Phys.Rev.}\ }\textbf {\bibinfo
  {volume} {D84}},\ \bibinfo {pages} {123506} (\bibinfo {year} {2011})},\
  \Eprint {http://arxiv.org/abs/1107.6011} {arXiv:1107.6011 [astro-ph.CO]}
  \BibitemShut {NoStop}%
%%CITATION = ARXIV:1107.6011;%%
\bibitem [{\citenamefont {Kawasaki}\ \emph {et~al.}(2013)\citenamefont
  {Kawasaki}, \citenamefont {Kobayashi},\ and\ \citenamefont
  {Takahashi}}]{Kawasaki:2012gg}%
  \BibitemOpen
  \bibfield  {author} {\bibinfo {author} {\bibfnamefont {M.}~\bibnamefont
  {Kawasaki}}, \bibinfo {author} {\bibfnamefont {T.}~\bibnamefont {Kobayashi}},
  \ and\ \bibinfo {author} {\bibfnamefont {F.}~\bibnamefont {Takahashi}},\
  }\href {\doibase 10.1088/1475-7516/2013/03/016} {\bibfield  {journal}
  {\bibinfo  {journal} {JCAP}\ }\textbf {\bibinfo {volume} {1303}},\ \bibinfo
  {pages} {016} (\bibinfo {year} {2013})},\ \Eprint
  {http://arxiv.org/abs/1210.6595} {arXiv:1210.6595 [astro-ph.CO]} \BibitemShut
  {NoStop}%
%%CITATION = ARXIV:1210.6595;%%
\bibitem [{\citenamefont {Fonseca}\ and\ \citenamefont
  {Wands}(2012)}]{Fonseca:2012cj}%
  \BibitemOpen
  \bibfield  {author} {\bibinfo {author} {\bibfnamefont {J.}~\bibnamefont
  {Fonseca}}\ and\ \bibinfo {author} {\bibfnamefont {D.}~\bibnamefont
  {Wands}},\ }\href {\doibase 10.1088/1475-7516/2012/06/028} {\bibfield
  {journal} {\bibinfo  {journal} {JCAP}\ }\textbf {\bibinfo {volume} {1206}},\
  \bibinfo {pages} {028} (\bibinfo {year} {2012})},\ \Eprint
  {http://arxiv.org/abs/1204.3443} {arXiv:1204.3443 [astro-ph.CO]} \BibitemShut
  {NoStop}%
%%CITATION = ARXIV:1204.3443;%%
\bibitem [{\citenamefont {Linde}\ and\ \citenamefont
  {Mukhanov}(1997)}]{Linde:1996gt}%
  \BibitemOpen
  \bibfield  {author} {\bibinfo {author} {\bibfnamefont {A.~D.}\ \bibnamefont
  {Linde}}\ and\ \bibinfo {author} {\bibfnamefont {V.~F.}\ \bibnamefont
  {Mukhanov}},\ }\href {\doibase 10.1103/PhysRevD.56.R535} {\bibfield
  {journal} {\bibinfo  {journal} {Phys.Rev.}\ }\textbf {\bibinfo {volume}
  {D56}},\ \bibinfo {pages} {535} (\bibinfo {year} {1997})},\ \Eprint
  {http://arxiv.org/abs/astro-ph/9610219} {arXiv:astro-ph/9610219 [astro-ph]}
  \BibitemShut {NoStop}%
%%CITATION = ASTRO-PH/9610219;%%
\bibitem [{\citenamefont {Enqvist}\ and\ \citenamefont
  {Sloth}(2002)}]{Enqvist:2001zp}%
  \BibitemOpen
  \bibfield  {author} {\bibinfo {author} {\bibfnamefont {K.}~\bibnamefont
  {Enqvist}}\ and\ \bibinfo {author} {\bibfnamefont {M.~S.}\ \bibnamefont
  {Sloth}},\ }\href {\doibase 10.1016/S0550-3213(02)00043-3} {\bibfield
  {journal} {\bibinfo  {journal} {Nucl.Phys.}\ }\textbf {\bibinfo {volume}
  {B626}},\ \bibinfo {pages} {395} (\bibinfo {year} {2002})},\ \Eprint
  {http://arxiv.org/abs/hep-ph/0109214} {arXiv:hep-ph/0109214 [hep-ph]}
  \BibitemShut {NoStop}%
%%CITATION = HEP-PH/0109214;%%
\bibitem [{\citenamefont {Lyth}\ \emph {et~al.}(2003)\citenamefont {Lyth},
  \citenamefont {Ungarelli},\ and\ \citenamefont {Wands}}]{Lyth:2002my}%
  \BibitemOpen
  \bibfield  {author} {\bibinfo {author} {\bibfnamefont {D.~H.}\ \bibnamefont
  {Lyth}}, \bibinfo {author} {\bibfnamefont {C.}~\bibnamefont {Ungarelli}}, \
  and\ \bibinfo {author} {\bibfnamefont {D.}~\bibnamefont {Wands}},\ }\href
  {\doibase 10.1103/PhysRevD.67.023503} {\bibfield  {journal} {\bibinfo
  {journal} {Phys.Rev.}\ }\textbf {\bibinfo {volume} {D67}},\ \bibinfo {pages}
  {023503} (\bibinfo {year} {2003})},\ \Eprint
  {http://arxiv.org/abs/astro-ph/0208055} {arXiv:astro-ph/0208055 [astro-ph]}
  \BibitemShut {NoStop}%
%%CITATION = ASTRO-PH/0208055;%%
\bibitem [{\citenamefont {Assadullahi}\ \emph {et~al.}(2007)\citenamefont
  {Assadullahi}, \citenamefont {Valiviita},\ and\ \citenamefont
  {Wands}}]{Assadullahi:2007uw}%
  \BibitemOpen
  \bibfield  {author} {\bibinfo {author} {\bibfnamefont {H.}~\bibnamefont
  {Assadullahi}}, \bibinfo {author} {\bibfnamefont {J.}~\bibnamefont
  {Valiviita}}, \ and\ \bibinfo {author} {\bibfnamefont {D.}~\bibnamefont
  {Wands}},\ }\href {\doibase 10.1103/PhysRevD.76.103003} {\bibfield  {journal}
  {\bibinfo  {journal} {Phys.Rev.}\ }\textbf {\bibinfo {volume} {D76}},\
  \bibinfo {pages} {103003} (\bibinfo {year} {2007})},\ \Eprint
  {http://arxiv.org/abs/0708.0223} {arXiv:0708.0223 [hep-ph]} \BibitemShut
  {NoStop}%
%%CITATION = ARXIV:0708.0223;%%
\bibitem [{\citenamefont {Kim}\ \emph {et~al.}(2010)\citenamefont {Kim},
  \citenamefont {Liddle},\ and\ \citenamefont {Seery}}]{Kim:2010ud}%
  \BibitemOpen
  \bibfield  {author} {\bibinfo {author} {\bibfnamefont {S.~A.}\ \bibnamefont
  {Kim}}, \bibinfo {author} {\bibfnamefont {A.~R.}\ \bibnamefont {Liddle}}, \
  and\ \bibinfo {author} {\bibfnamefont {D.}~\bibnamefont {Seery}},\ }\href
  {\doibase 10.1103/PhysRevLett.105.181302} {\bibfield  {journal} {\bibinfo
  {journal} {Phys.Rev.Lett.}\ }\textbf {\bibinfo {volume} {105}},\ \bibinfo
  {pages} {181302} (\bibinfo {year} {2010})},\ \Eprint
  {http://arxiv.org/abs/1005.4410} {arXiv:1005.4410 [astro-ph.CO]} \BibitemShut
  {NoStop}%
%%CITATION = ARXIV:1005.4410;%%
\bibitem [{\citenamefont {Watanabe}(2012)}]{Watanabe:2011sm}%
  \BibitemOpen
  \bibfield  {author} {\bibinfo {author} {\bibfnamefont {Y.}~\bibnamefont
  {Watanabe}},\ }\href {\doibase 10.1103/PhysRevD.85.103505} {\bibfield
  {journal} {\bibinfo  {journal} {Phys.Rev.}\ }\textbf {\bibinfo {volume}
  {D85}},\ \bibinfo {pages} {103505} (\bibinfo {year} {2012})},\ \Eprint
  {http://arxiv.org/abs/1110.2462} {arXiv:1110.2462 [astro-ph.CO]} \BibitemShut
  {NoStop}%
%%CITATION = ARXIV:1110.2462;%%
\bibitem [{\citenamefont {Abbott}\ \emph {et~al.}(1982)\citenamefont {Abbott},
  \citenamefont {Farhi},\ and\ \citenamefont {Wise}}]{Abbott:1982hn}%
  \BibitemOpen
  \bibfield  {author} {\bibinfo {author} {\bibfnamefont {L.}~\bibnamefont
  {Abbott}}, \bibinfo {author} {\bibfnamefont {E.}~\bibnamefont {Farhi}}, \
  and\ \bibinfo {author} {\bibfnamefont {M.~B.}\ \bibnamefont {Wise}},\ }\href
  {\doibase 10.1016/0370-2693(82)90867-X} {\bibfield  {journal} {\bibinfo
  {journal} {Phys.Lett.}\ }\textbf {\bibinfo {volume} {B117}},\ \bibinfo
  {pages} {29} (\bibinfo {year} {1982})}\BibitemShut {NoStop}%
%%CITATION = PHLTA,B117,29;%%
\bibitem [{\citenamefont {Kofman}\ \emph {et~al.}(1994)\citenamefont {Kofman},
  \citenamefont {Linde},\ and\ \citenamefont {Starobinsky}}]{Kofman:1994rk}%
  \BibitemOpen
  \bibfield  {author} {\bibinfo {author} {\bibfnamefont {L.}~\bibnamefont
  {Kofman}}, \bibinfo {author} {\bibfnamefont {A.~D.}\ \bibnamefont {Linde}}, \
  and\ \bibinfo {author} {\bibfnamefont {A.~A.}\ \bibnamefont {Starobinsky}},\
  }\href {\doibase 10.1103/PhysRevLett.73.3195} {\bibfield  {journal} {\bibinfo
   {journal} {Phys.Rev.Lett.}\ }\textbf {\bibinfo {volume} {73}},\ \bibinfo
  {pages} {3195} (\bibinfo {year} {1994})},\ \Eprint
  {http://arxiv.org/abs/hep-th/9405187} {arXiv:hep-th/9405187 [hep-th]}
  \BibitemShut {NoStop}%
%%CITATION = HEP-TH/9405187;%%
\bibitem [{\citenamefont {Shtanov}\ \emph {et~al.}(1995)\citenamefont
  {Shtanov}, \citenamefont {Traschen},\ and\ \citenamefont
  {Brandenberger}}]{Shtanov:1994ce}%
  \BibitemOpen
  \bibfield  {author} {\bibinfo {author} {\bibfnamefont {Y.}~\bibnamefont
  {Shtanov}}, \bibinfo {author} {\bibfnamefont {J.~H.}\ \bibnamefont
  {Traschen}}, \ and\ \bibinfo {author} {\bibfnamefont {R.~H.}\ \bibnamefont
  {Brandenberger}},\ }\href {\doibase 10.1103/PhysRevD.51.5438} {\bibfield
  {journal} {\bibinfo  {journal} {Phys.Rev.}\ }\textbf {\bibinfo {volume}
  {D51}},\ \bibinfo {pages} {5438} (\bibinfo {year} {1995})},\ \Eprint
  {http://arxiv.org/abs/hep-ph/9407247} {arXiv:hep-ph/9407247 [hep-ph]}
  \BibitemShut {NoStop}%
%%CITATION = HEP-PH/9407247;%%
\bibitem [{\citenamefont {Lawrie}(2002)}]{Lawrie:2002wm}%
  \BibitemOpen
  \bibfield  {author} {\bibinfo {author} {\bibfnamefont {I.~D.}\ \bibnamefont
  {Lawrie}},\ }\href {\doibase 10.1103/PhysRevD.66.041702} {\bibfield
  {journal} {\bibinfo  {journal} {Phys.Rev.}\ }\textbf {\bibinfo {volume}
  {D66}},\ \bibinfo {pages} {041702} (\bibinfo {year} {2002})},\ \Eprint
  {http://arxiv.org/abs/hep-ph/0204184} {arXiv:hep-ph/0204184 [hep-ph]}
  \BibitemShut {NoStop}%
%%CITATION = HEP-PH/0204184;%%
\bibitem [{\citenamefont {Drewes}\ and\ \citenamefont
  {Kang}(2013)}]{Drewes:2013iaa}%
  \BibitemOpen
  \bibfield  {author} {\bibinfo {author} {\bibfnamefont {M.}~\bibnamefont
  {Drewes}}\ and\ \bibinfo {author} {\bibfnamefont {J.~U.}\ \bibnamefont
  {Kang}},\ }\href {\doibase 10.1016/j.nuclphysb.2013.07.009} {\bibfield
  {journal} {\bibinfo  {journal} {Nucl.Phys.}\ }\textbf {\bibinfo {volume}
  {B875}},\ \bibinfo {pages} {315} (\bibinfo {year} {2013})},\ \Eprint
  {http://arxiv.org/abs/1305.0267} {arXiv:1305.0267 [hep-ph]} \BibitemShut
  {NoStop}%
%%CITATION = ARXIV:1305.0267;%%
\bibitem [{\citenamefont {Mazumdar}\ and\ \citenamefont
  {Zaldivar}(2013)}]{Mazumdar:2013gya}%
  \BibitemOpen
  \bibfield  {author} {\bibinfo {author} {\bibfnamefont {A.}~\bibnamefont
  {Mazumdar}}\ and\ \bibinfo {author} {\bibfnamefont {B.}~\bibnamefont
  {Zaldivar}},\ }\href@noop {} {\  (\bibinfo {year} {2013})},\ \Eprint
  {http://arxiv.org/abs/1310.5143} {arXiv:1310.5143 [hep-ph]} \BibitemShut
  {NoStop}%
%%CITATION = ARXIV:1310.5143;%%
\bibitem [{\citenamefont {Sasaki}\ and\ \citenamefont
  {Stewart}(1996)}]{Sasaki:1995aw}%
  \BibitemOpen
  \bibfield  {author} {\bibinfo {author} {\bibfnamefont {M.}~\bibnamefont
  {Sasaki}}\ and\ \bibinfo {author} {\bibfnamefont {E.~D.}\ \bibnamefont
  {Stewart}},\ }\href {\doibase 10.1143/PTP.95.71} {\bibfield  {journal}
  {\bibinfo  {journal} {Prog.Theor.Phys.}\ }\textbf {\bibinfo {volume} {95}},\
  \bibinfo {pages} {71} (\bibinfo {year} {1996})},\ \Eprint
  {http://arxiv.org/abs/astro-ph/9507001} {arXiv:astro-ph/9507001 [astro-ph]}
  \BibitemShut {NoStop}%
%%CITATION = ASTRO-PH/9507001;%%
\bibitem [{\citenamefont {Sasaki}\ and\ \citenamefont
  {Tanaka}(1998)}]{Sasaki:1998ug}%
  \BibitemOpen
  \bibfield  {author} {\bibinfo {author} {\bibfnamefont {M.}~\bibnamefont
  {Sasaki}}\ and\ \bibinfo {author} {\bibfnamefont {T.}~\bibnamefont
  {Tanaka}},\ }\href {\doibase 10.1143/PTP.99.763} {\bibfield  {journal}
  {\bibinfo  {journal} {Prog.Theor.Phys.}\ }\textbf {\bibinfo {volume} {99}},\
  \bibinfo {pages} {763} (\bibinfo {year} {1998})},\ \Eprint
  {http://arxiv.org/abs/gr-qc/9801017} {arXiv:gr-qc/9801017 [gr-qc]}
  \BibitemShut {NoStop}%
%%CITATION = GR-QC/9801017;%%
\bibitem [{\citenamefont {Saffin}(2012)}]{Saffin:2012et}%
  \BibitemOpen
  \bibfield  {author} {\bibinfo {author} {\bibfnamefont {P.~M.}\ \bibnamefont
  {Saffin}},\ }\href {\doibase 10.1088/1475-7516/2012/09/002} {\bibfield
  {journal} {\bibinfo  {journal} {JCAP}\ }\textbf {\bibinfo {volume} {1209}},\
  \bibinfo {pages} {002} (\bibinfo {year} {2012})},\ \Eprint
  {http://arxiv.org/abs/1203.0397} {arXiv:1203.0397 [hep-th]} \BibitemShut
  {NoStop}%
%%CITATION = ARXIV:1203.0397;%%
\bibitem [{\citenamefont {Wands}\ \emph {et~al.}(2000)\citenamefont {Wands},
  \citenamefont {Malik}, \citenamefont {Lyth},\ and\ \citenamefont
  {Liddle}}]{Wands:2000dp}%
  \BibitemOpen
  \bibfield  {author} {\bibinfo {author} {\bibfnamefont {D.}~\bibnamefont
  {Wands}}, \bibinfo {author} {\bibfnamefont {K.~A.}\ \bibnamefont {Malik}},
  \bibinfo {author} {\bibfnamefont {D.~H.}\ \bibnamefont {Lyth}}, \ and\
  \bibinfo {author} {\bibfnamefont {A.~R.}\ \bibnamefont {Liddle}},\ }\href
  {\doibase 10.1103/PhysRevD.62.043527} {\bibfield  {journal} {\bibinfo
  {journal} {Phys.Rev.}\ }\textbf {\bibinfo {volume} {D62}},\ \bibinfo {pages}
  {043527} (\bibinfo {year} {2000})},\ \Eprint
  {http://arxiv.org/abs/astro-ph/0003278} {arXiv:astro-ph/0003278 [astro-ph]}
  \BibitemShut {NoStop}%
%%CITATION = ASTRO-PH/0003278;%%
\bibitem [{\citenamefont {Lyth}\ \emph {et~al.}(2005)\citenamefont {Lyth},
  \citenamefont {Malik},\ and\ \citenamefont {Sasaki}}]{Lyth:2004gb}%
  \BibitemOpen
  \bibfield  {author} {\bibinfo {author} {\bibfnamefont {D.~H.}\ \bibnamefont
  {Lyth}}, \bibinfo {author} {\bibfnamefont {K.~A.}\ \bibnamefont {Malik}}, \
  and\ \bibinfo {author} {\bibfnamefont {M.}~\bibnamefont {Sasaki}},\ }\href
  {\doibase 10.1088/1475-7516/2005/05/004} {\bibfield  {journal} {\bibinfo
  {journal} {JCAP}\ }\textbf {\bibinfo {volume} {0505}},\ \bibinfo {pages}
  {004} (\bibinfo {year} {2005})},\ \Eprint
  {http://arxiv.org/abs/astro-ph/0411220} {arXiv:astro-ph/0411220 [astro-ph]}
  \BibitemShut {NoStop}%
%%CITATION = ASTRO-PH/0411220;%%
\bibitem [{\citenamefont {Seery}\ and\ \citenamefont
  {Lidsey}(2005)}]{Seery:2005gb}%
  \BibitemOpen
  \bibfield  {author} {\bibinfo {author} {\bibfnamefont {D.}~\bibnamefont
  {Seery}}\ and\ \bibinfo {author} {\bibfnamefont {J.~E.}\ \bibnamefont
  {Lidsey}},\ }\href {\doibase 10.1088/1475-7516/2005/09/011} {\bibfield
  {journal} {\bibinfo  {journal} {JCAP}\ }\textbf {\bibinfo {volume} {0509}},\
  \bibinfo {pages} {011} (\bibinfo {year} {2005})},\ \Eprint
  {http://arxiv.org/abs/astro-ph/0506056} {arXiv:astro-ph/0506056 [astro-ph]}
  \BibitemShut {NoStop}%
%%CITATION = ASTRO-PH/0506056;%%
\bibitem [{\citenamefont {Vernizzi}\ and\ \citenamefont
  {Wands}(2006)}]{Vernizzi:2006ve}%
  \BibitemOpen
  \bibfield  {author} {\bibinfo {author} {\bibfnamefont {F.}~\bibnamefont
  {Vernizzi}}\ and\ \bibinfo {author} {\bibfnamefont {D.}~\bibnamefont
  {Wands}},\ }\href {\doibase 10.1088/1475-7516/2006/05/019} {\bibfield
  {journal} {\bibinfo  {journal} {JCAP}\ }\textbf {\bibinfo {volume} {0605}},\
  \bibinfo {pages} {019} (\bibinfo {year} {2006})},\ \Eprint
  {http://arxiv.org/abs/astro-ph/0603799} {arXiv:astro-ph/0603799 [astro-ph]}
  \BibitemShut {NoStop}%
%%CITATION = ASTRO-PH/0603799;%%
\bibitem [{\citenamefont {Ade}\ \emph {et~al.}(2013{\natexlab{b}})\citenamefont
  {Ade} \emph {et~al.}}]{Ade:2013zuv}%
  \BibitemOpen
  \bibfield  {author} {\bibinfo {author} {\bibfnamefont {P.}~\bibnamefont
  {Ade}} \emph {et~al.} (\bibinfo {collaboration} {Planck Collaboration}),\
  }\href@noop {} {\  (\bibinfo {year} {2013}{\natexlab{b}})},\ \Eprint
  {http://arxiv.org/abs/1303.5076} {arXiv:1303.5076 [astro-ph.CO]} \BibitemShut
  {NoStop}%
%%CITATION = ARXIV:1303.5076;%%
\bibitem [{\citenamefont {Ade}\ \emph {et~al.}(2013{\natexlab{c}})\citenamefont
  {Ade} \emph {et~al.}}]{Ade:2013ydc}%
  \BibitemOpen
  \bibfield  {author} {\bibinfo {author} {\bibfnamefont {P.}~\bibnamefont
  {Ade}} \emph {et~al.} (\bibinfo {collaboration} {Planck Collaboration}),\
  }\href@noop {} {\  (\bibinfo {year} {2013}{\natexlab{c}})},\ \Eprint
  {http://arxiv.org/abs/1303.5084} {arXiv:1303.5084 [astro-ph.CO]} \BibitemShut
  {NoStop}%
%%CITATION = ARXIV:1303.5084;%%
\bibitem [{\citenamefont {Cunha}\ \emph {et~al.}(2010)\citenamefont {Cunha},
  \citenamefont {Huterer},\ and\ \citenamefont {Dore}}]{Cunha:2010zz}%
  \BibitemOpen
  \bibfield  {author} {\bibinfo {author} {\bibfnamefont {C.}~\bibnamefont
  {Cunha}}, \bibinfo {author} {\bibfnamefont {D.}~\bibnamefont {Huterer}}, \
  and\ \bibinfo {author} {\bibfnamefont {O.}~\bibnamefont {Dore}},\ }\href
  {\doibase 10.1103/PhysRevD.82.023004} {\bibfield  {journal} {\bibinfo
  {journal} {Phys.Rev.}\ }\textbf {\bibinfo {volume} {D82}},\ \bibinfo {pages}
  {023004} (\bibinfo {year} {2010})},\ \Eprint {http://arxiv.org/abs/1003.2416}
  {arXiv:1003.2416 [astro-ph.CO]} \BibitemShut {NoStop}%
%%CITATION = ARXIV:1003.2416;%%
\bibitem [{\citenamefont {Lidz}\ \emph {et~al.}(2013)\citenamefont {Lidz},
  \citenamefont {Baxter}, \citenamefont {Adshead},\ and\ \citenamefont
  {Dodelson}}]{Lidz:2013tra}%
  \BibitemOpen
  \bibfield  {author} {\bibinfo {author} {\bibfnamefont {A.}~\bibnamefont
  {Lidz}}, \bibinfo {author} {\bibfnamefont {E.~J.}\ \bibnamefont {Baxter}},
  \bibinfo {author} {\bibfnamefont {P.}~\bibnamefont {Adshead}}, \ and\
  \bibinfo {author} {\bibfnamefont {S.}~\bibnamefont {Dodelson}},\ }\href
  {\doibase 10.1103/PhysRevD.88.023534} {\bibfield  {journal} {\bibinfo
  {journal} {Phys.Rev.}\ }\textbf {\bibinfo {volume} {D88}},\ \bibinfo {pages}
  {023534} (\bibinfo {year} {2013})},\ \Eprint {http://arxiv.org/abs/1304.8049}
  {arXiv:1304.8049 [astro-ph.CO]} \BibitemShut {NoStop}%
%%CITATION = ARXIV:1304.8049;%%
\bibitem [{\citenamefont {Huston}\ and\ \citenamefont
  {Christopherson}(2013)}]{Huston:2013kgl}%
  \BibitemOpen
  \bibfield  {author} {\bibinfo {author} {\bibfnamefont {I.}~\bibnamefont
  {Huston}}\ and\ \bibinfo {author} {\bibfnamefont {A.~J.}\ \bibnamefont
  {Christopherson}},\ }\href@noop {} {\  (\bibinfo {year} {2013})},\ \Eprint
  {http://arxiv.org/abs/1302.4298} {arXiv:1302.4298 [astro-ph.CO]} \BibitemShut
  {NoStop}%
%%CITATION = ARXIV:1302.4298;%%
\bibitem [{\citenamefont {Byrnes}\ \emph {et~al.}(2008)\citenamefont {Byrnes},
  \citenamefont {Choi},\ and\ \citenamefont {Hall}}]{Byrnes:2008wi}%
  \BibitemOpen
  \bibfield  {author} {\bibinfo {author} {\bibfnamefont {C.~T.}\ \bibnamefont
  {Byrnes}}, \bibinfo {author} {\bibfnamefont {K.-Y.}\ \bibnamefont {Choi}}, \
  and\ \bibinfo {author} {\bibfnamefont {L.~M.}\ \bibnamefont {Hall}},\ }\href
  {\doibase 10.1088/1475-7516/2008/10/008} {\bibfield  {journal} {\bibinfo
  {journal} {JCAP}\ }\textbf {\bibinfo {volume} {0810}},\ \bibinfo {pages}
  {008} (\bibinfo {year} {2008})},\ \Eprint {http://arxiv.org/abs/0807.1101}
  {arXiv:0807.1101 [astro-ph]} \BibitemShut {NoStop}%
%%CITATION = ARXIV:0807.1101;%%
\bibitem [{\citenamefont {Elliston}\ \emph {et~al.}(2011)\citenamefont
  {Elliston}, \citenamefont {Mulryne}, \citenamefont {Seery},\ and\
  \citenamefont {Tavakol}}]{Elliston:2011dr}%
  \BibitemOpen
  \bibfield  {author} {\bibinfo {author} {\bibfnamefont {J.}~\bibnamefont
  {Elliston}}, \bibinfo {author} {\bibfnamefont {D.~J.}\ \bibnamefont
  {Mulryne}}, \bibinfo {author} {\bibfnamefont {D.}~\bibnamefont {Seery}}, \
  and\ \bibinfo {author} {\bibfnamefont {R.}~\bibnamefont {Tavakol}},\ }\href
  {\doibase 10.1088/1475-7516/2011/11/005} {\bibfield  {journal} {\bibinfo
  {journal} {JCAP}\ }\textbf {\bibinfo {volume} {1111}},\ \bibinfo {pages}
  {005} (\bibinfo {year} {2011})},\ \Eprint {http://arxiv.org/abs/1106.2153}
  {arXiv:1106.2153 [astro-ph.CO]} \BibitemShut {NoStop}%
%%CITATION = ARXIV:1106.2153;%%
\bibitem [{\citenamefont {Mulryne}\ \emph {et~al.}(2010)\citenamefont
  {Mulryne}, \citenamefont {Seery},\ and\ \citenamefont
  {Wesley}}]{Mulryne:2009kh}%
  \BibitemOpen
  \bibfield  {author} {\bibinfo {author} {\bibfnamefont {D.~J.}\ \bibnamefont
  {Mulryne}}, \bibinfo {author} {\bibfnamefont {D.}~\bibnamefont {Seery}}, \
  and\ \bibinfo {author} {\bibfnamefont {D.}~\bibnamefont {Wesley}},\ }\href
  {\doibase 10.1088/1475-7516/2010/01/024} {\bibfield  {journal} {\bibinfo
  {journal} {JCAP}\ }\textbf {\bibinfo {volume} {1001}},\ \bibinfo {pages}
  {024} (\bibinfo {year} {2010})},\ \Eprint {http://arxiv.org/abs/0909.2256}
  {arXiv:0909.2256 [astro-ph.CO]} \BibitemShut {NoStop}%
%%CITATION = ARXIV:0909.2256;%%
\bibitem [{\citenamefont {Rigopoulos}\ and\ \citenamefont
  {Shellard}(2005)}]{Rigopoulos:2004gr}%
  \BibitemOpen
  \bibfield  {author} {\bibinfo {author} {\bibfnamefont {G.~I.}\ \bibnamefont
  {Rigopoulos}}\ and\ \bibinfo {author} {\bibfnamefont {E.}~\bibnamefont
  {Shellard}},\ }\href {\doibase 10.1088/1475-7516/2005/10/006} {\bibfield
  {journal} {\bibinfo  {journal} {JCAP}\ }\textbf {\bibinfo {volume} {0510}},\
  \bibinfo {pages} {006} (\bibinfo {year} {2005})},\ \Eprint
  {http://arxiv.org/abs/astro-ph/0405185} {arXiv:astro-ph/0405185 [astro-ph]}
  \BibitemShut {NoStop}%
%%CITATION = ASTRO-PH/0405185;%%
\bibitem [{\citenamefont {Rigopoulos}\ \emph {et~al.}(2006)\citenamefont
  {Rigopoulos}, \citenamefont {Shellard},\ and\ \citenamefont {van
  Tent}}]{Rigopoulos:2005xx}%
  \BibitemOpen
  \bibfield  {author} {\bibinfo {author} {\bibfnamefont {G.}~\bibnamefont
  {Rigopoulos}}, \bibinfo {author} {\bibfnamefont {E.}~\bibnamefont
  {Shellard}}, \ and\ \bibinfo {author} {\bibfnamefont {B.}~\bibnamefont {van
  Tent}},\ }\href {\doibase 10.1103/PhysRevD.73.083521} {\bibfield  {journal}
  {\bibinfo  {journal} {Phys.Rev.}\ }\textbf {\bibinfo {volume} {D73}},\
  \bibinfo {pages} {083521} (\bibinfo {year} {2006})},\ \Eprint
  {http://arxiv.org/abs/astro-ph/0504508} {arXiv:astro-ph/0504508 [astro-ph]}
  \BibitemShut {NoStop}%
%%CITATION = ASTRO-PH/0504508;%%
\bibitem [{\citenamefont {Leung}\ \emph {et~al.}(2013)\citenamefont {Leung},
  \citenamefont {Tarrant}, \citenamefont {Byrnes},\ and\ \citenamefont
  {Copeland}}]{Leung:2013rza}%
  \BibitemOpen
  \bibfield  {author} {\bibinfo {author} {\bibfnamefont {G.}~\bibnamefont
  {Leung}}, \bibinfo {author} {\bibfnamefont {E.~R.}\ \bibnamefont {Tarrant}},
  \bibinfo {author} {\bibfnamefont {C.~T.}\ \bibnamefont {Byrnes}}, \ and\
  \bibinfo {author} {\bibfnamefont {E.~J.}\ \bibnamefont {Copeland}},\ }\href
  {\doibase 10.1088/1475-7516/2013/08/006} {\bibfield  {journal} {\bibinfo
  {journal} {JCAP}\ }\textbf {\bibinfo {volume} {1308}},\ \bibinfo {pages}
  {006} (\bibinfo {year} {2013})},\ \Eprint {http://arxiv.org/abs/1303.4678}
  {arXiv:1303.4678 [astro-ph.CO]} \BibitemShut {NoStop}%
%%CITATION = ARXIV:1303.4678;%%
\bibitem [{\citenamefont {Abramowitz}\ and\ \citenamefont
  {Stegun}(1965)}]{Abramowitz1965Handbook}%
  \BibitemOpen
  \bibfield  {author} {\bibinfo {author} {\bibfnamefont {M.}~\bibnamefont
  {Abramowitz}}\ and\ \bibinfo {author} {\bibfnamefont {I.~A.}\ \bibnamefont
  {Stegun}},\ }\href@noop {} {\emph {\bibinfo {title} {{Handbook of
  Mathematical Functions: with Formulas, Graphs, and Mathematical Tables}}}},\
  \bibinfo {edition} {1st}\ ed.\ (\bibinfo  {publisher} {Dover Publications},\
  \bibinfo {year} {1965})\BibitemShut {NoStop}%
\bibitem [{\citenamefont {Mulryne}\ \emph {et~al.}(2011)\citenamefont
  {Mulryne}, \citenamefont {Seery},\ and\ \citenamefont
  {Wesley}}]{Mulryne:2010rp}%
  \BibitemOpen
  \bibfield  {author} {\bibinfo {author} {\bibfnamefont {D.~J.}\ \bibnamefont
  {Mulryne}}, \bibinfo {author} {\bibfnamefont {D.}~\bibnamefont {Seery}}, \
  and\ \bibinfo {author} {\bibfnamefont {D.}~\bibnamefont {Wesley}},\ }\href
  {\doibase 10.1088/1475-7516/2011/04/030} {\bibfield  {journal} {\bibinfo
  {journal} {JCAP}\ }\textbf {\bibinfo {volume} {1104}},\ \bibinfo {pages}
  {030} (\bibinfo {year} {2011})},\ \Eprint {http://arxiv.org/abs/1008.3159}
  {arXiv:1008.3159 [astro-ph.CO]} \BibitemShut {NoStop}%
%%CITATION = ARXIV:1008.3159;%%
\bibitem [{\citenamefont {Byrnes}\ and\ \citenamefont
  {Tasinato}(2009)}]{Byrnes:2009qy}%
  \BibitemOpen
  \bibfield  {author} {\bibinfo {author} {\bibfnamefont {C.~T.}\ \bibnamefont
  {Byrnes}}\ and\ \bibinfo {author} {\bibfnamefont {G.}~\bibnamefont
  {Tasinato}},\ }\href {\doibase 10.1088/1475-7516/2009/08/016} {\bibfield
  {journal} {\bibinfo  {journal} {JCAP}\ }\textbf {\bibinfo {volume} {0908}},\
  \bibinfo {pages} {016} (\bibinfo {year} {2009})},\ \Eprint
  {http://arxiv.org/abs/0906.0767} {arXiv:0906.0767 [astro-ph.CO]} \BibitemShut
  {NoStop}%
%%CITATION = ARXIV:0906.0767;%%
\bibitem [{\citenamefont {Dimopoulos}\ \emph {et~al.}(2008)\citenamefont
  {Dimopoulos}, \citenamefont {Kachru}, \citenamefont {McGreevy},\ and\
  \citenamefont {Wacker}}]{Dimopoulos:2005ac}%
  \BibitemOpen
  \bibfield  {author} {\bibinfo {author} {\bibfnamefont {S.}~\bibnamefont
  {Dimopoulos}}, \bibinfo {author} {\bibfnamefont {S.}~\bibnamefont {Kachru}},
  \bibinfo {author} {\bibfnamefont {J.}~\bibnamefont {McGreevy}}, \ and\
  \bibinfo {author} {\bibfnamefont {J.~G.}\ \bibnamefont {Wacker}},\ }\href
  {\doibase 10.1088/1475-7516/2008/08/003} {\bibfield  {journal} {\bibinfo
  {journal} {JCAP}\ }\textbf {\bibinfo {volume} {0808}},\ \bibinfo {pages}
  {003} (\bibinfo {year} {2008})},\ \Eprint
  {http://arxiv.org/abs/hep-th/0507205} {arXiv:hep-th/0507205 [hep-th]}
  \BibitemShut {NoStop}%
%%CITATION = HEP-TH/0507205;%%
\bibitem [{\citenamefont {Liddle}\ \emph {et~al.}(1998)\citenamefont {Liddle},
  \citenamefont {Mazumdar},\ and\ \citenamefont {Schunck}}]{Liddle:1998jc}%
  \BibitemOpen
  \bibfield  {author} {\bibinfo {author} {\bibfnamefont {A.~R.}\ \bibnamefont
  {Liddle}}, \bibinfo {author} {\bibfnamefont {A.}~\bibnamefont {Mazumdar}}, \
  and\ \bibinfo {author} {\bibfnamefont {F.~E.}\ \bibnamefont {Schunck}},\
  }\href {\doibase 10.1103/PhysRevD.58.061301} {\bibfield  {journal} {\bibinfo
  {journal} {Phys.Rev.}\ }\textbf {\bibinfo {volume} {D58}},\ \bibinfo {pages}
  {061301} (\bibinfo {year} {1998})},\ \Eprint
  {http://arxiv.org/abs/astro-ph/9804177} {arXiv:astro-ph/9804177 [astro-ph]}
  \BibitemShut {NoStop}%
%%CITATION = ASTRO-PH/9804177;%%
\bibitem [{\citenamefont {Copeland}\ \emph {et~al.}(1999)\citenamefont
  {Copeland}, \citenamefont {Mazumdar},\ and\ \citenamefont
  {Nunes}}]{Copeland:1999cs}%
  \BibitemOpen
  \bibfield  {author} {\bibinfo {author} {\bibfnamefont {E.~J.}\ \bibnamefont
  {Copeland}}, \bibinfo {author} {\bibfnamefont {A.}~\bibnamefont {Mazumdar}},
  \ and\ \bibinfo {author} {\bibfnamefont {N.}~\bibnamefont {Nunes}},\ }\href
  {\doibase 10.1103/PhysRevD.60.083506} {\bibfield  {journal} {\bibinfo
  {journal} {Phys.Rev.}\ }\textbf {\bibinfo {volume} {D60}},\ \bibinfo {pages}
  {083506} (\bibinfo {year} {1999})},\ \Eprint
  {http://arxiv.org/abs/astro-ph/9904309} {arXiv:astro-ph/9904309 [astro-ph]}
  \BibitemShut {NoStop}%
%%CITATION = ASTRO-PH/9904309;%%
\bibitem [{\citenamefont {Adams}\ \emph {et~al.}(1993)\citenamefont {Adams},
  \citenamefont {Bond}, \citenamefont {Freese}, \citenamefont {Frieman},\ and\
  \citenamefont {Olinto}}]{Adams:1992bn}%
  \BibitemOpen
  \bibfield  {author} {\bibinfo {author} {\bibfnamefont {F.~C.}\ \bibnamefont
  {Adams}}, \bibinfo {author} {\bibfnamefont {J.~R.}\ \bibnamefont {Bond}},
  \bibinfo {author} {\bibfnamefont {K.}~\bibnamefont {Freese}}, \bibinfo
  {author} {\bibfnamefont {J.~A.}\ \bibnamefont {Frieman}}, \ and\ \bibinfo
  {author} {\bibfnamefont {A.~V.}\ \bibnamefont {Olinto}},\ }\href {\doibase
  10.1103/PhysRevD.47.426} {\bibfield  {journal} {\bibinfo  {journal}
  {Phys.Rev.}\ }\textbf {\bibinfo {volume} {D47}},\ \bibinfo {pages} {426}
  (\bibinfo {year} {1993})},\ \Eprint {http://arxiv.org/abs/hep-ph/9207245}
  {arXiv:hep-ph/9207245 [hep-ph]} \BibitemShut {NoStop}%
%%CITATION = HEP-PH/9207245;%%
\bibitem [{\citenamefont {Battefeld}\ and\ \citenamefont
  {Kawai}(2008)}]{Battefeld:2008bu}%
  \BibitemOpen
  \bibfield  {author} {\bibinfo {author} {\bibfnamefont {D.}~\bibnamefont
  {Battefeld}}\ and\ \bibinfo {author} {\bibfnamefont {S.}~\bibnamefont
  {Kawai}},\ }\href {\doibase 10.1103/PhysRevD.77.123507} {\bibfield  {journal}
  {\bibinfo  {journal} {Phys.Rev.}\ }\textbf {\bibinfo {volume} {D77}},\
  \bibinfo {pages} {123507} (\bibinfo {year} {2008})},\ \Eprint
  {http://arxiv.org/abs/0803.0321} {arXiv:0803.0321 [astro-ph]} \BibitemShut
  {NoStop}%
%%CITATION = ARXIV:0803.0321;%%
\bibitem [{\citenamefont {Battefeld}\ \emph {et~al.}(2009)\citenamefont
  {Battefeld}, \citenamefont {Battefeld},\ and\ \citenamefont
  {Giblin}}]{Battefeld:2009xw}%
  \BibitemOpen
  \bibfield  {author} {\bibinfo {author} {\bibfnamefont {D.}~\bibnamefont
  {Battefeld}}, \bibinfo {author} {\bibfnamefont {T.}~\bibnamefont
  {Battefeld}}, \ and\ \bibinfo {author} {\bibfnamefont {J.~T.}\ \bibnamefont
  {Giblin}},\ }\href {\doibase 10.1103/PhysRevD.79.123510} {\bibfield
  {journal} {\bibinfo  {journal} {Phys.Rev.}\ }\textbf {\bibinfo {volume}
  {D79}},\ \bibinfo {pages} {123510} (\bibinfo {year} {2009})},\ \Eprint
  {http://arxiv.org/abs/0904.2778} {arXiv:0904.2778 [astro-ph.CO]} \BibitemShut
  {NoStop}%
%%CITATION = ARXIV:0904.2778;%%
\bibitem [{\citenamefont {Suyama}\ and\ \citenamefont
  {Yamaguchi}(2008)}]{Suyama:2007bg}%
  \BibitemOpen
  \bibfield  {author} {\bibinfo {author} {\bibfnamefont {T.}~\bibnamefont
  {Suyama}}\ and\ \bibinfo {author} {\bibfnamefont {M.}~\bibnamefont
  {Yamaguchi}},\ }\href {\doibase 10.1103/PhysRevD.77.023505} {\bibfield
  {journal} {\bibinfo  {journal} {Phys.Rev.}\ }\textbf {\bibinfo {volume}
  {D77}},\ \bibinfo {pages} {023505} (\bibinfo {year} {2008})},\ \Eprint
  {http://arxiv.org/abs/0709.2545} {arXiv:0709.2545 [astro-ph]} \BibitemShut
  {NoStop}%
%%CITATION = ARXIV:0709.2545;%%
\end{thebibliography}%

\end{document}